# Assessing Consistency and Reproducibility in the Outputs of Large Language Models: Evidence Across Diverse Finance and Accounting Tasks

**September 2025**

Julian Junyan Wang

University College, University of Oxford

julian.wang@univ.ox.ac.uk

Victor Xiaoqi Wang

College of Business, California State University Long Beach

victor.wang@csulb.edu

**Abstract:** This study provides the first comprehensive assessment of consistency and reproducibility in Large Language Model (LLM) outputs in finance and accounting tasks. We evaluate how consistently LLMs produce outputs given identical inputs through extensive experimentation with 50 independent runs across five common tasks: classification, sentiment analysis, summarization, text generation, and prediction. Using three OpenAI models (GPT-3.5-turbo, GPT-4o-mini, and GPT-4o), we generate over 3.4 million outputs using diverse financial source texts and data, covering MD&As, FOMC statements, finance news articles, earnings call transcripts, and financial statements. Our findings reveal substantial but task-dependent consistency, with binary classification and sentiment analysis achieving near-perfect reproducibility, while complex tasks show greater variability. More advanced models do not consistently demonstrate better consistency and reproducibility, with task-specific patterns emerging. LLMs significantly outperform expert human annotators in consistency and maintain high agreement even where human experts significantly disagree. We further find that simple aggregation strategies across 3-5 runs dramatically improve consistency. We also find that aggregation may come with an additional benefit of improved accuracy for sentiment analysis when using newer models. Simulation analysis reveals that despite measurable inconsistency in LLM outputs, downstream statistical inferences remain remarkably robust. These findings address concerns about what we term "G-hacking," the selective reporting of favorable outcomes from multiple generative AI runs, by demonstrating that such risks are relatively low for finance and accounting tasks.

**Assessing Consistency and Reproducibility in the Outputs of Large Language Models: Evidence Across Diverse Finance and Accounting Tasks**

## 1. Introduction

The integration of artificial intelligence, particularly Large Language Models (LLMs), into finance and accounting research has surged dramatically in recent years. These sophisticated models, exemplified by OpenAI's GPT series, have demonstrated remarkable capabilities in analyzing financial text and generating insights from vast amounts of unstructured data (Dong et al. 2024). From classifying financial text (Hansen and Kazinnik 2024), analyzing sentiment (Lopez-Lira and Tang 2024), summarizing corporate disclosures (Kim et al. 2024a), to predicting future earnings or stock returns (E. X. Li et al. 2024), LLMs promise to revolutionize how researchers approach financial text analysis (de Kok 2025) and how investors process information and make investment decisions (Blankespoor et al. 2024).

However, as with any emerging methodological innovation, questions of consistency and reproducibility remain paramount concerns for the research community. The foundation of scientific inquiry rests on the principle that research findings should be reproducible; yet, this fundamental tenet faces a unique challenge when applied to research utilizing LLMs. Unlike deterministic algorithms with fixed outputs for given inputs, LLMs generate outputs through probabilistic token selection processes, meaning that even identical inputs with identical prompts may yield different results across different runs. This inherent stochasticity raises questions about the validity and reproducibility of research findings that rely on LLM-generated outputs.

While a growing body of literature has explored LLMs' performance in finance and accounting applications, surprisingly little attention has been paid to this critical issue of reproducibility. Existing studies have primarily focused on comparing LLMs' performance to traditional methods, evaluating their capabilities in specific tasks, or addressing potential biases in their outputs (Levy 2024, Sarkar and Vafa 2024, Bini et al. 2025). The broader finance and accounting literature has grappled with reproducibility challenges long before the advent of LLMs (Harvey 2019, Menkveld et al. 2024, Gow 2025). However, the fundamental question of whether LLM-based analyses can produce consistent, reproducible results, a prerequisite for their integration into rigorous academic research, remains largely unexplored.

While our study focuses specifically on LLM consistency and reproducibility, it is worth noting that the broader machine learning literature in finance has developed various approaches to address interpretability and validation challenges. Deep learning techniques have shown success in asset pricing (Chen et al. 2024), mortgage risk prediction (Sadhwani et al. 2021), mutual fund performance evaluation (Kaniel et al. 2023), credit rating transitions (Kumar et al. 2025), among many other applications. Giesecke et al. (2025) address interpretability concerns across these applications by developing model-agnostic significance tests for feature importance. Cong and Yang (2024) combine human perception measures with language embeddings to analyze discrimination patterns. These important methodological advances provide valuable frameworks for understanding machine learning models in finance. LLMs, with their probabilistic text generation processes, introduce additional considerations for reproducibility that complement these existing validation approaches, partly motivating our systematic investigation of output consistency across multiple runs.

The reproducibility concerns extend beyond academic settings to practical applications in investment and financial decision-making contexts. Blankespoor et al. (2024) document the widespread adoption of Generative AI among retail investors, with nearly half of surveyed investors already using these technologies for tasks ranging from interpreting financial information to calculating ratios and assessing sentiment. For these users, inconsistent outputs could lead to conflicting interpretations of the same underlying information, potentially leading to mis-informed investment decisions.

Our study addresses this critical gap by conducting the first comprehensive assessment of the consistency and reproducibility of LLM outputs in common finance and accounting tasks. We choose three GPT models from OpenAI, namely, GPT-3.5-turbo, GPT-4o-mini, and GPT-4o, due to the current dominance of GPT models in both research (Dong et al. 2024) and among investors (Blankespoor et al. 2024). Our assessment covers five key tasks: classification, sentiment analysis, summarization, text generation, and prediction. We employ a rigorous experimental design involving 50 independent runs for each task conducted over 50 days, using different random seeds to capture the models' inherent stochasticity. The analysis covers diverse financial text sources, including MD&As, FOMC statements, financial news, earnings conference call transcripts, and financial statements. We generate more than 3.4 million outputs through approximately 750,000 API calls, creating one of the most comprehensive assessments of LLM reproducibility to date, based on more than 80 million pairwise comparisons across different runs.

For each task, we implement a zero-shot prompting approach, focusing solely on the models' inherent capabilities without task-specific examples or fine-tuning. This methodological choice provides a more stringent test of the models' fundamental understanding and generalization capabilities while minimizing potential biases from few-shot examples or Chain-of-Thought (CoT) prompting. To assess consistency, we employ a comprehensive set of metrics tailored to each task's output type, including inter-rater agreement measures (Fleiss' Kappa, Krippendorff's Alpha), correlation metrics (intra-class, Pearson, Spearman, and Concordance), variation metrics (mean relative differences), and other run-level and document-level agreement metrics, for comprehensive analysis of both categorical and continuous outputs. We also assess the consistency of text outputs in semantic similarity and in characteristics such as tone and length. We measure semantic similarity using cosine similarities calculated from text embeddings generated by a state-of-the-art embedding model (*jina-embeddings-v3*), and we measure tone using both traditional word lists (Loughran and McDonald 2011) and FinBERT (Huang et al. 2023).

Our findings reveal substantial, but task-dependent, consistency across models and tasks. For binary classification tasks like identifying forward-looking statements, both GPT-3.5-turbo and GPT-4o-mini achieve near-perfect consistency (Fleiss' Kappa > 0.93), with identical classifications across all 50 runs for over 92% of sentences. Sentiment analysis similarly demonstrates high consistency (Fleiss' Kappa > 0.94) across all text types, although with notable differences in sentiment distribution between models. More complex tasks show greater variability, with multi-class classification of FOMC statements exhibiting moderate consistency (Fleiss' Kappa = 0.86-0.91), while numerical predictions of future earnings display substantial variation in point estimates from less advanced models.

Importantly, we find that more advanced models do not consistently outperform their predecessors in consistency. While GPT-4o demonstrates superior consistency in numerical predictions relative to earlier models, GPT-3.5-turbo unexpectedly outperforms GPT-4o-mini in multi-class classification of FOMC statements. These findings suggest that model sophistication does not necessarily translate to increased consistency, and task-specific considerations should guide model selection.

For summarization and text generation, we observe high semantic consistency (mean cosine similarity > 0.94) but moderate variation in length and tone. The semantic content of summaries and generated answers remains highly stable across runs, but word counts can vary by 8-15% on average, with substantially higher variation for some documents. These results suggest that while LLMs maintain consistent core messaging, researchers should exercise caution when analyzing structural or stylistic features of generated texts.

We conduct further analyses to address three additional important research questions. First, we investigate whether LLMs exhibit similar inconsistency patterns to expert human annotators in classification tasks. Using the Financial PhraseBank dataset (Malo et al. 2013, Yoo 2024), we compare consistency in sentiment classification between LLMs and human experts. We find that LLMs demonstrate higher consistency than human experts in over 50% of cases, with human experts outperforming LLMs in fewer than 9% of cases. Most strikingly, LLMs maintain high agreement levels (96.8-98.3%) even for cases where human expert annotators significantly disagree. This suggests that LLMs develop more deterministic classification boundaries that remain stable regardless of the inherent ambiguity in texts that causes human disagreement, making them potentially valuable tools for tasks requiring consistent classification.

Second, we examine whether aggregation strategies across multiple runs can enhance consistency. For both categorical and continuous outputs, we find that even a modest level of aggregation substantially improves consistency.[1] The most dramatic improvements occur at an aggregation level of three to five runs, with majority class strength (i.e., consensus) for classification tasks increasing from approximately 86% to 94% for GPT-3.5-turbo and from 82% to 91% for GPT-4o-mini. For continuous variables like text length, the variation between model runs decreases dramatically from approximately 8.5% with a single run to about 3% with just five aggregated runs. We also find that aggregation can improve the accuracy of sentiment analysis when using newer models. These findings provide practical guidance for users seeking to balance computational costs with consistency requirements, as most consistency benefits can be captured with minimal additional computational overhead, possibly even with the added benefit of improved accuracy.

Finally, we evaluate how LLM output variation affects downstream statistical analyses. Through extensive simulation with 10,000 iterations for totally 10 million regressions, we examine how variability in LLM-generated document lengths propagates into regression results. Despite measurable inconsistency in summary lengths, we find that regression coefficients appear to be unbiased, t-statistics maintain their expected distributions, and inference validity is preserved at remarkably high rates: 97.35% of all inferences drawn about the significance of length effects are correct, with Type I and Type II errors occurring at only 1.28% and 1.38% of the time, respectively. Most notably, when significant effects are detected, they almost never show an incorrect sign direction. These results provide strong reassurance that LLM-based textual analysis can be reliably incorporated into quantitative research frameworks, even when analyzing subtle effects where measurement precision is crucial.

This paper makes several important contributions to the emerging literature on LLMs in finance and accounting research (e.g., Cong and Zhu 2024, Jha et al. 2024, 2025, Sheng et al. 2025). First, it provides robust empirical evidence on the degree of consistency in LLM outputs for common research tasks, offering researchers essential guidance on the reliability of these models when incorporated into their methodological toolkits. Our findings establish benchmark expectations for reproducibility across different tasks, allowing researchers to anticipate potential variability and implement appropriate validation procedures.

Second, we provide comprehensive benchmarking of LLM consistency across multiple model versions (GPT-3.5-turbo, GPT-4o-mini, and GPT-4o) and common research tasks. Our analysis reveals that advanced models do not necessarily demonstrate improved reproducibility alongside their enhanced capabilities, offering users essential guidance for model selection that considers both performance and output stability.

[1] For aggregating multiple runs, majority voting is used for categorical outputs, while averaging is used for continuous outputs like text length. For example, if three runs are aggregated, then the length is the average from three runs.

Third, our additional analyses offer important insights into the incorporation of LLMs into data processing pipelines. We document that LLMs exhibit consistency patterns significantly different from human experts, shedding light on their distinct role in financial text annotation. We demonstrate that simple aggregation strategies effectively mitigate inconsistency, providing users with practical implementation guidance. Through extensive simulation, we further establish that LLM output variability has negligible impact on downstream statistical inference, offering crucial validation for incorporating these tools into quantitative research.

Taken together, these contributions have significant implications for LLM deployment in finance and accounting research. They provide researchers with both validation of LLM reliability and actionable strategies for implementation, addressing key methodological concerns in this emerging field. Our findings also address emerging concerns about what we term "G-hacking" in LLM-based research, analogous to p-hacking in statistical analyses. We define G-hacking as occurring when researchers generate multiple outputs from generative LLMs for the same task and cherry-pick results that support their hypotheses while discarding contradictory outcomes. Our analysis suggests that the risk of G-hacking is relatively low for most finance and accounting applications.

The remainder of this paper proceeds as follows: Section 2 describes background, related studies, and research questions. Section 3 details our experimental design, covering task selection, model choice, prompting strategies, and evaluation metrics. Section 4 presents results for each task. Section 5 provides additional analyses on human-LLM comparison, aggregation strategies, and downstream effects. Section 6 discusses implications. Online Appendices A and B provide detailed experimental designs and comprehensive result interpretations, respectively.

## 2. Background, Related Work, and Research Questions

The application of artificial intelligence (AI) and machine learning (ML) in finance and accounting research has gained significant traction in recent years (Bochkay et al. 2022). However, the use of LLMs in these fields is still at an early stage (Dong et al. 2024), particularly regarding issues of consistency and reproducibility. This section reviews relevant literature to contextualize our study.

### 2.1 LLMs in Finance and Accounting

The emergence of sophisticated LLMs has revolutionized how researchers process and analyze financial texts, offering unprecedented capabilities for tasks ranging from classification to prediction (Dong et al. 2024, de Kok 2025). For example, Hansen and Kazinnik (2024) demonstrate LLMs' superior performance in analyzing Federal Open Market Committee (FOMC) statements. Multiple studies have established LLMs' ability to extract sentiment from financial texts with greater nuance than dictionary-based and other ML-based approaches (e.g., Bond et al. 2025, J. Chen et al. 2023, Lopez-Lira and Tang 2024).

Recent studies demonstrate LLMs' diverse applications in finance and accounting research. These include classifying ESG disclosures and tax discussions (Li et al. 2023, Choi and Kim 2023), extracting data from unstructured text (H. Li et al. 2024, Wang and Wang 2024, Levich and Knust 2025), summarizing MD&As and assessing risk from earnings calls (Kim et al. 2024a, b), forecasting financial performance (E. X. Li et al. 2024), estimating core earnings (Shaffer and Wang 2024), analyzing employee reviews (Brown et al. 2025), and extracting firm investment and macroeconomic indicators (Jha et al. 2024, 2025). These findings demonstrate generative AI's growing relevance to finance and accounting research and decision-making.

### 2.2 Reproducibility Concerns in Finance and Accounting Research

Reproducibility and replicability have been a longstanding concern in finance and accounting research, long before the introduction of LLMs. Harvey (2019) discusses p-hacking and replicability issues in finance research. Menkveld et al. (2024) introduce the concept of "nonstandard errors" to characterize the variability in research outcomes when different research teams test the same hypotheses. Using this framework, Pérignon et al. (2024) find only 52% of finance results could be precisely reproduced even with provided code and documentation.

Reproducibility success varies substantially across finance subfields. While Jensen et al. (2023) offer optimistic findings in asset pricing and Chen and Zimmermann (2022) present positive assessments, Hou, Xue, and Zhang (2020) and Linnainmaa and Roberts (2018) find numerous documented effects fail under stringent statistical tests or out-of-sample testing.

In accounting, Hail et al. (2020) survey 136 researchers, finding 52% considered irreproducibility a major problem, with respondents estimating only 46% of published results were replicable and 59% reproducible. Gow (2025) documents p-hacking prevalence in accounting research, suggesting replication could significantly improve empirical work. Cross-disciplinarily, Fišar et al. (2024) find reproducibility rates improved after a premium business journal implemented data and code policies, although data accessibility and documentation remain barriers.

These reproducibility challenges in finance and accounting take on heightened significance with the integration of LLMs into research workflows. LLMs introduce additional layers of complexity through their stochastic generation processes, potentially exacerbating existing concerns. The stochastic nature of LLMs creates particular vulnerabilities to G-hacking, as the inherent randomness in model outputs provides researchers with plausible deniability for selective reporting. Unlike traditional p-hacking, where manipulation often leaves statistical traces, the opacity of LLM generation processes could more easily mask selective reporting practices. This emerging concern underscores the importance of establishing baseline expectations for LLM output consistency across different tasks and model types.

## 2.3 LLM Consistency and Reproducibility

Recent studies have explored various aspects of LLMs and their applications in finance, accounting, and many other disciplines. However, the consistency and reproducibility of LLMs' performance has not been extensively investigated, particularly in the context of finance and accounting research or applications.[2]

Aguda et al. (2024) investigate the potential of LLMs as data annotators for extracting relations in financial documents. They emphasize the importance of customizing prompts for each relation group and introduce a reliability index to identify outputs that may require expert attention. This work provides valuable insights into using LLMs for structured information extraction but does not examine output consistency across multiple runs. Reiss (2023) examines the consistency of ChatGPT's zero-shot capabilities for text annotation and classification, finding that even minor wording alterations in prompts or repeating identical inputs can lead to varying outputs. The study underscores the need for thorough validation against human-annotated data and represents one of the few works directly addressing LLM consistency, although with a narrower scope than our comprehensive assessment.

Gilardi et al. (2023) and Törnberg (2023) demonstrate that ChatGPT outperforms crowd-workers and experts in annotating political Twitter messages with zero-shot learning, achieving higher accuracy, reliability, and equal or lower bias. Exploring another dimension of LLM applications, Rossi et al. (2024) investigate the use of LLM-generated synthetic data in social science and design research. Their work raises important questions about the validity of LLM-generated content for research purposes but does not systematically evaluate output consistency.

Addressing a critical concern in financial applications, Sarkar and Vafa (2024) discuss the issue of temporal lookahead bias in empirical analysis that uses outputs from pretrained language models. Yoo (2024) evaluates the reliability of self-reported confidence scores from LLMs in sentiment analysis of finance news. This work highlights important limitations in LLMs' self-assessment capabilities but does not address the fundamental question of output reproducibility. Adding to these insights, Levy (2024) aims to provide rationale for the documented superior performance of generative AI systems in accounting and finance-related tasks, highlighting the poor numerical reasoning of LLMs and the presence of look-ahead bias in prediction tasks. Bini et al. (2025) find that LLMs exhibit systematic behavioral biases in economic decisions and propose ways to mitigate such biases.

While these studies have provided valuable insights into the capabilities and limitations of LLMs, they have primarily focused on aspects such as performance, overconfidence, look-ahead bias, numerical reasoning abilities, and behavioral bias. In contrast, our study aims to fill a critical gap in the literature by specifically assessing the consistency and reproducibility of LLMs' performance across multiple runs for various natural language processing tasks in finance and accounting. This contributes to a more comprehensive understanding of their reliability and robustness, which is essential for their practical applications in research settings where reproducibility is a fundamental requirement.

[2] Several studies have proposed methods for assessing and ensuring reproducibility in AI and ML research in general.(e.g., Gundersen and Kjensmo 2018, Raff 2019, Pineau et al. 2021, Ahmed and Lofstead 2022).

### 2.4 Practical Implications for Financial Decision-Making

The reproducibility concern extends beyond academia to practical investment and financial decision-making. Blankespoor et al. (2024) find nearly half of retail investors already use generative AI for explaining financial information (44.2%), defining terms (41.2%), searching company mentions (28.7%), and retrieving data (28.2%). Advanced applications include calculating ratios (21.4%), summarizing documents (21.0%), identifying trends (16.4%), and sentiment analysis (14.9%), with 27.4% seeking investment recommendations.

The reproducibility challenges of LLMs are particularly concerning for these specific applications. For investors relying on LLMs to explain financial concepts, identify trends, calculate ratios, or assess sentiment, inconsistent outputs could lead to conflicting interpretations of the same underlying information. This variability creates potential risks for decision-makers who rely on these tools for time-sensitive or high-stakes financial decisions, especially the more than one-quarter of users who directly seek investment advice from these systems.

### 2.5 Research Questions

Given the critical importance of this issue and the absence of systematic evidence, this study investigates the consistency and reproducibility of LLMs for common finance and accounting tasks. We systematically assess output variability across multiple runs and compare performance across different model versions. The following five research questions guide our investigation:

*RQ1: How consistent are LLM outputs for common finance and accounting tasks across multiple runs?*

This question evaluates the fundamental consistency of LLMs when provided with identical inputs and prompts across multiple executions. By subjecting the models to repeated runs on the same tasks, we aim to quantify output variability and assess their ability to produce consistent, reproducible results. This analysis is essential for establishing baseline expectations regarding LLM robustness in finance and accounting applications, where consistency is a prerequisite for scientific validity and reliable decision-making.

*RQ2: How do different versions of LLMs compare in terms of output consistency?*

This question examines whether more advanced model versions demonstrate improved consistency alongside their enhanced capabilities. By comparing output reliability across different models, we test whether consistency improves with model sophistication or follows different patterns. These insights are critical for researchers and practitioners weighing trade-offs between cost, performance, and consistency when selecting appropriate models for specific applications.

*RQ3: Do LLMs exhibit consistency patterns similar to human annotators?*

This question investigates whether LLMs mimic human annotation patterns, particularly in contexts where human annotators demonstrate significant disagreements. By analyzing model outputs on datasets with documented human inconsistencies, we assess whether LLMs struggle with the same ambiguous cases or manifest distinct variability patterns. These findings have important implications for determining the role of LLMs in financial text analysis.

*RQ4: Can aggregation strategies across multiple runs enhance LLM output consistency?*

This question explores practical mitigation strategies for LLM inconsistency through output aggregation. By implementing and evaluating various aggregation techniques, including majority voting for categorical outputs and averaging for continuous predictions, we assess their effectiveness in enhancing result consistency. Additionally, we examine the effect of aggregation on accuracy. These findings provide actionable guidance for researchers and practitioners seeking to increase consistency in LLM-based analyses while balancing computational costs.

*RQ5: How does LLM output variation affect downstream statistical analyses?*

This question examines both the propagation of LLM output variation into subsequent analytical workflows and addresses concerns about selective reporting of favorable LLM outputs. We assess how variability in model outputs influences statistical significance, effect sizes, and substantive conclusions when these outputs serve as variables in regression analyses. This quantifies the potential implications of incorporating LLM-generated variables into broader research frameworks. This analysis also allows us to evaluate the severity of G-hacking concerns in finance and accounting applications.

## 3. Experimental Design

In this section, we describe our experimental design, covering task selection, model choice, prompting strategies, and evaluation metrics. Specific design choices for each task, such as sample selection, prompts, and other implementation details, are discussed in Online Appendix A.

### 3.1 Task Selection

Our task selection is guided by a comprehensive literature review conducted by Dong et al. (2024), who find that numerous studies have utilized LLMs to carry out tasks such as classification, sentiment analysis, summarization, text generation, and prediction. We choose these tasks for our study. To assess the consistency and reproducibility of these tasks, we use a variety of text commonly employed in finance and accounting research, including FOMC statements, financial news articles, earnings conference call transcripts, and MD&As. For the prediction task, we supply LLMs with condensed comparative financial statements in addition to MD&A summaries.

Table 1 summarizes these tasks as well as their data sources and sample sizes. For classification tasks, the source data includes sentences from MD&As of 10-Ks and sentences from FOMC statements, with sample sizes being 10,000 and 1,096. Sentiment analysis tasks involve news articles, prepared presentations and Q&A sections from earnings call transcripts, and MD&As from 10-Ks, with sample sizes ranging from 2,000 to 10,000 paragraphs or sentences. The summarization task uses MD&As from 10-Ks and presentation sections from earnings calls transcripts, with sample sizes of both 1,000. The text generation task involves answering 1,000 questions from Q&A sections of earnings call transcripts. The prediction task requires the models to predict future earnings of 1,000 firm-years based on MD&A summaries and comparative financial statements. In what follows, we discuss related literature and provide additional details about each task.

#### 3.1.1 Classification and Sentiment Analysis

Classification involves categorizing data into predefined classes. Recent studies demonstrate LLMs' effectiveness in classifying financial texts, e.g., identifying ESG disclosures (Li et al. 2023), classifying management presentations as facts/opinions (Kuroki et al. 2023), determining earnings call discussion types (Jia et al. 2024), assessing firm characteristics (Bernard et al. 2023, Dasgupta et al. 2023), identifying tax audit disclosures (Choi and Kim 2023), and classifying mutual fund manager statements (Wang 2023).

For binary classification, we examine forward-looking statements in MD&As, which are useful for predicting future earnings and cash flows. Prior research used dictionary-based or machine learning approaches (e.g., Li 2010, Muslu et al. 2014, Bozanic et al. 2018, Brown et al. 2024). For multi-class classification, we analyze FOMC statements, where Hansen and Kazinnik (2024) have demonstrated LLMs' superiority in this task.

Fundamentally a classification task, sentiment analysis classifies emotional tone as positive, negative, or neutral using dictionary-based methods (Loughran and McDonald 2011) or ML approaches like FinBERT (Huang et al. 2023). Numerous studies use LLMs to extract sentiment from various financial texts (Breitung et al. 2023, Alonso-Robisco and Carbó 2023, Smales 2023, Bae et al. 2024), and apply LLM-enabled sentiment analysis to forecast stock trends and evaluate trading strategies (Álvarez-Díez et al. 2024, Bond et al. 2025, J. Chen et al. 2023, B. Chen et al. 2023, Kirtac and Germano 2024, Lopez-Lira and Tang 2024). LLMs consistently outperform traditional methods (Bond et al. 2025, J. Chen et al. 2023, Hu et al. 2023, Kirtac and Germano 2024, Lopez-Lira and Tang 2024), although studies note potential biases requiring careful implementation (Leippold 2023, Hu et al. 2023, Glasserman and Lin 2023).

For this common task, we use multiple text sources, as prior studies have done. The first source is sentences from the Financial PhraseBank (Malo et al. 2013, Yoo 2024), which consists of sentences from financial news articles. The benefit of this dataset is that it not only has human labels but also provides information about disagreements among expert human raters. This information allows us to additionally test whether LLM outputs are more varied for sentences where humans tend to disagree. We also include sentences from MD&As and paragraphs from earnings conference call transcripts, covering both prepared presentation and Q&A sessions, to assess the consistency of LLMs in processing and analyzing various types of financial text.

#### 3.1.2 Summarization and Text Generation

Summarization involves generating a concise summary of a longer text while preserving its key information. Most LLMs can now create reliable summaries that capture the depth of the original text. Kim, Muhn, and Nikolaev (Kim et al. 2024a, b) use GPT-3.5-turbo to summarize MD&As and conference call transcripts. For this task, we ask the models to summarize MD&As and earnings call presentations.

Text generation involves creating new text based on a given prompt or context, often aiming to mimic human-like writing or response. Bai et al. (2023) use LLMs to answer questions raised by equity analysts during earnings call Q&A sessions, while Wu, Dong, Li, and Shi (2023) employ ChatGPT to generate analyses of textual loan assessments. For this task, we ask the models, serving as executives of the company to answer questions posed by financial analysts at earnings conferences.

#### 3.1.3 Prediction

Prediction involves forecasting future events or outcomes, such as stock price movements or a company's future earnings, based on historical data and patterns. Li, Tu, and Zhou (2023) find that earnings forecasts generated by GPT-4 exhibit greater forecast errors than analyst consensus, but its performance improves for firms with better information environments or higher-quality disclosures. Comlekci et al. (2023) use ChatGPT to forecast future financial performance and dividends of public companies. For this task, we ask the models to predict future earnings.

### 3.2 Model Choice

In this study, we focus on GPT models provided by OpenAI, specifically GPT-3.5-turbo, GPT-4o-mini, and GPT-4o. This choice is informed by Dong et al.'s (2024) finding that most existing studies on LLMs in accounting and finance use these models, as well as OpenAI's dominant market share among retail investors (Blankespoor et al. 2024). We also consider cost-effectiveness in our model selection. GPT-3.5 and GPT-4o-mini are relatively less expensive, while GPT-4o is more capable but also significantly more costly (more than 15 times as expensive). For most tasks, we employ more than one model, so that we can compare the consistency across models and test whether newer or more advanced models exhibit greater consistency. Table 2 presents the key characteristics and pricing information for the selected models, as well as the models chosen for each task.

For our classification and sentiment analysis tasks, which are considered to have low difficulty, we use the capable and cost-effective GPT-3.5-turbo and GPT-4o-mini models. For summarization, we use only GPT-4o-mini, which has a large context window and is cost-effective. For text generation, we use both GPT-3.5-turbo and GPT-4o-mini. Our prediction task, considered to have the highest complexity due to its requirement for mathematical calculations and logical reasoning, is performed using GPT-4o, GPT-4o-mini, and GPT-3.5-turbo, allowing us to compare the performance of these models on a challenging task.

#### 3.2.1 Prompting Strategies and Parameter Setting

This study employs zero-shot prompting across all tasks to evaluate the inherent capabilities of LLMs without task-specific examples or fine-tuning. This methodological choice is motivated by several considerations. First, zero-shot prompting provides a more stringent test of the model's fundamental understanding and generalization capabilities, as it requires the model to perform tasks solely based on instructions without exemplars. Second, this approach minimizes potential biases that might be introduced through few-shot examples or Chain-of-thought (CoT) prompting, which could inadvertently guide the model toward specific patterns or responses. Third, zero-shot prompting offers greater operational consistency across different tasks, facilitating more direct comparisons of model performance across diverse applications. For parameters, we set the temperature to zero for maximum reproducibility, a choice commonly made by most researchers, while keeping other parameters at their defaults.

#### 3.2.2 Experimental Runs and Experiment Period

We choose to conduct 50 runs for each task to balance cost, statistical power, and the inherent randomness of LLMs. This choice aligns with our sample size selection for each run, detailed in the task-specific design provided in Online Appendix A. By incorporating large sample sizes across numerous runs, we capture both within-run and across-run variation, strengthening the generalizability of our findings.

These experiments span 50 days between December 27, 2024, and February 15, 2025, conducted at varying times to ensure diverse conditions. Each run utilizes a different seed to maintain independence between runs. We intentionally avoid using identical seeds across runs, as our goal is not to assess reproducibility given the same seed, but rather to evaluate consistency under varying conditions and the inherent randomness fundamental to generative AI models.

### 3.3 Evaluation Metrics

We evaluate consistency using a comprehensive set of metrics tailored to each task's output type: categorical variables, continuous variables, or texts. For text outputs, we examine the consistency of characteristics such as similarity, length, and tone, where similarity and length are continuous variables and tone is categorical.

As shown in Panel A of Table 3, we use a wide variety of metrics covering overall inter-rater agreements, run-level agreements, and document-level agreements. The overall inter-rater agreement metrics provide an overall summary metric of consistency, and they include Fleiss' Kappa (Fleiss 1971) and Krippendorff's Alpha (Krippendorff 2011). These metrics quantify the overall agreement across all runs, with higher values indicating stronger agreement.

The run-level agreements provide summary metrics for pairwise consistency between runs, including Cohen's Kappa scores and agreement percentages. Cohen's Kappa adjusts for chance agreement, making it more stringent than simple percentage agreement, with values from -1 to 1 where higher indicates stronger agreement. The run-pair agreement percentage shows the proportion of documents classified identically between runs, offering an intuitive consistency measure.

At the document level, we measure percentage of perfect agreement, document-wise agreement, majority class strength, and classification uncertainty. These metrics focus on individual documents rather than run-pairs, showing how consistently each document is classified across all runs. Perfect agreement identifies documents with complete consensus across 50 runs, while document-wise agreement measures the percentage of run-pairs yielding identical classifications for a document. Majority class strength quantifies the dominance of the most frequent classification, and classification uncertainty uses entropy to measure variability where disagreement exists.

Appendix A illustrates calculation methods for both run-level and document-level metrics. From 50 runs yielding 1,225 possible run-pairs, we report aggregated statistics (mean, median, standard deviation) to characterize the distribution. For document-level metrics, each document has an agreement measure, and we similarly provide summary statistics (mean, median, standard deviation) across all documents in each task. See Appendix B for detailed metric definitions.

For continuous outputs (Panel B, Table 3), we use metrics designed to assess numerical consistency across runs, organized into four categories: overall inter-run reliability, correlation metrics, run-level variation, and document-level variation. The Intraclass Correlation Coefficient (ICC2, type 2) measures overall inter-run reliability, assessing absolute agreement rather than just relative consistency, with values near one indicating strong reliability. ICC2 accounts for both systematic and random differences between runs, making it suitable for determining if runs can be used interchangeably. We also employ Concordance, Pearson, and Spearman correlations.

We assess run-level variation using run-pair MARD (Mean Absolute Relative Difference), which measures the typical magnitude of differences between run pairs as a proportion of their average value, with lower percentages indicating smaller relative variations. At the document level, we measure documents with identical output (%) and document-wise MARD (%). The identical output percentage quantifies how often runs produce exactly the same numerical value for a document, while document-wise MARD calculates the average relative difference across all run pairs for each document. We provide descriptive statistics for both run-level and document-level metrics across all run pairs and documents.

For summarization and text generation tasks, the model outputs are texts. To evaluate the consistency of these outputs, we examine several common textual attributes, including length (word counts), tone (positive, negative, or neutral), and semantic similarity (cosine similarity), as shown in Panel C of Table 3.

Since length and semantic similarity are continuous variables, their results are evaluated using the metrics for continuous variables. To determine the tone of generated texts, we use both a lexicon-based approach (Loughran and McDonald 2011) and FinBERT (Huang et al. 2023). For each text, we classify its tone and evaluate the result using our chosen metrics for categorical variables.

To evaluate semantic similarity, we utilize a state-of-the-art embedding model, *jina-embeddings-v3*, which offers a good balance between computational efficiency and performance (Sturua et al. 2024). The process involves converting each generated text into high-dimensional embeddings and computing pairwise cosine similarities between runs. This approach allows us to quantify how consistently the model maintains core semantic content across multiple generations. We assess the consistency of the similarity scores using our chosen metrics for continuous variables.

## 4. Experimental Results

As shown in Table A1 (Online Appendix A), our analysis covers over 3.4 million outputs from approximately 750,000 API calls. Each task involves 50 independent runs, generating 1,225 run pairs per document and over 84 million pairwise comparisons.[3] We summarize key findings here. See Online Appendix B for detailed task-specific results.

We distinguish between run-pair and document-level agreement measures. Run-pair metrics assess consistency between run pairs across all documents, revealing overall classification pattern similarity. They are particularly relevant for academic research analyzing large samples where aggregate patterns matter most. Document-wise agreement examines how consistently each document is classified across runs, providing granular stability insights crucial for investors and practitioners who analyze small sets of company documents where individual classification reliability directly impacts decisions.

### 4.1 Key Findings

Our analysis reveals remarkably high consistency across most financial text analysis tasks, although with notable variations by task complexity and model. For classification tasks, both GPT-3.5-turbo and GPT-4o-mini demonstrate exceptional reproducibility, with GPT-4o-mini achieving Fleiss' Kappa values of 0.97 for binary classification and 0.86 for multi-class classification, as shown in Table 4. Interestingly, for the multi-class FOMC task, GPT-3.5-turbo slightly outperforms GPT-4o-mini with a Kappa of 0.91. Perfect agreement rates (identical classification across all 50 runs) reach 96.18% for binary tasks with GPT-4o-mini, although they drop to 55.66% for the more complex five-class FOMC statement classification. Krippendorff's Alpha scores and mean run-pair Cohen's Kappa scores are identical to Fleiss' Kappa scores when rounded to two decimal places, with minimal differences at higher precision levels. The mean classification uncertainty, which is the entropy for the documents where not all runs agree, is small, ranging from 0.34 to 0.40.

[3] For each document, we generate (50 × 49)/2 = 1,225 unique run pairs. With 3.44 million total outputs (documents × 50 runs), this yields more than 84 million pairwise comparisons (3.44 million × 49/2).

These values indicate "almost perfect" agreement according to multiple established scales. Krippendorff (2004) suggests Alpha values above 0.80 indicate "substantial" reliability. Similarly, McHugh (2012) classifies Cohen's Kappa scores of 0.80-0.90 as "strong" and above 0.9 as "almost perfect" agreement. These interpretations are widely accepted in computational linguistics studies that assess inter-rater reliability in classification tasks (e.g., Artstein and Poesio 2008, Reiss 2023). Fleiss' Kappa is an adapted version of Cohen's Kappa designed for more than two raters.

It is worth noting that the mean run-pair agreement (%) and mean document-wise agreement share identical mean values. This equivalence is not coincidental but by design, as both agreement metrics are derived from the same underlying pairwise comparisons and, at the highest aggregation level, converge to equal values. We present both metrics for convenience of interpretation. However, their distributions do differ. For their detailed distributions, please see relevant tables in Online Appendix B.

As shown in Table 5, sentiment analysis demonstrates high consistency across all text types (news articles, MD&As, earnings call presentations, and Q&As), with GPT-4o-mini achieving mean run-pair agreement exceeding 97.5% across all document categories. All commonly used inter-rater agreement metrics such as Krippendorff's Alpha and Kappas have values well above 0.9, indicating near-perfect agreement across runs. Mean majority class strength percentages consistently approach maximum values, while mean classification uncertainty metrics remain exceptionally low, suggesting that even in rare instances where different runs produce disagreements, these discrepancies are minimal. This remarkable consistency across multiple measures reinforces the robustness of LLM-based sentiment analysis for financial text applications.

For generative tasks like summarization, semantic similarity remains consistently high (mean cosine similarity of 0.98) across both MD&As and earnings call presentations, as shown in Table 6. However, summary length shows moderate variability with Mean Absolute Relative Differences (MARD) of approximately 8.5% between runs. Text generation tasks demonstrate strong semantic consistency (0.97 for GPT-4o-mini vs. 0.94 for GPT-3.5-turbo), although with greater length variability (8.18% vs. 14.91% MARD), as shown in Table 7. Tone consistency shows interesting patterns, with GPT-4o-mini maintaining moderate consistency (Fleiss' Kappa of 0.68 for MD&A summaries in Table 6 and 0.66 for text generation when measured by FinBERT in Table 7). Earnings call presentations exhibit substantially higher tone consistency (run-pair agreement of 95.55%) compared to MD&As (84.40%) in Table 6, likely reflecting that presentations at earnings conference calls tend to be predominantly positive.[4]

In earnings prediction tasks, all models maintain high directional consistency (Fleiss' Kappa ≥ 0.97), but point estimate consistency varies significantly, as shown in Table 8. GPT-4o achieves near-perfect consistency with run-pair MARD of just 2.05%, substantially outperforming GPT-4o-mini (6.18%) and GPT-3.5-turbo (9.89%). GPT-4o also demonstrates superior consistency across all metrics, including ICC2 (0.997), concordance correlation (0.995), and percentage of documents with identical point estimates (29.20%). Despite the variation in other metrics, average Spearman correlations remain close to one between different runs for all three models, although GPT-4o-mini and GPT-3.5-turbo show lower values in the mid-70s to high-80s range for other correlation measures.

### 4.2 Comparison Across Model Versions

To address our second research question regarding inter-model consistency differences, we provide a more in-depth analysis and discussion of the related results in this section. Overall, we find that the relationship between model sophistication and output consistency follows task-dependent patterns rather than a universal trend. This has important implications for model selection in financial text analysis applications.

For most classification tasks, GPT-4o-mini demonstrates superior consistency compared to GPT-3.5-turbo, with statistically significant improvements in binary classification (Fleiss' Kappa of 0.97 vs. 0.93) and sentiment analysis across all document types. However, this pattern reverses for multi-class classification of FOMC statements, where GPT-3.5-turbo outperforms GPT-4o-mini (Kappa: 0.91 vs. 0.86). This suggests that newer models may not always provide more consistent outputs, particularly for tasks involving nuanced categorical distinctions.

For text generation tasks, we observe a clearer correlation between model advancement and consistency. GPT-4o-mini generates responses with significantly higher semantic similarity (0.97 vs. 0.94) and lower length variability (MARD: 8.18% vs. 14.91%) compared to GPT-3.5-turbo. The minimum document-level similarity shows particularly striking differences (0.76 vs. 0.46, as shown in Table B7 in Online Appendix B), indicating that GPT-4o-mini avoids the extreme inconsistencies occasionally produced by GPT-3.5-turbo.

The most pronounced model-dependent consistency differences appear in numerical prediction tasks. For earnings point estimates, consistency improves dramatically with model sophistication: GPT-4o achieves near-perfect consistency (MARD: 2.05%) compared to GPT-4o-mini (6.18%) and GPT-3.5-turbo (9.89%). This stepwise improvement is visually evident in the non-overlapping error distributions, with

[4] In Online Appendix B, we provide the agreement metrics for tone measured using the finance word list (Loughran and McDonald 2011). These metrics exhibit lower agreements than those from FinBERT, consistent with findings that FinBERT better captures sentiment as demonstrated by multiple studies (Huang et al. 2023).

GPT-4o showing remarkably narrow dispersion centered around 2%, as shown in Panel B of Figure B3 in Online Appendix B.

Overall, our results indicate that while newer models generally offer improved consistency, this advantage is not uniform across all tasks. The selection of appropriate models should therefore consider both the specific task requirements and the relative importance of consistency versus other factors such as cost and latency. For applications requiring precise numerical outputs or complex text generation, more advanced models provide substantial consistency benefits that may justify their higher computational costs.[5]

## 5. Additional Analyses

### 5.1 GPT Models vs Human Experts

In this sub-section, we report the results for RQ3 regarding whether LLMs outperform expert human annotators in classification tasks in consistency and whether they exhibit similar patterns in cases where human annotators tend to disagree. For this analysis, we leverage our sentiment analysis task that utilizes the Financial PhraseBank dataset, which consists of more than 4,000 sentences with human labels and human agreement levels. The dataset comes with four levels of agreements, namely, 50%, 66%, 75%, and 100%, where 50% means 50% of human annotators agree on a label, e.g., a positive sentiment. We measure the level of agreement in the same manner whereby we calculate the agreement over the majority class for each document.

As shown in Panel A of Figure 1, GPT-3.5-turbo demonstrates greater consistency than human annotators in 51.8% of sentences. Human annotators outperform GPT-3.5-turbo in only 8.4% of cases, while ties occur in 39.8% of sentences. GPT-4o-mini shows even stronger performance, being more consistent than humans in 52.5% of sentences, while humans outperform the model in just 4.8% of cases, with ties accounting for 42.7% of cases. These results indicate that both models, particularly GPT-4o-mini, achieve higher consistency than human annotators in most cases.

Next, we assess whether LLMs struggle with sentences that challenge human annotators. Panel B of Figure 1 shows that while 46.7% of sentences achieve perfect human agreement, over half exhibit disagreement (24.6% at 75% agreement, 15.8% at 66%, and 13.0% at only 50%). Notably, when human agreement is unanimous (100%), both models achieve near-perfect consistency (98.1% and 98.8%) as well. However, even for sentences with low human agreement (50-75%), model agreement remains remarkably high (96.8-98.3%), showing no significant decline as human consensus decreases. This suggests that LLMs maintain consistent classifications even for sentences that human annotators find ambiguous or challenging to classify. Unlike human experts, the models appear to have more deterministic classification boundaries that remain stable regardless of the inherent ambiguity in the text that causes human disagreement.

[5] Additionally, we observe that model outputs can exhibit systematic differences in classification distributions, suggesting difference in accuracy, even though output accuracy is not the focus of this study.

### 5.2 Aggregation Across Multiple Runs

This sub-section reports results for RQ4 examining whether aggregating outputs from multiple runs of the same model increases consistency. We evaluate this in two contexts: classification tasks and summary length consistency.

For classification tasks, we focus on multi-class classification of FOMC statements, as distinguishing among five distinct classes presents a more challenging scenario for evaluating model consistency. Our sample includes sentences where runs disagreed on labels. We randomly aggregate 2-20 outputs from 50 runs, using majority voting with tie-breaking via median rank (classes coded 0-4 from 'Dovish' to 'Hawkish'). This process generates 50 synthetic runs for calculating document-level metrics.

Figure 2 shows clear improvements with aggregation. Panel A reveals that majority class strength increases substantially with minimal aggregation. The steepest gains occur with 3-5 runs, where GPT-3.5-turbo improves from around 86% to 94% and GPT-4o-mini from 82% to 91%. At 20 aggregated runs, these reach 98% and 95.5%, respectively. Classification uncertainty (entropy) drops most sharply in the first few aggregated runs, with GPT-3.5-turbo maintaining lower uncertainty throughout.

Panel B demonstrates enhanced consistency: perfect agreement jumps dramatically with just 5 aggregated runs (GPT-3.5-turbo: 25% to 65%; GPT-4o-mini: 0% to 55%). At 20 runs, GPT-3.5-turbo achieves around 85% perfect agreement versus around 78% for GPT-4o-mini. Document-wise agreement scores similarly improve, exceeding 96% and 94% respectively at 20 runs.

We apply similar aggregation methodology to MD&A summary length, averaging sampled runs rather than using majority voting for this continuous variable. Figure 3 demonstrates substantial benefits across all metrics. Correlation metrics (ICC2, CCC, Spearman, Pearson) show dramatic improvement within the first 5 aggregated runs. For example, CCC increases from 0.43 to 0.8, while others improve from 0.65-0.7 to around 0.9. At 20 runs, all metrics converge between 0.95-0.98. Mean Run-pair MARD decreases from 8.5% to 3% with 5 runs, reaching around 1.5% at 20 runs.

These findings reveal that users do not need to implement extensive aggregation to enhance LLM consistency. Modest aggregation over 3-5 runs captures substantial benefit in improved consistency, offering a valuable insight for balancing efficiency against reliability requirements. The results provide strong evidence that run aggregation effectively enhances LLM consistency across both classification and continuous output types, offering a practical guidance for researchers and practitioners.

### 5.3 Impact of Aggregation on Classification Accuracy

Building on our consistency findings, we examine how aggregation affects classification accuracy. Using the Financial PhraseBank dataset from Section 5.1, we focus on sentences with ≥75% expert annotator agreement as ground truth, following Yoo (2024), specifically selecting cases where 50 model runs disagreed on labels. We generate 50 synthetic runs at each aggregation level and merge them with human labels. Given the sample's class imbalance (predominantly neutral labels), we apply SMOTE to create balance before calculating weighted F1 scores against expert labels.

Panel C of Figure 2 reveals striking model-specific differences. GPT-3.5-turbo shows unexpected F1 score decline with aggregation, dropping from 0.29 to below 0.23 and falling beneath the 0.28 baseline. Conversely, GPT-4o-mini demonstrates remarkable improvement, jumping from 0.75 to 0.84 with minimal aggregation before stabilizing. Error bars indicate GPT-3.5-turbo's higher performance variability compared to GPT-4o-mini's consistency.

These results have significant practical implications. Despite comparable consistency levels across models, their responses to aggregation diverge markedly. While LLMs can surpass human annotator consistency, the efficacy of aggregation proves highly model-dependent. GPT-4o-mini's improvements suggest that recent architectures effectively synthesize collective predictions, whereas earlier models like GPT-3.5-turbo may form inflexible decision boundaries that resist refinement through aggregation. This model-specific relationship between consistency and accuracy merits careful consideration when designing aggregation strategies.

### 5.4 Impact on Downstream Tasks

Studies in finance and accounting increasingly incorporate LLM-generated variables in regression analyses. Our final research question (RQ5) examines how LLM output variation affects statistical analysis reliability. We implement a simulation-based methodology quantifying how length variability in LLM-generated MD&A summaries propagates into regression analyses, evaluating impacts on coefficient estimates, statistical significance, and inference reliability. We scale summary length by MD&A length, generating an inverse bloatedness measure similar to Kim et al. (2024a).

We employ Monte Carlo simulation with bootstrapping and leave-one-out validation to evaluate how LLM inconsistency affects downstream statistical analysis (Efron and Tibshirani 1994, Shao and Tu 2012, Hammersley 2013). Using the 50 runs from our consistency analyses, we generate 1,001 synthetic runs by randomly sampling summary lengths for each document. From these synthetic runs: (1) we designate one run as "ground truth" to define the summary length variable; (2) we generate an outcome variable (y) with a known relationship to this length variable; (3) we estimate the true coefficient by regressing “y” on the ground truth length variable; (4) We estimate 1,000 separate regressions for the remaining 1,000 synthetic runs and compare their length coefficients to the true coefficient. This entire process is repeated 10,000 times, yielding 10 million total regressions that allow us to assess estimation accuracy and bias.

Our simulation employs the following specifications designed to mirror challenging but realistic empirical finance and accounting research conditions. We set the length effect at 0.005, a deliberately small effect size reflecting realistic detection challenges. The control variable (X) effect is set at 0.2 to ensure substantial control variable influence, with a confounding correlation of 0.5 that provides moderate correlation while avoiding multicollinearity. The signal-to-noise ratio is 0.5, meaning the error standard deviation is twice the combined signal strength. These parameters generate R-squared values of approximately 0.2, consistent with typical empirical studies in accounting and finance.

Figure 4 presents our simulation results. Panel A demonstrates that coefficient estimates remain stable and unbiased despite length variability introduced by LLM processing. The estimated coefficient distribution closely aligns with true values, with points clustering tightly around the perfect agreement line. Panel B reveals similar stability for t-statistics, where the distributions of estimated and true values overlap substantially. The fitted trend line tracks the perfect agreement line almost exactly, indicating minimal bias in significance testing.

Panel C shows that statistical inference remains highly reliable. The significance of length effects is correctly identified in 97.35% of cases, with Type I and Type II error rates of only 1.28% and 1.38%, respectively. The reliability heatmap indicates 86.15% correct sign identification for non-significant effects, while significant effects maintain proper direction virtually always (0% incorrect signs). Overall correct coefficient sign identification reaches 92.40%.[6]

These findings provide important reassurance for researchers. While LLM summarization introduces measurable length inconsistency, the downstream statistical impact remains negligible. Regression coefficients appear unbiased, t-statistics maintain their expected distributions, and inference validity is preserved at high rates. This suggests that LLM-based document processing can be safely incorporated into quantitative research pipelines, even when analyzing subtle effects that require substantial measurement precision.

## 6. Discussion and Conclusion

This study provides the first comprehensive assessment of the consistency and reproducibility of outputs from commonly used LLMs across common finance and accounting research tasks. Through extensive experimentation involving 50 independent runs for each task and generating over 3.4 million outputs, we have established several key findings regarding LLM reliability in financial text analysis.

Our results demonstrate that LLM output consistency varies substantially across tasks. Binary classification and sentiment analysis achieve near-perfect reproducibility, while more complex tasks exhibit greater variability. Contrary to expectations, more advanced models do not consistently demonstrate superior reproducibility. For textual outputs, we observe high semantic consistency (mean cosine similarity

[6] In an untabulated anlaysis, we find that correct inference improves to 98.49% and correct sign identification to 95.45% when we aggregate over three runs.

> 0.94) alongside moderate variation in structural features such as length and tone. These findings have significant methodological implications. Researchers should consider task complexity when applying these models, and conduct task-specific evaluations of both performance and consistency when selecting appropriate models.

Our additional analysis shows that LLMs outperform human annotators in consistency in sentiment analysis. LLMs maintain high agreement levels (96.8-98.3%) even for sentences where human annotators significantly disagree. Additionally, aggregation strategies like majority voting for categorical outputs and averaging for continuous predictions can substantially improve consistency, with the most dramatic gains occurring within aggregation levels of 3-5 runs. Our simulation of downstream regression analyses shows that despite output variability, statistical inference remains highly reliable (97.35% correct inferences) with minimal Type I and Type II errors (approximately 1.3% each).

Our findings have important implications for addressing what we term "G-hacking," i.e., the selective reporting of favorable outcomes from multiple runs of a generative LLM. The high consistency we observe across most tasks suggests that G-hacking risks are relatively low for most finance and accounting applications, particularly for classification and sentiment analysis.

While comprehensive, our study has limitations that suggest future research directions. We focus exclusively on OpenAI models using zero-shot prompting. Future studies can extend our analysis to other LLMs and prompting strategies. Additional finance and accounting applications beyond our five chosen tasks, such as stock return prediction, risk assessment, and fraud detection, also warrant investigation. Furthermore, examining how consistency patterns evolve as models are updated would provide insights into the trajectory of reliability improvements in LLM technology. We also call for further research into whether aggregation over multiple runs can increase accuracy in addition to consistency.

**Declaration of Generative AI and AI-Assisted Technologies in the Writing Process:**

In preparing this manuscript, the author(s) used ChatGPT and Claude to enhance language and readability. Following the use of these tools, the author(s) carefully reviewed and edited the content to ensure accuracy and take full responsibility for the final version of the manuscript.

## Appendix A
## Agreement Metric Calculation Illustration

This appendix illustrates the calculation of two key agreement metrics used in our analysis: run-pair agreement and document-wise agreement. Both metrics are derived from pairwise comparisons but differ in their unit of analysis.

**Panel A: Run-pair agreement**

Comparing Run 1 vs Run 2:

| Document | Run 1 | Run 2 | Match? |
|---|---|---|---|
| Document 1 | F | F | ✓ |
| Document 2 | N | N | ✓ |
| Document 3 | F | N | ✗ |
| Document 4 | N | N | ✓ |
| Document 5 | F | F | ✓ |
| Document 6 | N | N | ✓ |

Agreement Rate = (5 matching docs / 6 total docs) × 100 = 83%

F = Forward-looking   N = Non-Forward-looking

Run-pair agreement measures how consistently two runs classify the same documents. The figure shows Run 1 vs Run 2 across six documents. Both runs agree for Documents 1, 2, 4, 5, 6, while the two runs disagree for Document 3.

Since 5 documents match out of 6, the agreement rate = (5/6) × 100 = 83%. This calculation is repeated for all 1,225 possible run pairs. The mean run-pair agreement is the average of all these agreement rates, showing overall classification consistency.

## Appendix A
## Agreement Metric Calculation Illustration (Continued)

**Panel B: Document-wise agreement**

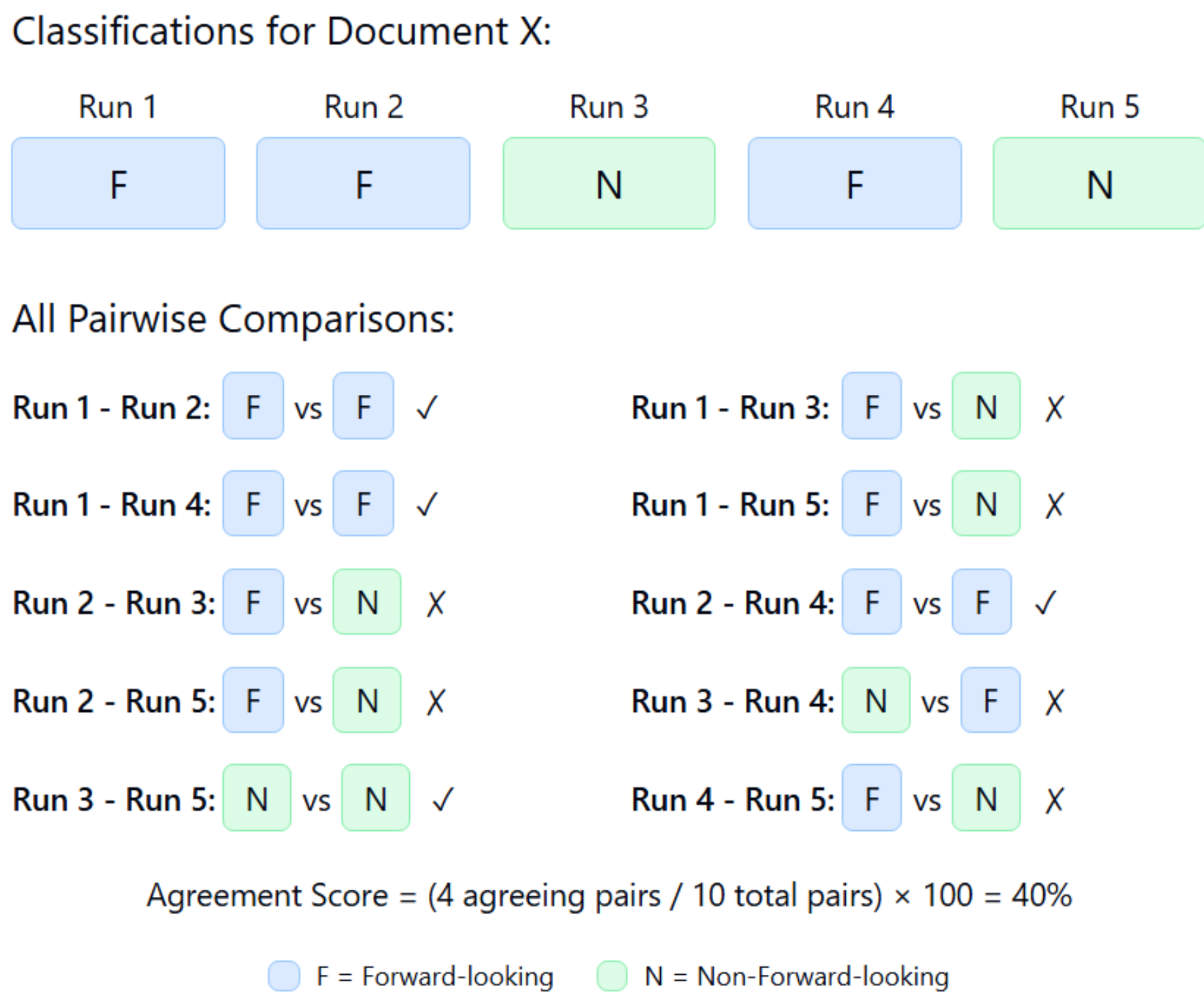

Document-wise agreement measures classification consistency for individual documents across multiple runs. For each document, we compare all possible pairs of runs. The figure shows Document X classified by five runs. Runs 1, 2, and 4 classify it as Forward-looking (F), while Runs 3 and 5 classify it as Non-Forward-looking (N). Since only 4 pairs agree out of 10 total pairs, the agreement score = (4/10) × 100 = 40%.In our full analysis with 50 runs, each document has 1,225 run pairs, and its agreement score is calculated across all these pairs. The mean document-wise agreement is the average of these scores across all documents.

## Appendix B
## Metric Definitions

**Panel A: Sample size metrics**

**Number of documents**: The total number of documents used for a task. A document can be a sentence, paragraph, longer text (e.g., a full MD&A section), or a combination of multiple data types as in the case of Q&A answering and future earnings prediction. In our terminology, a document represents a single unit of the input provided to an LLM, which generates an output based on this input according to the prompt.

**Number of runs**: The number of independent attempts performed by an LLM on each document using different random seeds (50 in our experiments).

**Number of run pairs**: The total number of unique pairs that can be formed from the runs, calculated as (n × (n-1))/2, where "n" is the number of runs. With 50 runs, this yields 1,225 pairs. Each pair represents a unique comparison between two different runs' outputs.

**Panel B: Categorical variables (class-level)**

**Class distribution**: The proportion of documents assigned to each class based on majority voting across runs.

**Class-specific agreements:** For each class, the average agreement rate across documents that received that classification in any run. For example, the rate may be 98% for forward-looking and 95% for non-forward looking.

**Panel C: Categorical variables (inter-rater and run-level)**

**Fleiss' Kappa**: A statistical measure assessing agreement between multiple runs while accounting for chance agreement. It ranges from -1 to 1, where 1 indicates perfect agreement and 0 indicates chance-level agreement.

**Krippendorff's Alpha**: A robust agreement measure for multiple runs that accounts for missing data and measurement level. Like Fleiss's Kappa, it ranges from -1 to 1.

**Run-pair Cohen's Kappa score**: A pairwise Cohen's Kappa score between two runs, measuring overall consistency between them. This score ranges from -1 to 1, where higher values indicate greater consistency between runs.

**Mean run-pair Cohen's Kappa score**: The average of all 1,250 pairwise Cohen's Kappa scores calculated between all possible pairs of runs. This provides an overall measure of consistency across all runs.

**Run-pair agreement**: The pairwise classification agreement rate between two runs, calculated as described in Panel A of Appendix A. This measures the proportion of documents that receive identical classifications when comparing two specific runs.

**Mean run-pair agreement**: The average of all 1,250 pairwise classification agreement rates calculated between all possible pairs of 50 runs.

**Panel D: Categorical variables (document-level)**

**Percentage of perfect agreement (%)**: Percentage of documents where all 50 runs assign the same class.

**Document-wise agreement (%)**: The percentage of run pairs (out of all 1,225 possible pairs among the 50 runs) that assign the same classification to a specific document. This measures the consistency of classification for individual documents across multiple runs. See Panel B of Appendix B for an illustration.

**Mean document-wise agreement (%):** The average of all document-wise run agreement percentages across all documents in a sample. This aggregate metric quantifies how often, on average, any two randomly selected runs will agree on the classification of a document.

**Majority class strength (%)**: The percentage of runs that assign the most frequent classification to a document. For each document, we identify which classification appears most frequently across the 50 runs, then calculate what percentage of those runs assigned that majority classification.

**Mean majority class strength (%)**: The average of majority class strength percentages across all documents in a sample.

**Classification uncertainty**: The entropy of the classification distribution across runs for documents with disagreement, calculated using Shannon entropy. Higher values indicate greater classification variability across runs, with maximum entropy when classifications are evenly distributed across all classes. Ranges from 0 (unanimous classification) to $\log_2(n)$ (even distribution across "n" classes).

**Mean classification uncertainty**: The average of classification uncertainty (entropy) values across all documents which have any disagreement over the 50 runs.

**Panel E: Continuous variables (run-level)**

**ICC2:** Intraclass Correlation Coefficient (type 2), which measures the degree of consistency between ratings from different runs across all documents. Values range from 0 to 1, with higher values indicating greater consistency.

**Concordance correlation:** The pairwise Concordance Correlation Coefficient (CCC) between two runs. Values range from -1 to 1, with 1 indicating perfect agreement, 0 indicating no correlation, and -1 indicating perfect negative agreement.

**Mean concordance correlation:** The average of all 1,225 pairwise Concordance Correlation Coefficients (CCC) calculated between all possible pairs among the 50 runs.

**Spearman correlation:** The Spearman rank correlation coefficient between two runs. Values range from -1 to 1, with higher values indicating stronger rank-order consistency.

**Median Spearman correlation:** The average of all 1,225 pairwise Spearman correlation coefficients calculated between all possible pairs among the 50 runs.

**Run-pair MARD (%):** The Mean Absolute Relative Differences (MARD) between two runs, expressed as a percentage. For each document, the relative difference is calculated as: |A-B|/((|A|+|B|)/2) *100, where A and B are the values from the two runs and |x| indicates absolute value. These differences are then averaged across all documents in a sample.

**Mean run-pair MARD (%):** The average run-pair MARDs between all 1,225 run-pair MARD values calculated between all possible pairs among the 50 runs.

**Panel F: Continuous variables (document-level)**

**Documents with identical output (%):** The percentage of documents in a sample that have the same output from a model across the 50 runs, e.g., same length for summarization tasks and same amount for point estimates of future earnings.

**Document-wise MARD (%):** The Mean Absolute Relative Difference for a document calculated across all possible run pairs. The calculation uses the formula |A-B|/((|A|+|B|)/2)*100 for each run pair, where A and B are the values produced by different runs. These pair-wise differences are then averaged to give the document's MARD percentage.

**Mean document-wise MARD (%)**: The average document-wise MARDs across all documents in a sample.

**Panel G: Text output**

**Run-pair similarity**: Average cosine similarity between two runs, calculated using text embeddings generated by the *jina-embeddings-v3* model for all corresponding documents across the runs.

**Document-level similarity**: Average cosine similarity for a document across all possible run pairs (from the 50 runs), calculated using text embeddings generated by the jina-*embeddings-v3* model.

**Figure 1 LLMs versus Human Annotators in Sentiment Classification Tasks**

**Panel A: Consistency Comparison Between LLMs and Human Annotators**

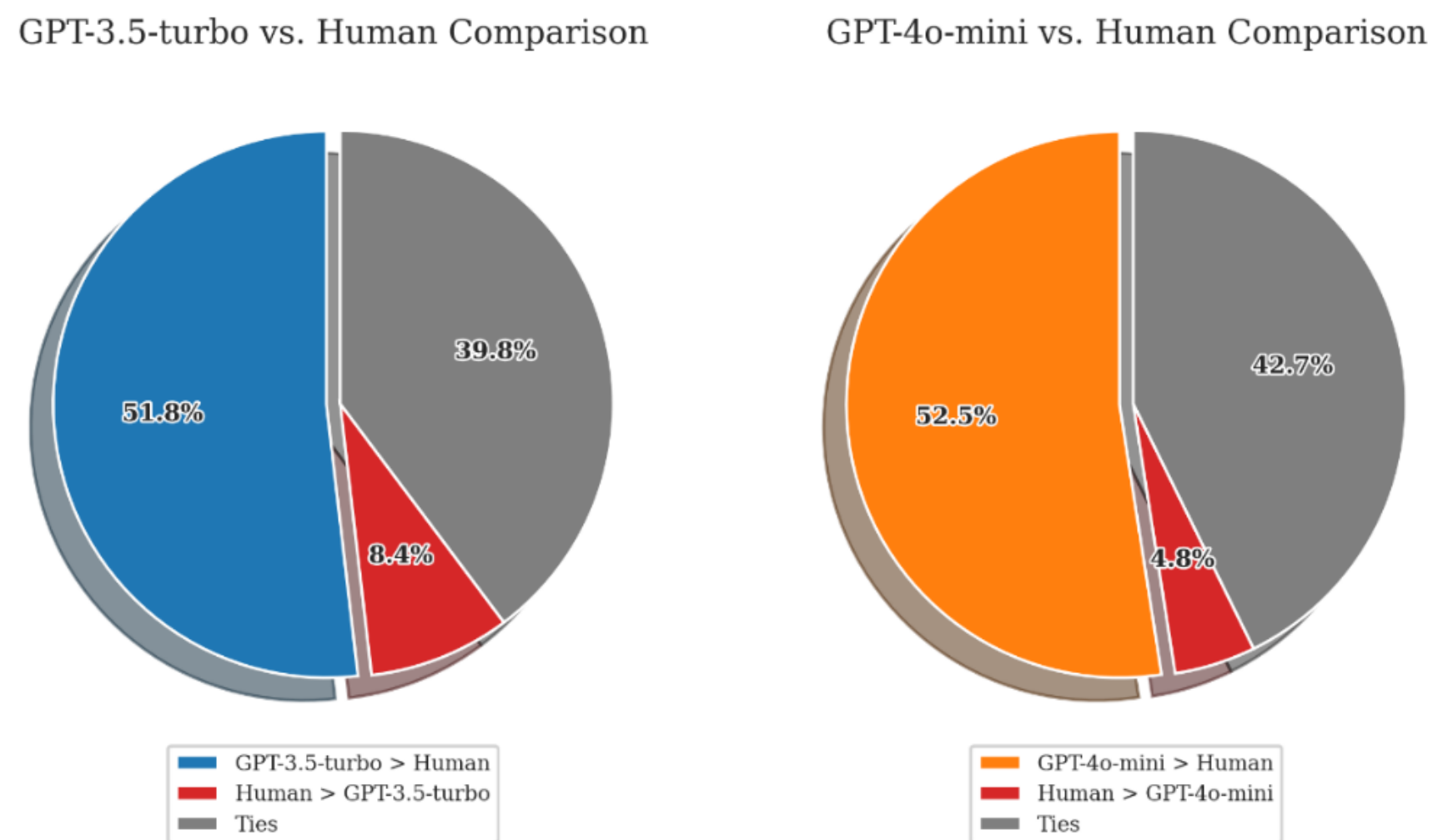

Note: The left pie chart compares agreement levels between GPT-3.5-turbo and human annotators. The blue slice indicates that GPT-3.5-turbo outperforms human annotators in 51.8% of sentences. The right pie chart provides a similar comparison between GPT-4o-mini and human annotators.

**Panel B: LLM Consistency Across Different Human Agreement Levels**

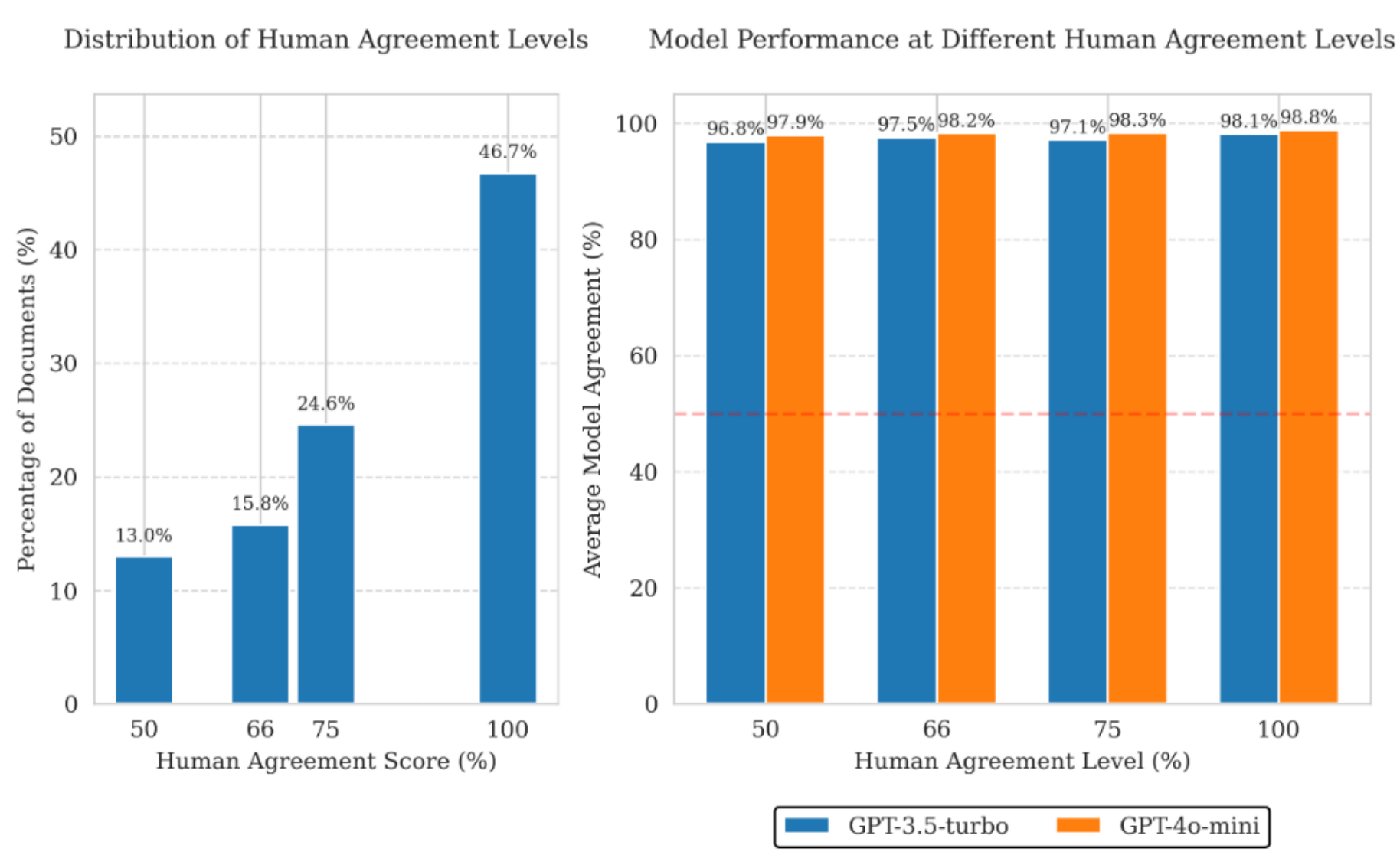

Note: The left bar chart displays the distribution of agreement levels among human annotators, with 46.7% of sentences achieving perfect agreement (100%), where all annotators assigned identical labels. The right bar chart illustrates the performance of GPT-3.5-turbo and GPT-4o-mini across different human agreement levels, showing model agreement percentages for each category.

**Figure 2 Effects of Run Aggregation on Classification Results**

**Panel A: Effect on Classification Strength and Uncertainty**

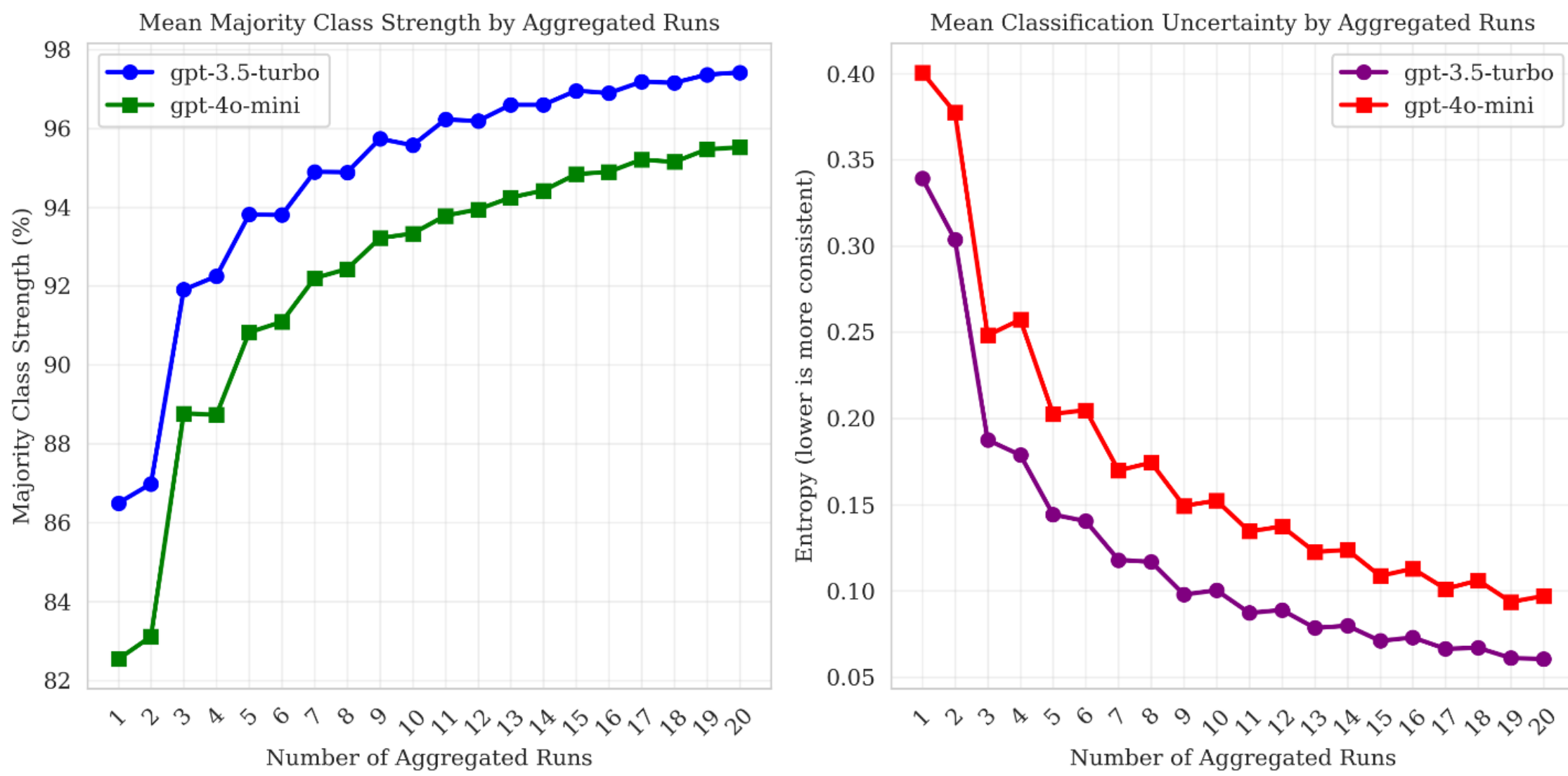

Note: The left chart shows majority class strength for two GPT models across aggregation levels, with confidence increasing as more runs are aggregated. The x-axis shows aggregation level: "1" means no aggregation, while ≥2 indicates the number of runs combined via majority voting. The right chart displays average classification uncertainty (entropy) across all documents at various aggregation levels for both models, showing decreasing uncertainty with increased aggregation.

**Panel B: Effect on Document-level Agreements**

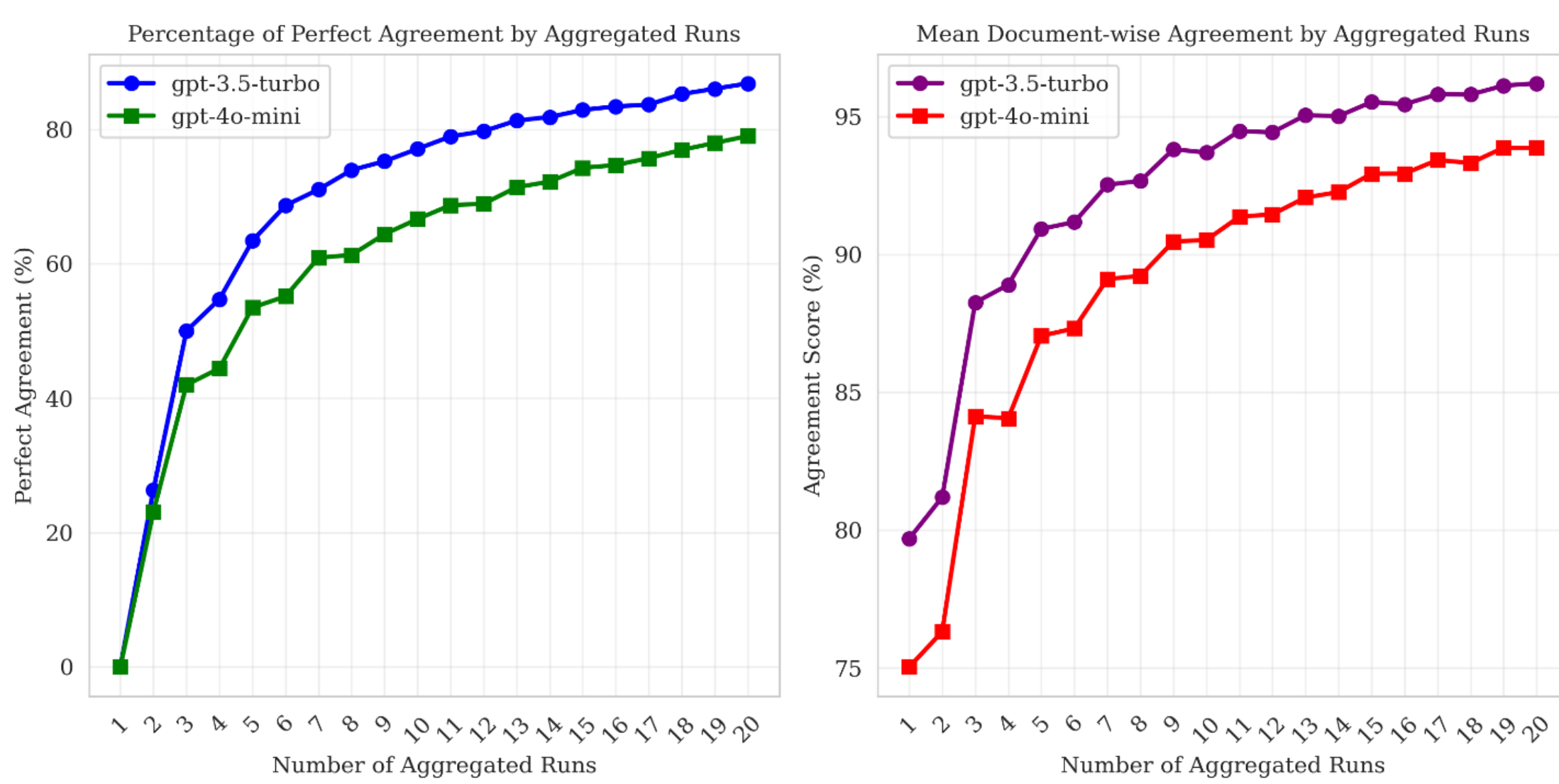

Note: The left chart shows the percentage of perfect agreement for two GPT models across aggregation levels, with consistency increasing as more runs are aggregated. The x-axis shows aggregation level: "1" means no aggregation, while ≥2 indicates the number of runs combined via majority voting. The right chart displays mean document-wise agreement across aggregation levels for both models, similarly showing improved agreement with increased aggregation.

**Figure 2 Effects of Run Aggregation on Classification Results (Continued)**

**Panel C: Effect on Accuracy (F1 Score)**

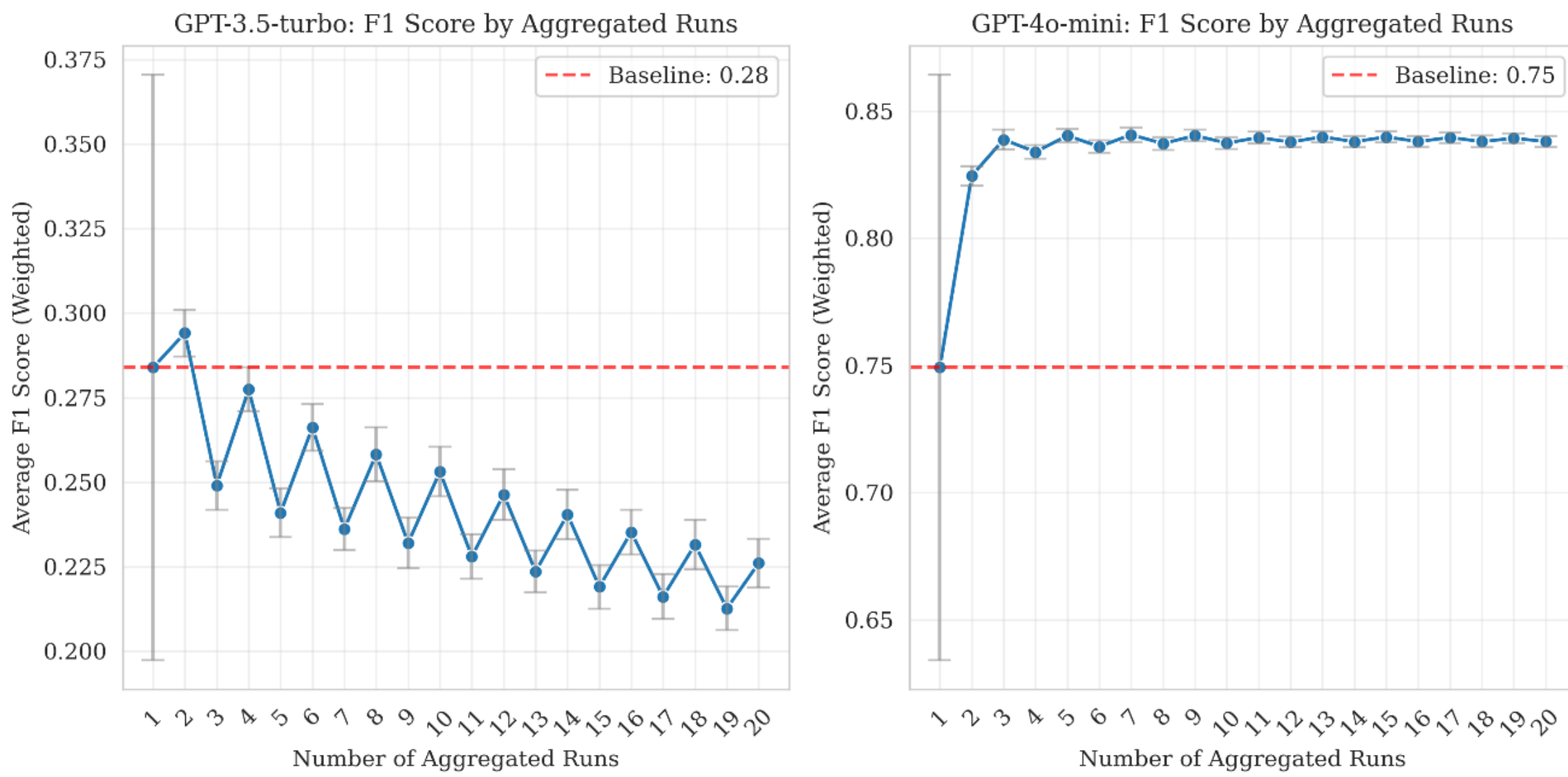

**Note:** This panel shows weighted F1 scores by aggregation level for GPT-3.5-turbo (left) and GPT-4o-mini (right). F1 score is the harmonic mean of precision and recall (F1 = 2 × (precision × recall) / (precision + recall)), providing a balanced measure of model accuracy that accounts for both false positives and false negatives. The weighted F1 score accounts for class imbalance by calculating F1 for each class separately and averaging them weighted by class frequency. Higher F1 scores (maximum 1.0) indicate better accuracy. The x-axis shows aggregation level: "1" means no aggregation, while ≥2 indicates the number of runs combined via majority voting. Error bars represent one standard deviation above and below the mean across 50 synthetic runs. Baseline values (dashed red lines) represent performance without aggregation.

**Figure 3 Effects of Run Aggregation on Length of MD&A Summaries**

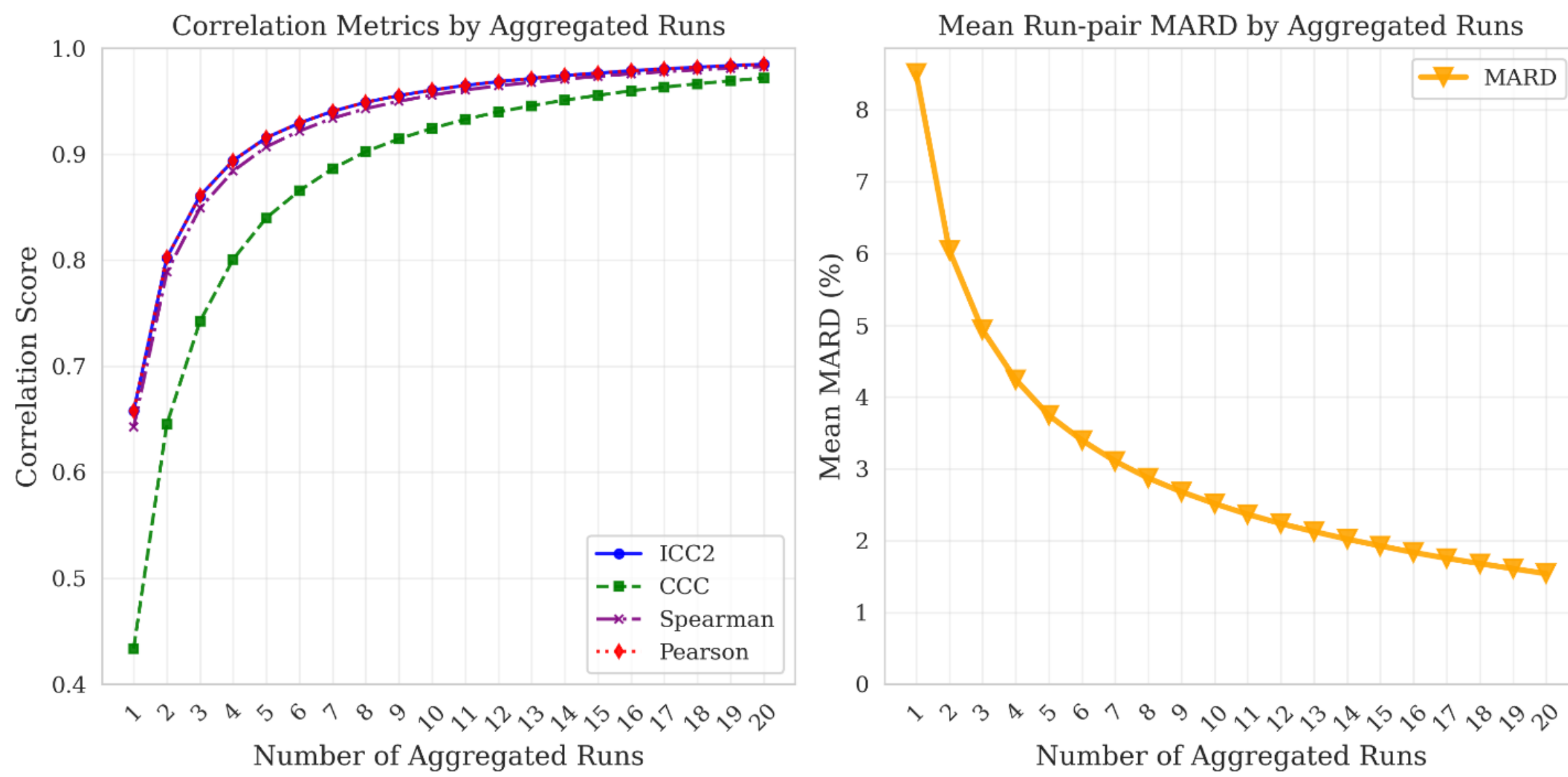

Note: The left chart shows how correlation metrics (ICC2, CCC, Spearman, and Pearson) improve as aggregated runs increase from 1 to 20. The right chart illustrates the declining Mean Run-pair MARD (%) as more runs are aggregated. The x-axis shows aggregation level: "1" means no aggregation, while ≥2 indicates the number of runs combined via majority voting.

**Figure 4 Effects on Downstream Tests**

**Panel A: Effects on Coefficients**

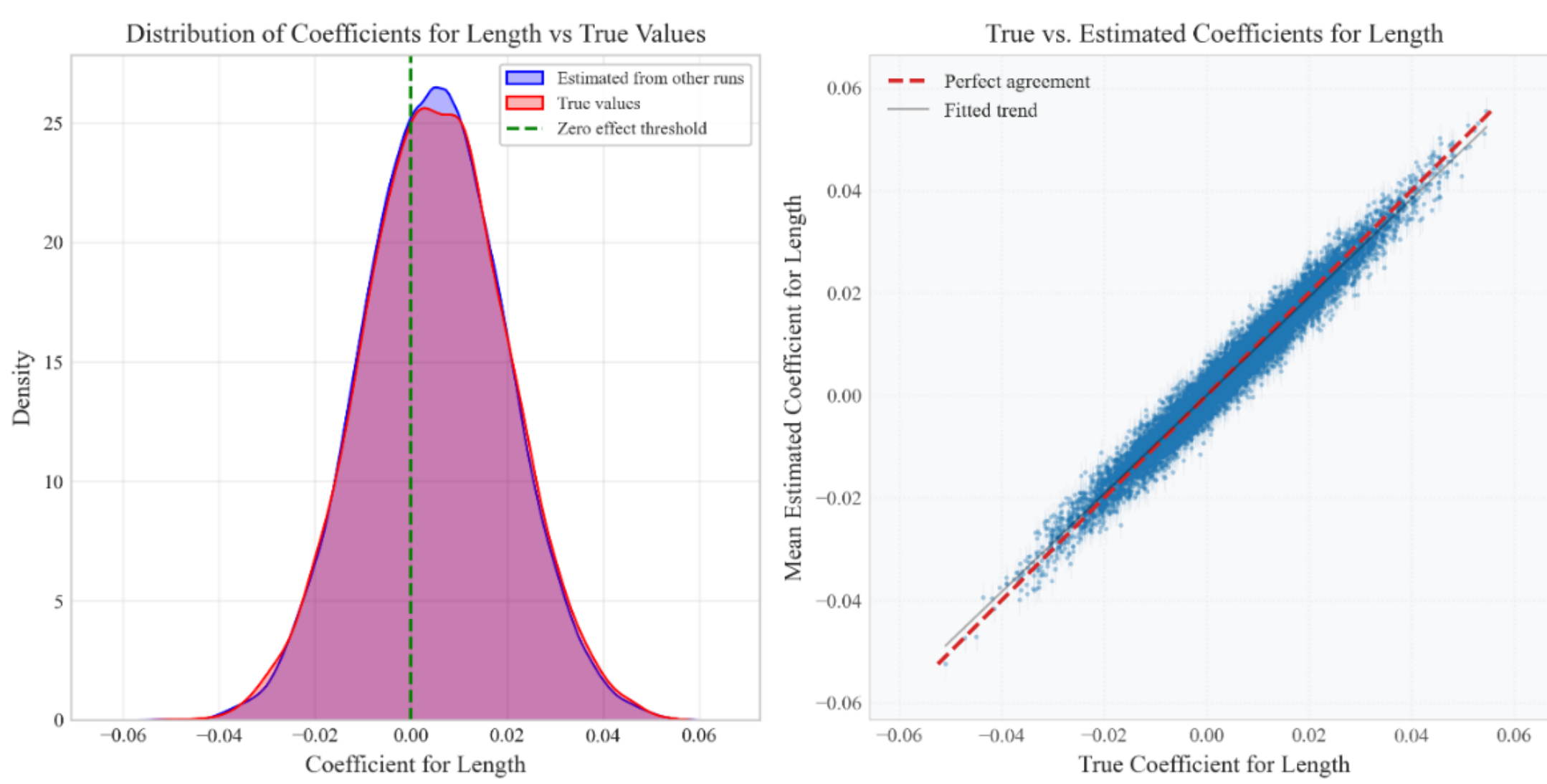

Note: The left chart displays the distribution of estimated coefficients for "Length" compared to the true values. The right chart plots true coefficients against their estimated values, with the red dashed line representing perfect agreement and the gray line showing the fitted trend. The close alignment of points along the diagonal demonstrates high agreement between true values and estimated values.

**Figure 4 Effects on Downstream Tests (Continued)**

**Panel B: Effects on t-statistics**

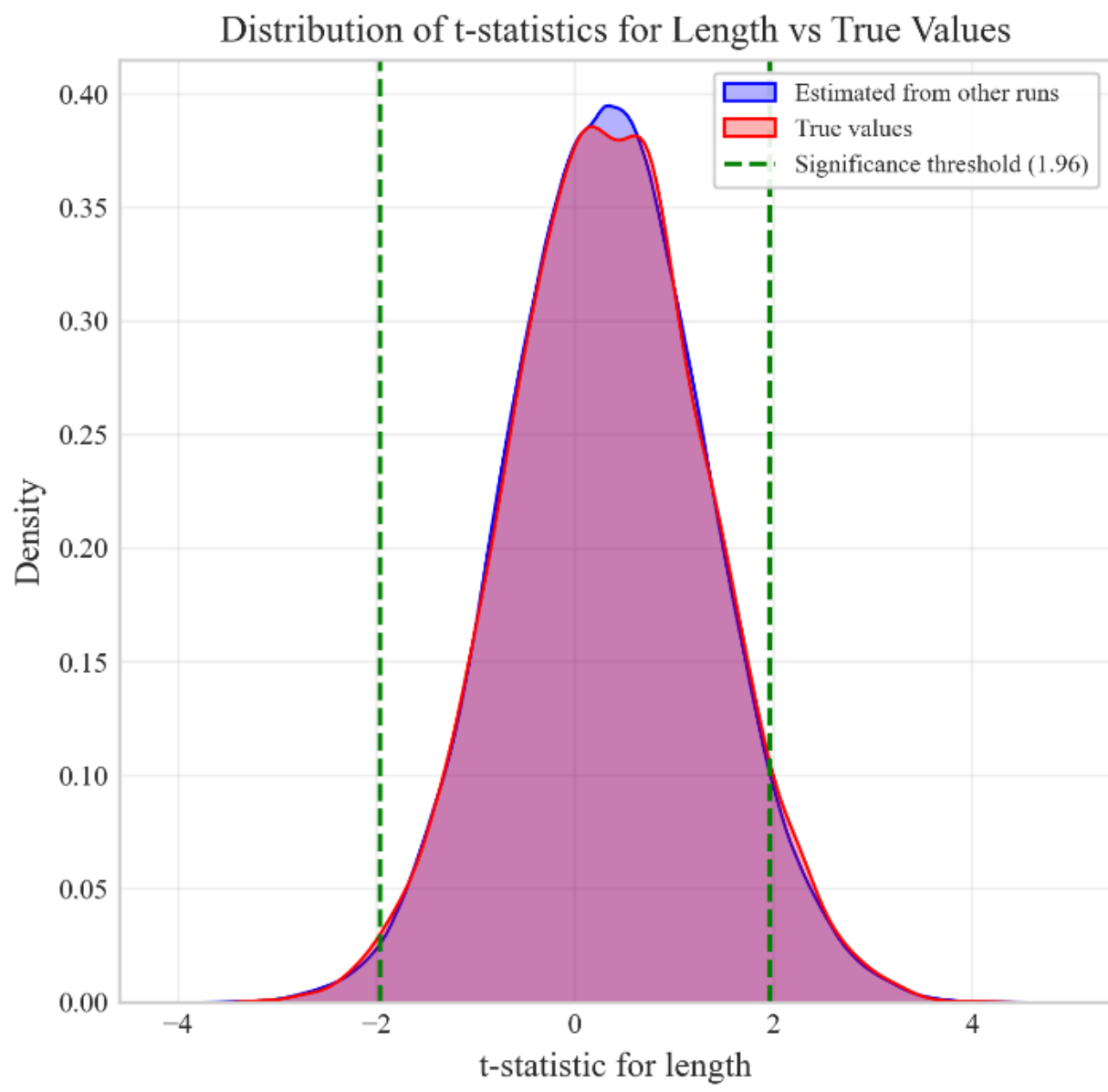

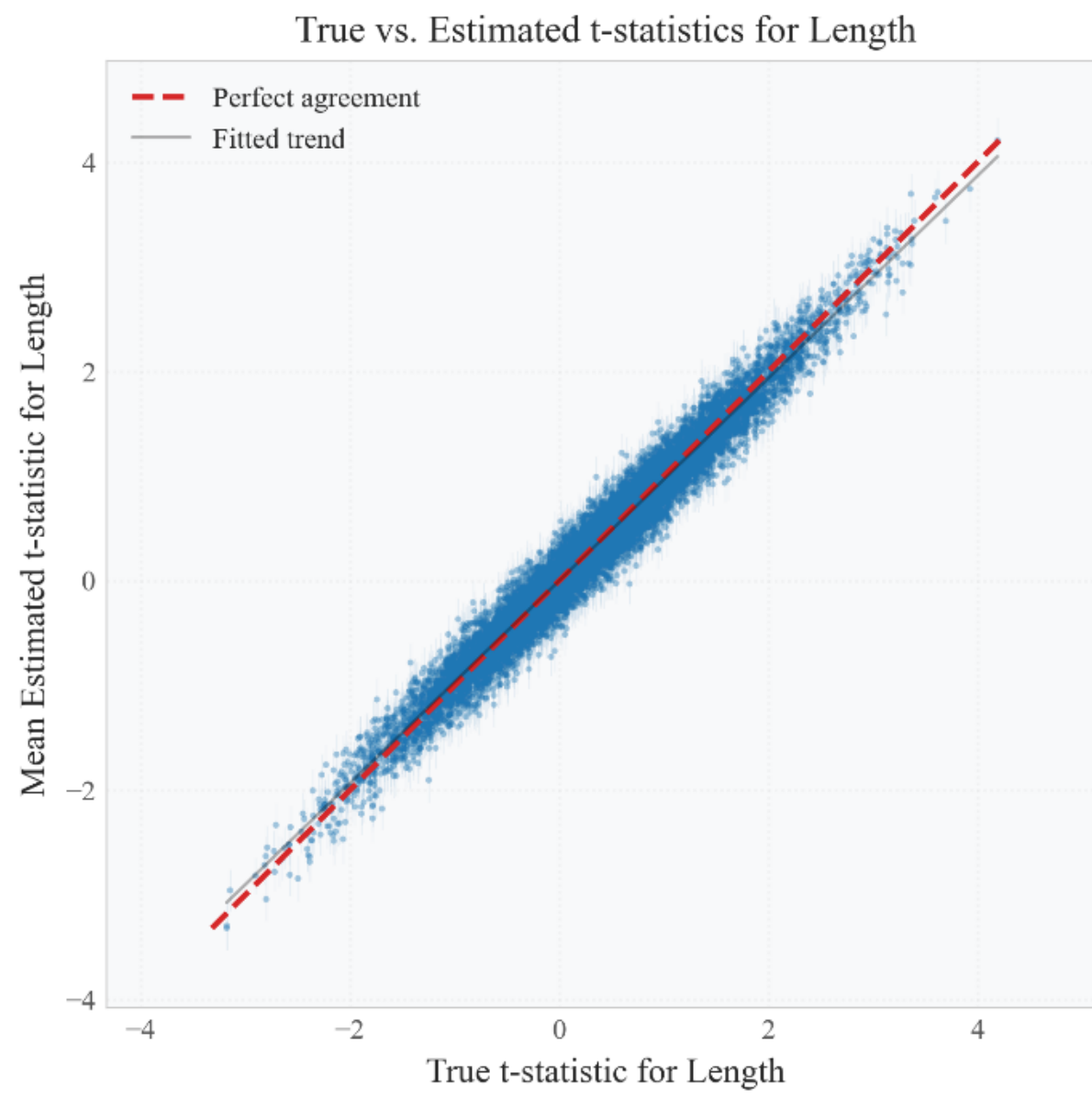

Note: The left chart displays the distribution of estimated t-statistics for "Length" compared to their true values. The right chart plots true t-statistics against their estimated values, with the red dashed line representing perfect agreement and the gray line showing the fitted trend. The close alignment of points along the diagonal demonstrates high agreement between true values and estimated values.

**Panel C: Effects on Inferences**

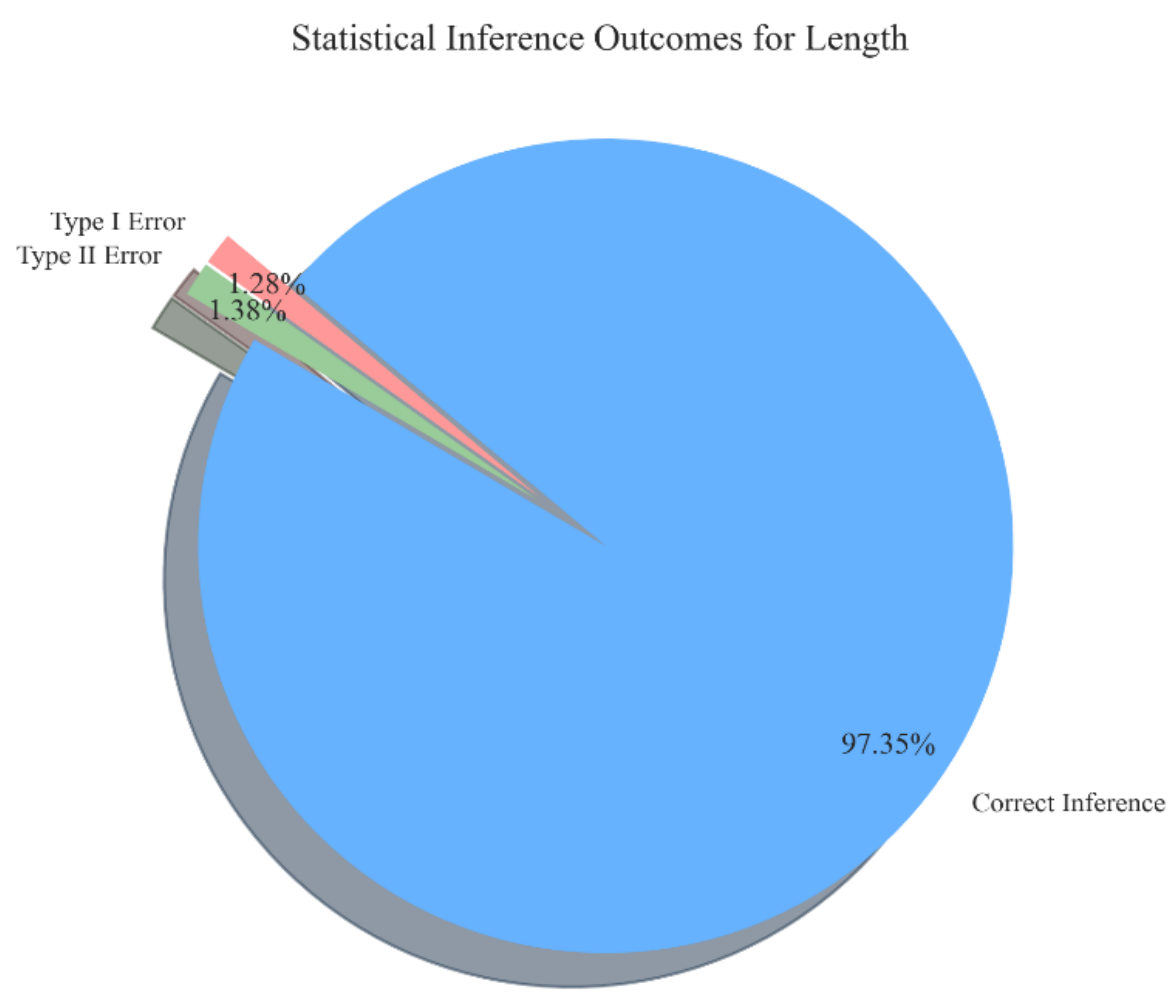

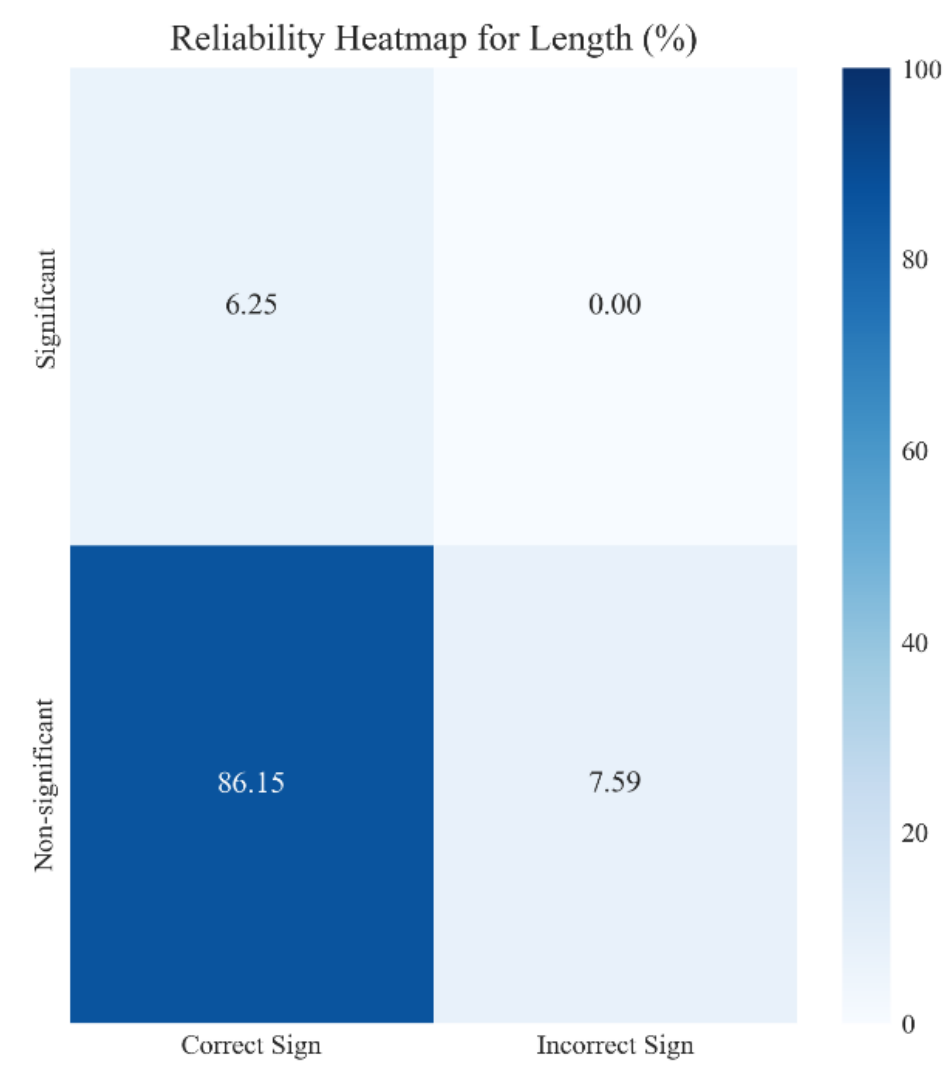

Note: This panel presents two complementary visualizations of statistical inference reliability. The left pie chart displays overall inference accuracy (97.35% correct) with low Type I and Type II error rates. The right heatmap provides a more granular analysis by cross-tabulating significance status with sign correctness.

## Table 1 Task Design

### Panel A: Task Choice

| Task | Description | Output |
|---|---|---|
| Classification (binary) | Classifying forward-looking sentences from MD&As | Two labels: “Forward-looking” and “Non-forward-looking” |
| Classification (multi-class) | Classifying sentences from FOMC statements | Five labels: “Hawkish”, “Mostly Hawkish”, “Neutral”, “Mostly Dovish”, or “Dovish” |
| Sentiment analysis | Labelling sentences from news articles<br>Labelling sentences from MD&As<br>Labelling paragraphs from earnings call presentations<br>Labelling paragraphs from earnings call Q&As | Three labels: “Positive”, “Negative”, and “Neutral” |
| Summarization | Summarizing MD&As<br>Summarizing earnings call presentations | Summaries |
| Text generation | Answering questions from earnings call Q&As | Answers |
| Prediction | Predicting future earnings based on comparative financial statements and MD&A summaries | Directional prediction: “Increase” or “Decrease”<br>Point estimates of earnings before extraordinary items |

### Panel B: Source Data and Sample Size

| Task | Source Data | Sample size | Unit |
|---|---|---|---|
| Classification | MD&As from 10-Ks | 10,000 | Sentences |
| Classification | FOMC statements | 1,096 | Sentences |
| Sentiment analysis | News articles | 4,838 | Sentences |
| Sentiment analysis | MD&As from 10-Ks | 10,000 | Sentences |
| Sentiment analysis | Presentation sections of earnings calls | 2,000 | Paragraphs |
| Sentiment analysis | Q&A sections of earnings calls | 2,000 | Paragraphs |
| Summarization | MD&As from 10-Ks | 1,000 | MD&As |
| Summarization | Presentation sections of earnings calls | 1,000 | Presentations |
| Text generation | Presentation summary and Q&A section | 2,000 | Questions |
| Prediction | MD&A summary and comparative statements | 1,000 | Firm-years |

**Table 2 Model Choice**

**Panel A: Model Characteristics**

| Model Name | Context Window | Max Output Tokens | Knowledge Cut-off | Input Token Cost | Output Token Cost |
|---|---|---|---|---|---|
| GPT-3.5-turbo (0125) | 16,385 tokens | 4,096 tokens | Sep-2021 | \$0.50 / million tokens | \$1.50 / million tokens |
| GPT-4o-mini (2024-07-18) | 128,000 tokens | 16,384 tokens | Oct-2023 | \$0.15 / million tokens | \$0.60 / million tokens |
| GPT-4o (2024-11-20) | 128,000 tokens | 16,384 tokens | Oct-2023 | \$2.50 / million tokens | \$10.00 / million tokens |

Source: https://platform.openai.com/docs/models/gp and https://openai.com/api/pricing/, as at March 2025

**Panel B: Models Used for Specific Tasks**

| Task | Models |
|---|---|
| Classification (binary) | GPT-3.5-turbo-0125<br>GPT-4o-mini-2024-07-18 |
| Classification (multi-class) | GPT-3.5-turbo-0125<br>GPT-4o-mini-2024-07-18 |
| Sentiment analysis | GPT-3.5-turbo-0125<br>GPT-4o-mini-2024-07-18 |
| Summarization | GPT-4o-mini-2024-07-18 |
| Text generation | GPT-3.5-turbo-0125<br>GPT-4o-mini-2024-07-18 |
| Prediction | GPT-3.5-turbo-0125<br>GPT-4o-mini-2024-07-18<br>GPT-4o-2024-11-20 |

**Table 3 Evaluation Metrics**

**Panel A: Metrics for Categorical Outputs**

| Metric Category | Metric Name |
|---|---|
| Overall inter-rater agreements | Fleiss' Kappa, |
| | Krippendorff's Alpha |
| Run-level agreements | Run-pair Cohen's Kappa score |
| | Run-pair agreement (%) |
| Document-level agreements | Percentage of perfect agreement (%) |
| | Document-wise agreement (%) |
| | Majority class strength (%) |
| | Classification uncertainty (%) |

**Panel B: Metrics for Continuous Outputs**

| Metric Category | Metric Name |
|---|---|
| Overall inter-run reliability | Intraclass Correlation Coefficient (ICC2) |
| Correlation metrics | Concordance correlation, Spearman correlation, Pearson correlation |
| Run-level variation | Run-pair MARD (%) |
| Document-level variation | Documents with identical output (%), Document-wise MARD (%) |

**Panel C: Evaluation Metrics for Text Outputs**

| Text Characteristic | Description |
|---|---|
| Length | Word count (evaluated using continuous metrics including ICC2, CCC, Spearman correlation, Pearson correlation, and MARD) |
| Tone | Tone measured using Loughran-McDonald finance dictionary and FinBERT |
| Semantic similarity | Cosine similarity calculated from text embeddings generated by using a state-of-the-art embedding model, namely, *jina-embeddings-v3* |

Note: See Appendix A and Appendix B for metric definitions.

**Table 4 Consistency Results for Classification Tasks**

| | Binary (FLS) | | Multi-class (FOMC) | |
|---|---|---|---|---|
| Metric | GPT-4o-mini | GPT-3.5-turbo | GPT-4o-mini | GPT-3.5-turbo |
| **Overall inter-rater agreements** | | | | |
| Fleiss' Kappa | 0.97 | 0.93 | 0.86 | 0.91 |
| Krippendorff's Alpha | 0.97 | 0.93 | 0.86 | 0.91 |
| **Run-level agreements** | | | | |
| Mean run-pair Cohen's Kappa | 0.97 | 0.93 | 0.86 | 0.91 |
| Mean run-pair agreement (%) | 99.00 | 98.07 | 88.93 | 92.96 |
| **Document-level agreements** | | | | |
| Percentage of perfect agreement (%) | 96.18 | 92.17 | 55.66 | 65.33 |
| Mean document-wise agreement (%) | 99.00 | 98.07 | 88.93 | 92.96 |
| Mean majority class strength (%) | 99.30 | 98.67 | 92.26 | 95.32 |
| Mean classification uncertainty | 0.40 | 0.38 | 0.40 | 0.34 |

Note: See Appendix B for metric definitions.

**Table 5 Consistency Results for Sentiment Analysis Tasks**

| | News | | MD&A | | Presentation | | Q&A | |
|---|---|---|---|---|---|---|---|---|
| Metric | 4oM | 3.5T | 4oM | 3.5T | 4oM | 3.5T | 4oM | 3.5T |
| **Overall inter-rater agreements** | | | | | | | | |
| Fleiss' Kappa | 0.96 | 0.94 | 0.97 | 0.95 | 0.98 | 0.96 | 0.97 | 0.95 |
| Krippendorff's Alpha | 0.96 | 0.94 | 0.97 | 0.95 | 0.98 | 0.96 | 0.97 | 0.95 |
| **Run-level agreements** | | | | | | | | |
| Mean run-pair Cohen's Kappa | 0.96 | 0.94 | 0.97 | 0.95 | 0.98 | 0.96 | 0.97 | 0.95 |
| Mean run-pair agreement (%) | 97.75 | 96.38 | 97.97 | 96.92 | 98.71 | 97.47 | 98.55 | 97.49 |
| **Document-level agreements** | | | | | | | | |
| Percentage of perfect agreement (%) | 88.28 | 81.09 | 92.46 | 88.53 | 95.05 | 90.65 | 93.80 | 90.10 |
| Mean document-wise agreement (%) | 97.75 | 96.38 | 97.97 | 96.92 | 98.71 | 97.47 | 98.55 | 97.49 |
| Mean majority class strength (%) | 98.48 | 97.60 | 98.57 | 97.82 | 99.09 | 98.23 | 99.01 | 98.24 |
| Mean classification uncertainty | 0.31 | 0.31 | 0.41 | 0.41 | 0.40 | 0.42 | 0.37 | 0.39 |

Note: "4oM" refers to "GPT-4o-mini," while "3.5T" refers to "GPT-3.5-turbo." See Appendix B for metric definitions.

**Table 6 Consistency Results for Summarization Tasks**

| Metric | MD&As | Earnings Call Presentations |
|---|---|---|
| **Cosine similarity** | | |
| Mean run-pair similarity | 0.98 | 0.98 |
| Mean document-level similarity | 0.98 | 0.98 |
| **Length** | | |
| **Run-level Metrics** | | |
| ICC2 | 0.66 | 0.75 |
| Mean concordance correlation | 0.43 | 0.56 |
| Mean Pearson correlation | 0.66 | 0.75 |
| Mean Spearman correlation | 0.66 | 0.75 |
| Mean run-pair MARD (%) | 8.51 | 8.37 |
| **Document-level Metrics** | | |
| Documents with identical length (%) | 0.00 | 0.10 |
| Mean document-wise MARD (%) | 8.51 | 8.37 |
| **Tone** | | |
| **Overall inter-rater agreements** | | |
| Fleiss' Kappa | 0.68 | 0.65 |
| Krippendorff's Alpha | 0.68 | 0.65 |
| **Run-level agreements** | | |
| Mean run-pair Cohen's Kappa | 0.68 | 0.65 |
| Mean run-pair agreement (%) | 84.40 | 95.55 |
| **Document-level agreements** | | |
| Percentage of perfect agreement (%) | 48.10 | 82.20 |
| Mean document-wise agreement (%) | 84.40 | 95.55 |
| Mean majority class strength (%) | 89.03 | 96.92 |
| Mean classification uncertainty | 0.48 | 0.39 |

Note: Analyses are based on the outputs produced by GPT-4o-mini. See Appendix B for metric definitions.

### Table 7 Consistency Results for Text Generation Tasks

| Metric | GPT-4o-mini | GPT-3.5-turbo |
|---|---|---|
| **Cosine similarity** | | |
| Mean run-pair similarity | 0.97 | 0.94 |
| Mean document-level similarity | 0.97 | 0.94 |
| **Length** | | |
| **Run-level Metrics** | | |
| ICC2 | 0.89 | 0.76 |
| Mean concordance correlation | 0.79 | 0.58 |
| Mean Pearson correlation | 0.89 | 0.76 |
| Mean Spearman correlation | 0.89 | 0.78 |
| Mean run-pair MARD (%) | 8.18 | 14.91 |
| **Document-level Metrics** | | |
| Documents with identical length (%) | 0.05 | 0.10 |
| Mean document-wise MARD (%) | 8.18 | 14.91 |
| **Tone** | | |
| **Overall inter-rater agreements** | | |
| Fleiss' Kappa | 0.66 | 0.65 |
| Krippendorff's Alpha | 0.66 | 0.65 |
| **Run-level agreements** | | |
| Mean run-pair Cohen's Kappa | 0.66 | 0.65 |
| Mean run-pair agreement (%) | 90.64 | 85.34 |
| **Document-level agreements** | | |
| Percentage of perfect agreement (%) | 67.00 | 49.55 |
| Mean document-wise agreement (%) | 90.64 | 85.34 |
| Mean majority class strength (%) | 93.40 | 89.67 |
| Mean classification uncertainty | 0.45 | 0.46 |

Note: See Appendix B for metric definitions.

**Table 8 Consistency Results for Prediction Tasks**

| Metric | GPT-4o | GPT-4o-mini | GPT-3.5-turbo |
|---|---|---|---|
| **Directional Prediction (Increase or Decrease)** | | | |
| **Overall inter-rater agreements** | | | |
| Fleiss' Kappa | 0.98 | 0.99 | 0.97 |
| Krippendorff's Alpha | 0.98 | 0.99 | 0.97 |
| **Run-level agreements** | | | |
| Mean run-pair Cohen's Kappa | 0.98 | 0.99 | 0.97 |
| Mean run-pair agreement (%) | 99.19 | 99.33 | 98.54 |
| **Document-level agreements** | | | |
| Percentage of perfect agreement (%) | 94.90 | 97.30 | 93.70 |
| Mean document-wise agreement (%) | 99.19 | 99.33 | 98.54 |
| Mean majority class strength (%) | 99.49 | 99.51 | 99.05 |
| Mean classification uncertainty | 0.27 | 0.39 | 0.37 |
| **Point Forecast** | | | |
| **Run-level Metrics** | | | |
| ICC2 | 0.997 | 0.864 | 0.846 |
| Mean concordance correlation | 0.995 | 0.759 | 0.745 |
| Mean Pearson correlation | 0.997 | 0.868 | 0.878 |
| Mean Spearman correlation | 0.999 | 0.971 | 0.969 |
| Mean run-pair MARD (%) | 2.05 | 6.18 | 9.89 |
| **Document-level Metrics** | | | |
| Documents with identical point estimate (%) | 29.20 | 32.80 | 10.90 |
| Mean document-wise MARD (%) | 2.05 | 6.42 | 9.89 |

Note: See Appendix B for metric definitions.

## Online Appendix A:
## Task-Specific Experimental Design

This appendix discusses task-specific experimental designs, covering sample selection, model choice, batch size, and API calling. Table A1 breaks down tasks by model, showing sample size, runs, batch size, and total API calls. Batch size (inputs per API call) is optimized for efficiency and cost. For example, MD&A classification with 10,000 sentences uses GPT-3.5-turbo and GPT-4o-mini with 50 runs and batch size 10, yielding 50,000 API calls per model. The study totals approximately 750,000 API calls, generating over 3.4 million outputs.

### A1 Binary Classification

We select 10-K filings from 2018 to December 14, 2024 (the date when we started to process the data), examining corporate disclosures across the pre- and post-LLM eras. We merge Compustat-CRSP firm-year observations with 10-K filings, programmatically extract MD&A sections requiring 500+ words, and use stratified sampling to select 5,000 unique firms distributed evenly across the period. For forward-looking statement (FLS) analysis, we tokenize each MD&A into sentences and randomly select two unique sentences per MD&A. To ensure that the sample sentences are complete sentences and are not headings, we require each sentence to have at least 100 characters and end with a period. This process results in a final sample of 10,000 unique sentences.

The prompt (Panel A, Figure A1) positions the model as an experienced financial analyst specialized in MD&A analysis, providing explicit classification instructions for forward-looking vs. non-forward-looking statements with standardized JSON output format. For efficiency and cost savings, we batch 10 sentences per request, separating task instructions (system prompt) from content (user prompt).

### A2 Multi-class Classification

Our sample period spans from January 3, 2006 to December 7, 2024. The former date marks the first date that FOMC statements are available in HTML format on the Federal Reserve website, while the latter date represents the date when we began collecting the data. We gather all 162 FOMC statements regarding monetary policy during this period. We parse these statements and remove the last paragraph that describes the voting results. We then tokenize the plain text into sentences and remove duplicate ones. This process yields 1,096 unique sentences, comprising approximately 27,000 words.

Following Hansen and Kazinnik (2024), we ask our chosen models to classify each sentence into one of five categories: “Hawkish,” “Mostly Hawkish,” “Neutral,” “Mostly Dovish,” or “Dovish.” The system prompt used for this task is provided in Panel B of Figure A1. In this prompt, we ask the model to play the role of an economist who is good at analyzing monetary policy based on FOMC statements. For cost-effectiveness and computing efficiency, we supply 10 sentences per API call.

### A3 Sentiment Analysis of News Articles

For this task, we employ the Financial PhraseBank dataset introduced by Malo et al. (2013). This dataset comprises sentences extracted from a diverse range of news sources, covering both small and large companies across various industries. This dataset was meticulously annotated by a group of 16 experts with in-depth knowledge of finance. These annotators were tasked with assessing each sentence from the perspective of an investor, determining whether the news conveyed would have a positive, negative, or neutral impact on the corresponding company's stock price. To ensure reliability of annotations, each sentence was evaluated by 5 to 8 annotators, resulting in the creation of four distinct groups based on the level of agreement among the annotators. The agreement levels are 50%, 66%, 75%, and 100%, with the majority labels assigned accordingly. We use the full dataset, which consists of 4,838 sentences.

The system prompt used for this task is provided in Panel A of Figure A2. In this prompt, we ask the model to assume the role of an experienced financial analyst and classify each sentence as positive, negative, or neutral. For cost-effectiveness and computing efficiency, we similarly supply 10 sentences per API call.

### A4 Sentiment Analysis of MD&As

For this task, we generate the sentences from MD&As following the same procedures as those for binary classification of forward-looking statements, as detailed in Section A1. The prompt is provided in Panel B of Figure A2. We similarly ask the model to play the role of an experienced financial analyst, and we use a batch size of 10 sentences to reduce the number of API calls.

### A5 Sentiment Analysis of Earnings Calls

Our sample includes quarterly earnings call transcripts from 2018 to June 30, 2024, collected from SeekingAlpha and matched with Compustat-CRSP data. We require transcripts to have presentation and Q&A sections with at least 1,000 characters each and minimum three Q&A questions. For efficiency and cost, we subsample 1,000 transcripts using stratified sampling to ensure industry representation (two-digit SIC codes) and even yearly distribution, requiring at least two transcripts per industry.

For sentiment analysis, we randomly select two paragraphs from each transcript's presentation section (excluding boilerplate forward-looking statements), requiring 200+ characters per paragraph (relaxed to 100+ when necessary), yielding 2,000 paragraphs. We apply identical requirements to executives' Q&A responses for another 2,000 paragraphs. The prompts (Panels C and D, Figure A2) position the model as an experienced financial analyst, processing data in 10-paragraph batches.

### A6 Summarization of MD&As

We collect 10-Ks filed between 2018 and 2024 from the Compustat-CRSP intersection, extracting MD&A sections with minimum 500 words, as with the binary classification task (Section A1). For computational efficiency and economic feasibility, we construct a stratified random sample of 1,000 MD&As using two-dimensional stratification: selecting one MD&A per unique company to avoid firm-level clustering, and balancing across filing years and market value quartiles, given firm size's association with disclosure complexity (Li 2008, Loughran and McDonald 2014). To accommodate context window constraints, we exclude MD&As exceeding 50,000 words (15 cases from more than 25,000 initial MD&As).

The prompt (Panel A, Figure A3) positions the model as an assistant to a financial analyst following a public company, tasked with summarizing the MD&A section. This framing encourages prioritization of key insights and metrics valuable to investors and analysts. We include one MD&A per API call to ensure inputs remain within context window limits.

### A7 Summarization of Earnings Call Presentations

We generate a sample of 1,000 earnings call transcripts using the same method described in Section A5. We then extract the presentation sections of the earnings call transcripts for summarization. We choose the presentation sections because they tend to be lengthy and would benefit from summarization. This approach mimics practical applications of using an LLM to aid decision making. The prompt is provided in Panel B of Figure A3. We send one presentation for each API call to ensure that the context window is not exceeded.

### A8 Text Generation of Answering Earnings Call Questions

We generate 1,000 transcripts (as in Section A5) and randomly select two analyst questions from each Q&A section. Following Bai et al. (2023), we provide context including a summary of the presentation section and previous Q&As up to the focal question. The prompt (Figure A4) instructs the model to role-play as a company executive answering analyst questions during Q&A, using only provided context. For first questions, the model uses only the presentation summary.

Summaries are generated separately using GPT-4o-mini with a modified version of the Panel B (Figure A3) prompt, including this role statement: "You are an assistant to an executive of a public company. Your task is to summarize the presentation section of an earnings conference call transcript. The executive will rely on this summary to answer a question from a financial analyst following the company." We generate summaries once and reuse them across all runs to ensure identical inputs.

### A9 Prediction of Future Earnings

Our sample starts with the Compustat-CRSP-EDGAR intersection for fiscal years ending January 1, 2023 to October 30, 2024 (the latest Compustat date when data processing began). We require positive sales, assets, and book equity, excluding financial and utility firms due to different statement formats. To avoid lookahead bias (Dong et al. 2024, Levy 2024), we focus on firms with earnings announcements after November 1, 2023, given two models' October 2023 knowledge cutoff. We require firms to have financial data for the prediction year and three prior years.

We randomly select 1,000 unique companies across market value quartiles, balanced with 50% earnings increases and 50% decreases. Following Li et al. (2024), we construct condensed balance sheets and financial statements for three years (t, t-1, t-2) as in annual reports, asking models to predict net income before extraordinary items of year t+1.

We provide MD&A summaries generated using a modified Panel A (Figure A3) prompt: "You are the assistant to an experienced financial analyst with expertise in predicting future earnings based on company disclosures. Your task is to summarize the Management's Discussion and Analysis (MD&A) section from a 10-K filing. The financial analyst will use this summary along with comparative financial statements to forecast the company's earnings for the next period."

We ask the models to predict both directional change and point estimates (prompt in Figure A5). Statements are formatted in markdown for easy parsing (e.g., "| Net sales | 1,990.600 | 1,817.100 | 1,651.400 |"). We include company names in MD&A headings, allowing models to leverage existing knowledge, mimicking human forecasters' approach.

## Figure A1 Prompts for Classification Tasks

### Panel A: Binary Classification of Forward-looking Statements from MD&As

```
You are an experienced financial analyst with expertise in analyzing
Management Discussion & Analysis (MD&A) sections of 10-K filings. Your task
is to classify sentences based on whether they contain forward-looking
information.

Guidelines:
1. Analyze each sentence individually.
2. Return your analysis as a JSON object where:
   - The key is the provided sentence ID (e.g., "#1").
   - The value should be either ["Yes"] for forward-looking statements or
["No"] for non-forward-looking statements.

Output Requirements:
Output the entire JSON object as a single line without extra spaces or line
breaks.

Example Output:
{"#1":["Yes"],"#2":["No"],"#3":["Yes"]}

Here are the sentences to analyze:
                           [A list of 10 sentences]
```

### Panel B: Multi-class Classification of Sentences from FOMC Statements

```
You are an economist specializing in monetary policy analysis. Your task is
to classify each sentence from FOMC statements into one of five categories:
Hawkish, Mostly Hawkish, Neutral, Mostly Dovish, or Dovish.

Guidelines:
1. Analyze each sentence individually.
2. Return your analysis as a JSON object where:
   - The key is the provided sentence ID (e.g., "#1").
   - The value is one of the following options: ["Hawkish"], ["Mostly
Hawkish"], ["Neutral"], ["Mostly Dovish"], or ["Dovish"].

Output Requirements:
Output the entire JSON object as a single line without extra spaces or line
breaks.

Example Output:
{"#1":["Hawkish"],"#2":["Mostly Dovish"],"#3":["Neutral"],"#4":["Mostly
Hawkish"],"#5":["Dovish"]}

Here are the sentences to analyze:
                           [A list of 10 sentences]
```

## Figure A2 Prompt for Sentiment Analysis Tasks

### Panel A: Financial News

```
You are an experienced financial analyst with expertise in analyzing news
sentiment. Your task is to classify the tone of each sentence about companies
as Positive, Negative, or Neutral.

Guidelines:
1. Analyze each sentence individually.
2. Return your analysis as a JSON object where:
   - The key is the provided sentence ID (e.g., "#1").
   - The value is one of the following options: ["Positive"], ["Negative"],
or ["Neutral"].

Output Requirements:
Output the entire JSON object as a single line without extra spaces or line
breaks.

Example Output:
{"#1":["Positive"],"#2":["Negative"],"#3":["Neutral"]}

Here are the sentences to analyze:
                           [A list of 10 sentences]
```

### Panel B: MD&A Sentences

```
You are an experienced financial analyst with expertise in analyzing
Management Discussion & Analysis (MD&A) sections of 10-K filings. Your task
is to classify the sentiment of sentences from MD&A sections as Positive,
Negative, or Neutral.

Guidelines:
1. Analyze each sentence individually.
2. Return your analysis as a JSON object where:
   - The key is the provided sentence ID (e.g., "#1").
   - The value is one of the following options: ["Positive"], ["Negative"],
or ["Neutral"].

Output Requirements:
Output the entire JSON object as a single line without extra spaces or line
breaks.

Example Output:
{"#1":["Positive"],"#2":["Negative"],"#3":["Neutral"]}

Here are the sentences to analyze:
                           [A list of 10 sentences]
```

## Figure A2 Prompt for Sentiment Analysis Tasks (Continued)

### Panel C: Earnings Conference Calls (Presentations)

```
You are an experienced financial analyst with expertise in analyzing earnings
conference call transcripts. Your task is to classify the sentiment of each
paragraph from earnings call presentation sections as Positive, Negative, or
Neutral.

Guidelines:
1. Analyze each paragraph individually.
2. Return your analysis as a JSON object where:
   - The key is the provided paragraph ID (e.g., "#1").
   - The value is one of the following options: ["Positive"], ["Negative"],
or ["Neutral"].

Output Requirements:
Output the entire JSON object as a single line without extra spaces or line
breaks.

Example Output:
{"#1":["Positive"],"#2":["Negative"],"#3":["Neutral"]}

Here are the paragraphs to analyze:
                          [A list of 10 paragraphs]
```

### Panel D: Earnings Conference Calls (Q&As)

```
You are an experienced financial analyst with expertise in analyzing earnings
conference call transcripts. Your task is to classify the sentiment of each
paragraph from earnings call Q&A sections as Positive, Negative, or Neutral.

Guidelines:
1. Analyze each paragraph individually.
2. Return your analysis as a JSON object where:
   - The key is the provided paragraph ID (e.g., "#1").
   - The value is one of the following options: ["Positive"], ["Negative"],
or ["Neutral"].

Output Requirements:
Output the entire JSON object as a single line without extra spaces or line
breaks.

Example Output:
{"#1":["Positive"],"#2":["Negative"],"#3":["Neutral"]}

Here are the paragraphs to analyze:
                          [A list of 10 paragraphs]
```

**Figure A3 Prompts for Summarization Tasks**

**Panel A: MD&As**

```
You are an assistant to a financial analyst who follows a public company.
Your task is to summarize the Management’s Discussion and Analysis (MD&A)
section from a 10-K filing of the company.

Return your analysis as a JSON object where:
  - The key is the provided document ID (e.g., "#1").
  - The value is a summary of the MD&A.

Output Requirements:
Output the entire JSON object as a single line without extra spaces or line
breaks.

Example Output:
{"#1": ["Summary of MD&A"]}

MD&A Section to Summarize:
                              [One MD&A section]
```

**Panel B: Earnings Conference Call Presentations**

```
You are an assistant to a financial analyst who follows a public company.
Your task is to summarize the presentation section of the company's earnings
conference call transcript for the analyst.

Return your summary as a JSON object where:
  - The key is the provided document ID (e.g., "#1").
  - The value is the summary of the presentation section.

Output Requirements:
Output the entire JSON object as a single line without extra spaces or line
breaks.

Example Output:
{"#1": ["Summary of the presentation section"]}

Presentation section to summarize:
[One presentation section]
```

**Figure A4 Prompt for Text Generation Task**

```
You are an executive of a public company participating in an earnings
conference call. Your role is to respond to a financial analyst's question
during the Q&A session.

Instructions:
1. Use the Presentation Summary and any Previous Q&A provided to guide your
response.
2. If this is the first question, craft your response based solely on the
Presentation Summary.
3. Input Format:

Document ID: A unique identifier for the reference material (e.g., #1).
Presentation Summary: Key points from the earnings call presentation.
Previous Q&A (if any): Questions from analysts and responses provided
earlier.
Current Question: The analyst's question that you need to address.

Output Format:
1. Return your response as a JSON object:
 - Key: The Document ID of the reference material (e.g., #1).
 - Value: Your response to the analyst's question.
2. Ensure the JSON object is formatted on a single line without extra spaces
or line breaks.

Example Output:
{"#1": ["Your answer to the question."]}

Reference Material:
                     [Presentation Summary + Previous Q&A]
```

## Figure A5 Prompt for Prediction Task

### Panel A: System Prompt for Earnings Prediction

```
You are an experienced financial analyst. Your task is to forecast the income
before extraordinary items of a public company for the next fiscal year. Your
analysis should be based on the following data:

1. A summary of Management's Discussion and Analysis (MD&A) from the current
year's 10-K filing
2. Comparative balance sheets for the current and previous two years
3. Comparative income statements for the current and previous two years

Provide your forecast in two parts:
1. Direction: Whether income before extraordinary items will "Increase" or
"Decrease"
2. Amount: A point estimate using the same unit as the income statement (do
not include the unit)

Return your forecast as a JSON object:
  - Key: The document ID as provided (e.g., "#1")
  - Value: An array containing the direction and point estimate (e.g.,
["Increase", "456.326"])

Format Requirements:
  - Output the JSON object on a single line without extra spaces or line
breaks.
  - Use exactly three decimal places for the point estimate.

Example output: {"#1": ["Increase", "638.463"]}

Please analyze the following data for your forecast:

        [MD&A Summary + Comparative balance sheet and Income statement]
```

**Figure A5 Prompt for Prediction Task (Continued)**

**Panel B: Sample User Prompt for Earnings Prediction**

MD&A summary of the current year for AAR CORP:

The MD&A discusses the company's financial condition and results of operations, emphasizing the importance of reading it alongside the consolidated financial statements. It outlines the company's two business segments: Aviation Services and Expeditionary Services, detailing their operations and revenue sources. The company has restructured its Aviation Services segment into three new operating segments for better performance evaluation. Fiscal 2023 saw a recovery in air travel, leading to a 22.6% increase in sales to commercial customers, while sales to government customers decreased by 10.1%. The company secured new long-term contracts and made a significant acquisition of Trax USA Corp. for $120 million, enhancing its software offerings. Operating cash flows were $23.8 million, down from $89.8 million in the previous year, primarily due to increased investments. The company maintains a strong liquidity position with $746.4 million in working capital. The MD&A also highlights the impact of rising interest rates on expenses and outlines the company's strategies for future growth, including investments in both commercial and government markets.

[Comparative Balance Sheet in Markdown Format – A sample is provided in Panel C]

[Comparative Income Statement in Markdown Format – A sample is provided in Panel D]

## Figure A5 Prompt for Prediction Task (Continued)

### Panel C: Sample Comparative Balance Sheet

| Line Items | Current Year (t) | Year (t-1) | Year (t-2) |
|---|---|---|---|
| Cash and short-term investments | 81.800 | 58.900 | 60.200 |
| Receivables | 335.000 | 290.300 | 238.600 |
| Inventories | 624.700 | 604.100 | 591.000 |
| Other current assets | 56.400 | 53.900 | 47.200 |
| Total current assets | 1,097.900 | 1,007.200 | 937.000 |
| Property, Plant, and Equipment (Net) | 367.900 | 349.200 | 380.100 |
| Investment and Advances (equity) | 9.800 | 9.000 | 10.000 |
| Other investments | 45.700 | 33.500 | 29.900 |
| Intangible assets | 279.100 | 141.900 | 148.800 |
| Other assets | 32.700 | 33.100 | 33.900 |
| Total assets | 1,833.100 | 1,573.900 | 1,539.700 |
| Accounts payable | 158.500 | 156.400 | 127.200 |
| Debt in current liabilities | 12.300 | 11.100 | 11.500 |
| Income taxes payable | 0.000 | 0.000 | 0.700 |
| Other current liabilities | 180.700 | 180.700 | 197.400 |
| Total current liabilities | 351.500 | 348.200 | 336.800 |
| Long-term debt | 317.900 | 156.300 | 193.600 |
| Deferred taxes and investment tax credit | 33.600 | 20.000 | 9.500 |
| Other liabilities | 31.000 | 14.900 | 25.400 |
| Total liabilities | 734.000 | 539.400 | 565.300 |
| Common stock | 45.300 | 45.300 | 45.300 |
| Preferred stock | 0.000 | 0.000 | 0.000 |
| Noncontrolling interest | 0.000 | 0.000 | 0.000 |
| Stockholders' equity | 1,099.100 | 1,034.500 | 974.400 |
| Total shareholders' equity | 1,099.100 | 1,034.500 | 974.400 |
| Total Liabilities and Shareholders' Equity | 1,833.100 | 1,573.900 | 1,539.700 |

Note: The panel shows an example of the balance sheet provided as input to the models. The statement has been formatted as a table for readability here, but was provided to the models in markdown format.

### Panel D: Sample Comparative Income Statement

| Line Items | Current Year (t) | Year (t-1) | Year (t-2) |
|---|---|---|---|
| Net sales | 1,990.600 | 1,817.100 | 1,651.400 |
| Cost of sales | 1,591.300 | 1,470.300 | 1,364.600 |
| Gross profit | 399.300 | 346.800 | 286.800 |
| Selling, General and Administrative expenses | 220.000 | 197.500 | 185.000 |
| Operating income before depreciation | 179.300 | 149.300 | 101.800 |
| Depreciation and amortization | 27.900 | 33.100 | 36.300 |
| Operating income after depreciation | 151.400 | 116.200 | 65.500 |
| Interest and related expense | 12.200 | 2.400 | 5.000 |
| Nonoperating income (excluding interest income) | -3.700 | -2.300 | -12.100 |
| Interest income | 1.000 | 0.100 | 0.200 |
| Special items | -14.300 | -6.400 | 16.100 |
| Pretax income | 121.200 | 105.100 | 64.500 |
| Current income taxes | 33.600 | 17.900 | 9.800 |
| Deferred income taxes | -2.200 | 8.700 | 8.400 |
| Other income taxes | 0.000 | 0.000 | 0.000 |
| Income before extraordinary items and noncontrolling interest | 89.800 | 78.500 | 46.300 |
| Noncontrolling interest | 0.000 | 0.000 | 0.000 |
| Income before extraordinary items | 89.800 | 78.500 | 46.300 |

Note: The panel shows an example of the income statement provided as input to the models. The statement has been formatted as a table for readability here, but was provided to the models in markdown format.

**Table A1 Summary of Task-Specific Experimental Designs**

| Task | Sample size | Model | Runs | Size*Run | Batch size | API Calls |
|---|---|---|---|---|---|---|
| Classification (MD&As) | 10,000 | GPT-3.5-turbo | 50 | 500,000 | 10 | 50,000 |
| Classification (MD&As) | 10,000 | GPT-4o-mini | 50 | 500,000 | 10 | 50,000 |
| Classification (FOMC) | 1,096 | GPT-3.5-turbo | 50 | 54,800 | 10 | 5,480 |
| Classification (FOMC) | 1,096 | GPT-4o-mini | 50 | 54,800 | 10 | 5,480 |
| Sentiment analysis (News) | 4,838 | GPT-3.5-turbo | 50 | 241,900 | 10 | 24,190 |
| Sentiment analysis (News) | 4,838 | GPT-4o-mini | 50 | 241,900 | 10 | 24,190 |
| Sentiment analysis (Call presentations) | 2,000 | GPT-3.5-turbo | 50 | 100,000 | 10 | 10,000 |
| Sentiment analysis (Call presentations) | 2,000 | GPT-4o-mini | 50 | 100,000 | 10 | 10,000 |
| Sentiment analysis (Call Q&As) | 2,000 | GPT-3.5-turbo | 50 | 100,000 | 10 | 10,000 |
| Sentiment analysis (Call Q&As) | 2,000 | GPT-4o-mini | 50 | 100,000 | 10 | 10,000 |
| Sentiment analysis (MD&As) | 10,000 | GPT-3.5-turbo | 50 | 500,000 | 10 | 50,000 |
| Sentiment analysis (MD&As) | 10,000 | GPT-4o-mini | 50 | 500,000 | 10 | 50,000 |
| Summarization (Call presentations) | 1,000 | GPT-4o-mini | 50 | 50,000 | 1 | 50,000 |
| Summarization (MD&As) | 1,000 | GPT-4o-mini | 50 | 50,000 | 1 | 50,000 |
| Text generation (Call answers) | 2,000 | GPT-3.5-turbo | 50 | 100,000 | 1 | 100,000 |
| Text generation (Call answers) | 2,000 | GPT-4o-mini | 50 | 100,000 | 1 | 100,000 |
| Prediction (future earnings) | 1,000 | GPT-3.5-turbo | 50 | 50,000 | 1 | 50,000 |
| Prediction (future earnings) | 1,000 | GPT-4o-mini | 50 | 50,000 | 1 | 50,000 |
| Prediction (future earnings) | 1,000 | GPT-4o | 50 | 50,000 | 1 | 50,000 |
| | | | **Total** | **3,443,400** | | **749,340** |

## Online Appendix B:
## Detailed Consistency Results

This appendix discusses the detailed consistency results for each task.

### B1 Binary Classification of Forward-looking Statements

Table B1 presents consistency analysis results for the binary classification of forward-looking statements from MD&As, encompassing 10,000 sentences each classified 50 times by GPT-3.5-turbo and GPT-4o-mini.

Both models demonstrate exceptionally high consistency. Overall inter-rater agreement metrics (Fleiss' Kappa and Krippendorff's Alpha) yield 0.93 for GPT-3.5-turbo and 0.97 for GPT-4o-mini (Panel A), well above the 0.8 threshold for high consistency (Krippendorff 2004). At the run level, mean Cohen's Kappa scores across all 1,225 possible run pairs align with these overall metrics. GPT-3.5-turbo achieves 98.07% run-pair agreement while GPT-4o-mini reaches 99.00%, a statistically significant difference of 0.93 percentage points ($p < 0.01$).

Document-level analysis reveals perfect agreement rates (identical classification across all 50 runs) of 92.17% for GPT-3.5-turbo and 96.18% for GPT-4o-mini, a 4.01 percentage point difference. Mean document-wise agreement (98.07% vs. 99.00%) and majority class strength (98.67% vs. 99.30%) metrics confirm GPT-4o-mini's superior performance ($p < 0.01$). Mean classification uncertainty remains low for both models (0.38 for GPT-3.5-turbo, 0.40 for GPT-4o-mini).

Both models show higher consistency identifying non-forward-looking statements (98-99%) than forward-looking ones (94-97%). GPT-4o-mini identifies significantly more forward-looking statements (24.30%) than GPT-3.5-turbo (17.76%), likely reflecting improved detection of subtle forward-looking language.

Panel B shows that GPT-3.5-turbo's Kappa scores average 0.93 (SD=0.01, range: 0.90-0.95) with mean run-pair agreement at 98.07% (SD=0.20%, IQR: 97.98%-98.19%). GPT-4o-mini shows superior performance with mean Kappa of 0.97 (SD=0.00) and run-pair agreement of 99.00% (SD=0.09%, IQR: 98.93%-99.06%). Panel C confirms that while GPT-3.5-turbo's mean document-wise agreement is 98.07% with 7.83% of documents showing any disagreement, GPT-4o-mini achieves 99.00% agreement with only 3.82% of documents receiving different classifications across runs.

Both models demonstrate remarkably high consistency for financial text analysis, with GPT-4o-mini showing statistically significant improvements across most metrics.

### B2 Multi-class Classification of FOMC Statements

Table B2 presents consistency analysis results for the multi-class classification of FOMC statements, examining 1,096 sentences classified 50 times by GPT-3.5-turbo and GPT-4o-mini into five categories (Hawkish, Mostly Hawkish, Neutral, Mostly Dovish, and Dovish).

Both models demonstrate high consistency, although lower than in the binary classification task. Fleiss' Kappa and Krippendorff's Alpha values are 0.91 for GPT-3.5-turbo and 0.86 for GPT-4o-mini (Panel A), with GPT-3.5-turbo showing 0.05 higher consistency in both metrics. Run-level metrics confirm this pattern. GPT-3.5-turbo achieves mean Kappa of 0.91 and run-pair agreement of 92.96%, compared to GPT-4o-mini's 0.86 and 88.93%, with these differences statistically significant ($p < 0.01$).

At the document level, GPT-3.5-turbo demonstrates 65.33% perfect agreement across all 50 runs versus GPT-4o-mini's 55.66%, a 9.67 percentage point difference. Mean document-wise agreement (92.96% vs. 88.93%) and majority class strength (95.32% vs. 92.26%) are significantly higher for GPT-3.5-turbo ($p < 0.01$). Mean classification uncertainty is lower for GPT-3.5-turbo (0.34 vs. 0.40, $p < 0.01$).

Class-specific analysis reveals both models struggle more with intermediate categories. GPT-3.5-turbo achieves highest agreement for "Neutral" (93.66%) and "Hawkish" (93.13%), with lower consistency for "Mostly Dovish" (91.34%) and "Dovish" (90.22%). GPT-4o-mini shows similar patterns with "Neutral" highest (91.83%) and "Mostly Hawkish" lowest (87.19%). GPT-3.5-turbo consistently outperforms GPT-4o-mini across all classes by 1.61 to 6.44 percentage points.

The models differ substantially in class distribution. GPT-4o-mini classifies more statements as "Dovish" (21.35% vs. 10.95%) and "Hawkish" (8.30% vs. 6.02%), while GPT-3.5-turbo assigns more to intermediate categories: "Mostly Dovish" (25.36% vs. 19.07%) and "Mostly Hawkish" (23.36% vs. 18.70%). Both classify approximately one-third as "Neutral" (34.31% vs. 32.57%).

Panel B details show GPT-3.5-turbo's Kappa scores average 0.91 (SD=0.06, range: 0.69-0.96) with run-pair agreement of 92.96% (SD=4.28%, IQR: 93.52%-94.62%). GPT-4o-mini shows lower consistency with mean Kappa of 0.86 (SD=0.06) and run-pair agreement of 88.93% (SD=4.51%), with minimum agreement dropping to 71.81% versus GPT-3.5-turbo's 77.28%. Panel C confirms GPT-3.5-turbo's superior document-level performance with mean agreement of 92.96% versus 88.93% and lower uncertainty in contested cases (0.34 vs. 0.40).

Unlike the binary classification task where GPT-4o-mini demonstrated superior consistency, GPT-3.5-turbo shows better performance across all metrics in this multi-class task. This reversal suggests GPT-4o-mini may be more sensitive to nuanced distinctions between closely related categories. While both models demonstrate high consistency, neither reaches the exceptional levels observed in binary classification, reflecting the inherently greater challenge of distinguishing between five closely related categories.

### B3 Sentiment Analysis

Table B3 presents sentiment analysis results across four financial text sources: news articles, MD&As, earnings call presentations, and earnings call Q&As.

Both models demonstrate exceptionally high consistency across all text types. GPT-4o-mini consistently outperforms GPT-3.5-turbo, with Fleiss' Kappa and Krippendorff's Alpha values ranging from 0.96-0.98 versus 0.94-0.96 (Panel A). The highest agreement metrics occur for earnings call presentations (0.98 for GPT-4o-mini), while news articles show the lowest (0.94 for GPT-3.5-turbo).

Run-level metrics confirm high consistency with mean run-pair agreement exceeding 96% for all text types and models. GPT-4o-mini achieves up to 98.71% for earnings call presentations. The difference between models is most pronounced for news articles (97.75% vs. 96.38%) and least notable for MD&A sections (97.97% vs. 96.92%), with all differences statistically significant.

At the document level, perfect agreement rates range from 81.09% (GPT-3.5-turbo on news) to 95.05% (GPT-4o-mini on presentations). GPT-4o-mini shows superior perfect agreement across all text types, with the largest advantage in news articles (88.28% vs. 81.09%). Majority class strength exceeds 97% for all combinations, indicating highly decisive classifications.

Class-specific analysis shows high agreement across all sentiment classes, with slightly higher consistency for negative sentiment (exceeding 96% except for Q&A with GPT-3.5-turbo). Neutral class agreement is consistently lowest among the three categories, although still above 95%, suggesting greater ambiguity in neutral classifications.

Class distribution varies significantly across text types and models. Earnings call presentations contain the highest proportion of positive statements (56.65-64.15%), while MD&A sections show the highest neutral content (51.38-55.93%). GPT-4o-mini consistently identifies more positive content than GPT-3.5-turbo across all text types, with the largest difference in Q&A sections (55.20% vs. 40.50%).

Panel B reveals narrow distributions of Kappa scores and run-pair agreements, particularly for GPT-4o-mini with standard deviations as low as 0.11 percentage points for MD&A run-pair agreement. Minimum agreement values remain high, with the lowest being 88.07% for news articles with GPT-3.5-turbo. Panel C confirms median document-wise agreement of 100% across all text types. Documents with any disagreement range from 99 (4.95%) for earnings call presentations with GPT-4o-mini to 915 (18.91%) for news articles with GPT-3.5-turbo. Classification uncertainty remains low across all text types and models.

Both models demonstrate exceptional consistency in sentiment analysis with agreement metrics well above conventional thresholds. GPT-4o-mini shows consistently superior reliability across all metrics and text types.

## B4 Summarization

### B4.1 Semantic Similarity

We use GPT-4o-mini to summarize MD&As from 10-Ks and presentation sections of earnings call transcripts. Since summarization aims to provide condensed versions that capture essential content, we assess consistency using cosine similarity based on context-aware embeddings from "jina-embeddings-v3."

Table B4 presents our semantic similarity analysis. GPT-4o-mini demonstrates remarkably high consistency, with mean similarity scores of 0.98 for both MD&A sections and earnings call presentations across run-pair and document-level measurements. The run-pair similarity (measuring consistency across 1,225 pairs of runs, each with 1,000 documents) shows minimal variation (standard deviation of 0.00), with minimum scores of 0.97 for both text types.

At the document level, we observe slightly more variation (standard deviations of 0.01 for MD&As and 0.02 for call presentations). Minimum similarity scores are 0.90 for MD&As and 0.88 for call presentations, still representing strong semantic consistency. These results indicate GPT-4o-mini produces highly consistent summaries for complex financial texts.

### B4.2 Length of Summaries

We assess consistency in summary word count, as prior studies use summary length to evaluate disclosure practices, with short summaries of long texts indicating bloated disclosure (Kim et al. 2024a).

Table B5 shows moderate consistency for MD&As (ICC2 of 0.66, mean Spearman correlation of 0.64), although concordance correlation is lower (0.43), indicating systematic differences between runs. The run-pair Mean Absolute Relative Differences (MARD) of 8.51% quantifies this variation. Earnings call presentations show higher consistency (ICC2 of 0.75, mean Spearman correlation of 0.76, run-pair MARD of 8.37%).

At the document level, virtually no MD&A documents (0.00%) receive identical word counts across runs, while 0.10% of earnings call presentations maintain identical lengths. Document-wise MARD values have notably higher standard deviations (3.38% for MD&As, 4.34% for earnings calls), indicating substantial variation across individual documents. While rank ordering remains stable, exact word counts vary moderately, suggesting that the models maintain similar judgments about relative information density but exhibit flexibility in conciseness.

### B4.3 Tone

We assess tone consistency using traditional keyword lists (Loughran and McDonald 2011) and FinBERT (Huang et al. 2023). Table B6 reports tone consistency for GPT-4o-mini summaries. For MD&As (Panel A), moderate consistency is observed, with FinBERT showing higher inter-rater agreement (Fleiss' Kappa and Krippendorff's Alpha of 0.68) compared to L&M keyword list (0.55). Mean run-pair agreement reaches 84.40% (FinBERT) and 73.15% (L&M).

For earnings call presentations (Panel B), tone consistency is significantly higher, with mean run-pair agreement of 95.55% (FinBERT) and 91.21% (L&M). Perfect agreement percentages rise to 82.20% (FinBERT) and 70.30% (L&M).

The class distribution reveals GPT-4o-mini-generated summaries are predominantly positive, especially for presentations (93.90% positive via FinBERT). This sentiment imbalance contributes to high consistency. Presentations show more stable tone classifications than MD&As due to both the model's positive framing tendency and more standardized language patterns. While GPT-4o-mini demonstrates reasonable tone consistency, content type and inherent positive framing may significantly influence consistency metrics in financial text analysis applications.

## B5 Text Generation

### B5.1 Semantic Similarity

We ask models to assume the role of company executives answering analyst questions during earnings calls. This is a more challenging consistency task than summarization due to greater response flexibility. We assess semantic consistency using cosine similarity with jina-embeddings-v3.

Table B7 shows GPT-4o-mini achieves higher mean similarity (0.97) than GPT-3.5-turbo (0.94) across metrics, with a significant 0.03-point difference. GPT-4o-mini demonstrates better minimum scores (0.96 vs. 0.93) and half the standard deviation at document level (0.02 vs. 0.04). Compared to summarization, lower mean similarities confirm question answering challenges consistency more, particularly for GPT-3.5-turbo (minimum document-level similarity: 0.46 vs. GPT-4o-mini's 0.76).

Figure B1 visualizes these differences: GPT-4o-mini's run-level similarity clusters tightly around 0.97 versus GPT-3.5-turbo's broader 0.94-0.95 distribution. The document-level comparison shows nearly all points above the diagonal, confirming GPT-4o-mini's consistently superior performance.

### B5.2 Length

Table B8 Panel B reveals GPT-4o-mini's superior length consistency (ICC2: 0.89, Spearman: 0.89, concordance: 0.79) versus GPT-3.5-turbo (ICC2: 0.76, Spearman: 0.78, concordance: 0.58). GPT-4o-mini's run-pair MARD is 8.18% compared to GPT-3.5-turbo's 14.91%, representing a 6.73 percentage point improvement. Document-wise MARD variability is also lower for GPT-4o-mini (SD: 3.52% vs. 8.50%).

Figure B2 shows non-overlapping MARD distributions, with GPT-4o-mini clustered around 8% and GPT-3.5-turbo near 15%. The scatter plot confirms GPT-4o-mini's consistently lower variation across documents.

### B5.3 Tone

Table B9 presents tone consistency using FinBERT and L&M dictionary. With FinBERT (Panel A), GPT-4o-mini achieves higher perfect agreement (67.00% vs. 49.55%) and mean run-pair agreement (90.64% vs. 85.34%) than GPT-3.5-turbo, despite similar inter-rater scores (0.66 vs. 0.65).

L&M dictionary (Panel B) shows lower consistency for both models (Fleiss' Kappa: 0.57 vs. 0.56), with perfect agreement dropping to 44.10% and 33.70%, respectively. Both models generate predominantly positive content, although GPT-4o-mini produces more (85.70% vs. 73.65% with FinBERT).

FinBERT provides more consistent measurements than fixed word lists, with higher agreement metrics indicating better reproducibility. Context-aware tools like FinBERT better capture LLMs' contextual nuances and semantic variations than predefined vocabulary lists.

## B6 Prediction of Future Earnings

We assess consistency of future earnings predictions by GPT-3.5-turbo, GPT-4o-mini, and GPT-4o, examining both directional predictions (increase/decrease) and point estimates for next fiscal year's earnings.

### B6.1 Directional Prediction

Table B10 presents consistency analysis for directional predictions across 1,000 firm-years, each classified 50 times. All models demonstrate high consistency with Fleiss' Kappa and Krippendorff's Alpha values of 0.97 (GPT-3.5-turbo), 0.99 (GPT-4o-mini), and 0.98 (GPT-4o), exceeding the 0.8 reliability threshold.

Run-pair agreement ranges from 98.54% (GPT-3.5-turbo) to 99.33% (GPT-4o-mini). Document-level perfect agreement rates are 93.70%, 97.30%, and 94.90% respectively. Class distribution varies: GPT-4o predicts increases 54.20% of the time, GPT-4o-mini 45.10%, and GPT-3.5-turbo 52.80%. Notably, GPT-4o-mini outperforms the more advanced GPT-4o in several consistency metrics, suggesting model advancement does not necessarily correlate with increased consistency for this task.

### B6.2 Point Estimates

Table B11 shows varying consistency degrees across models. GPT-4o exhibits exceptional consistency with perfect scores (1.00) across ICC2, concordance correlation, and Spearman correlation, with mean run-pair MARD of only 2.05%, substantially lower than GPT-4o-mini's 6.18% and GPT-3.5-turbo's 9.89%.

GPT-4o-mini demonstrates strong consistency (ICC2: 0.86, mean Spearman: 0.97) but lower concordance correlation (0.76) with standard deviation of 0.15. GPT-3.5-turbo maintains respectable metrics (ICC2: 0.85, mean Spearman: 0.97) but shows lowest concordance correlation (0.75) with highest standard deviation (0.28).

At document level, GPT-4o produces 29.20% identical estimates across runs versus GPT-4o-mini's 32.80% and GPT-3.5-turbo's 10.90%. Document-wise MARD shows substantial variability differences: GPT-3.5-turbo shows highest mean (9.89%) with substantial variability (SD: 20.69%), GPT-4o-mini moderate variability (SD: 18.10%), and GPT-4o remarkably low variability (SD: 7.75%).

Figure B3 visualizes these differences. Panel A shows GPT-4o's CCC values clustered near 1.0, GPT-4o-mini's wider distribution around 0.8, and GPT-3.5-turbo's bimodal distribution with peaks at 0.9 and 0.4. Panel B's run-level MARD distributions show clear separation: GPT-4o centered at 2%, GPT-4o-mini at 6%, and GPT-3.5-turbo at 9-10% with a right tail beyond 15%.

Panel C's ECDF reveals GPT-4o's stochastic dominance, where approximately 80% of its documents have MARD below 10%, versus 70% for GPT-4o-mini and 60% for GPT-3.5-turbo. This demonstrates strong correlation between model sophistication and consistency in numerical predictions.

**Figure B1 Comparative Analysis of Model Consistency in Financial Q&A Generation**

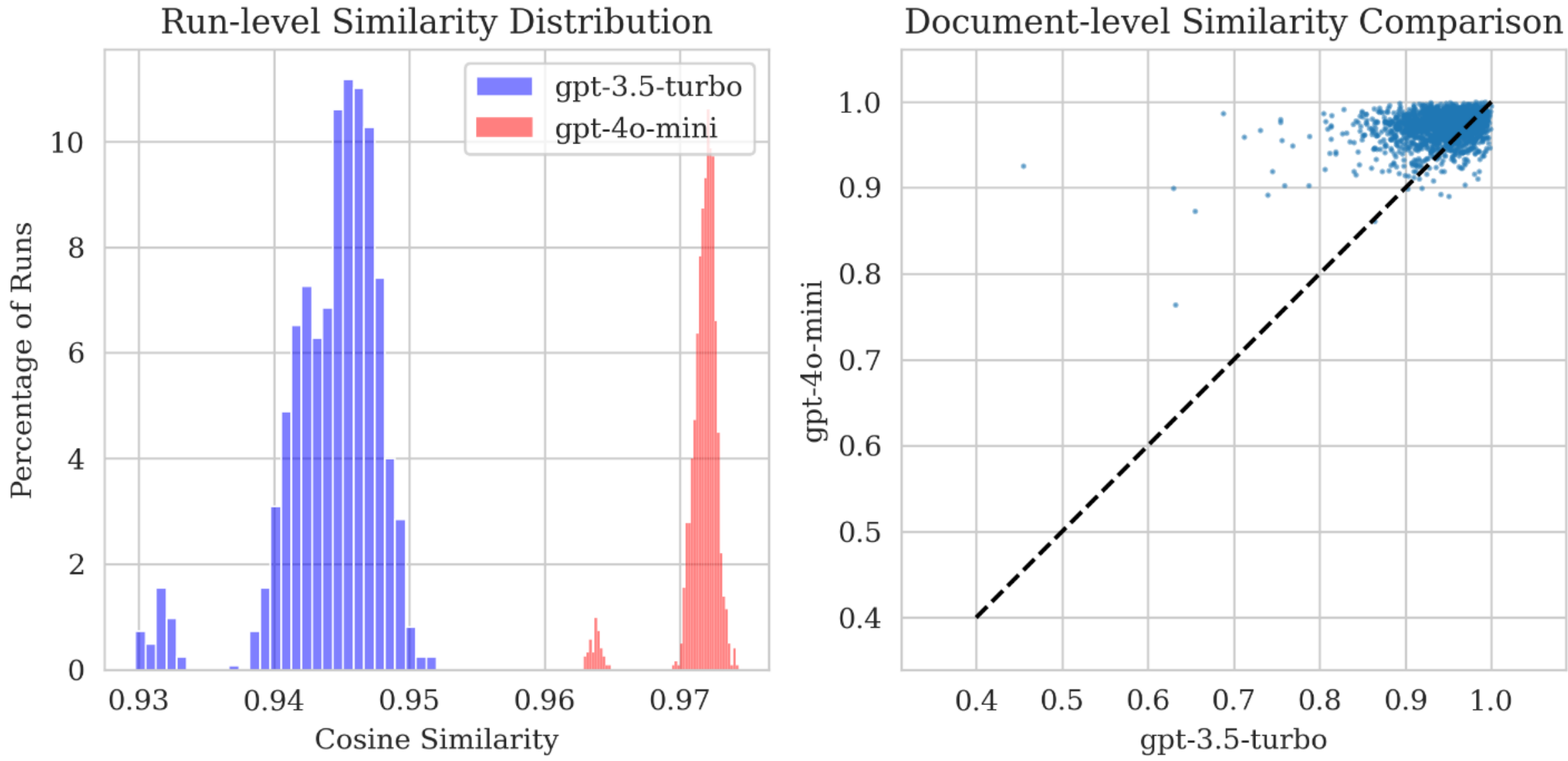

Note: The left panel displays the distribution of run-level similarity scores for both models. The right panel presents a document-by-document comparison where each point represents similarity scores for the same document generated by both models. Points above the dashed 45-degree line indicate higher consistency from GPT-4o-mini, which is the case for nearly all documents analyzed. See Appendix B for metric definitions.

**Figure B2 Comparative Analysis of Model Consistency in Length of Question Answering**

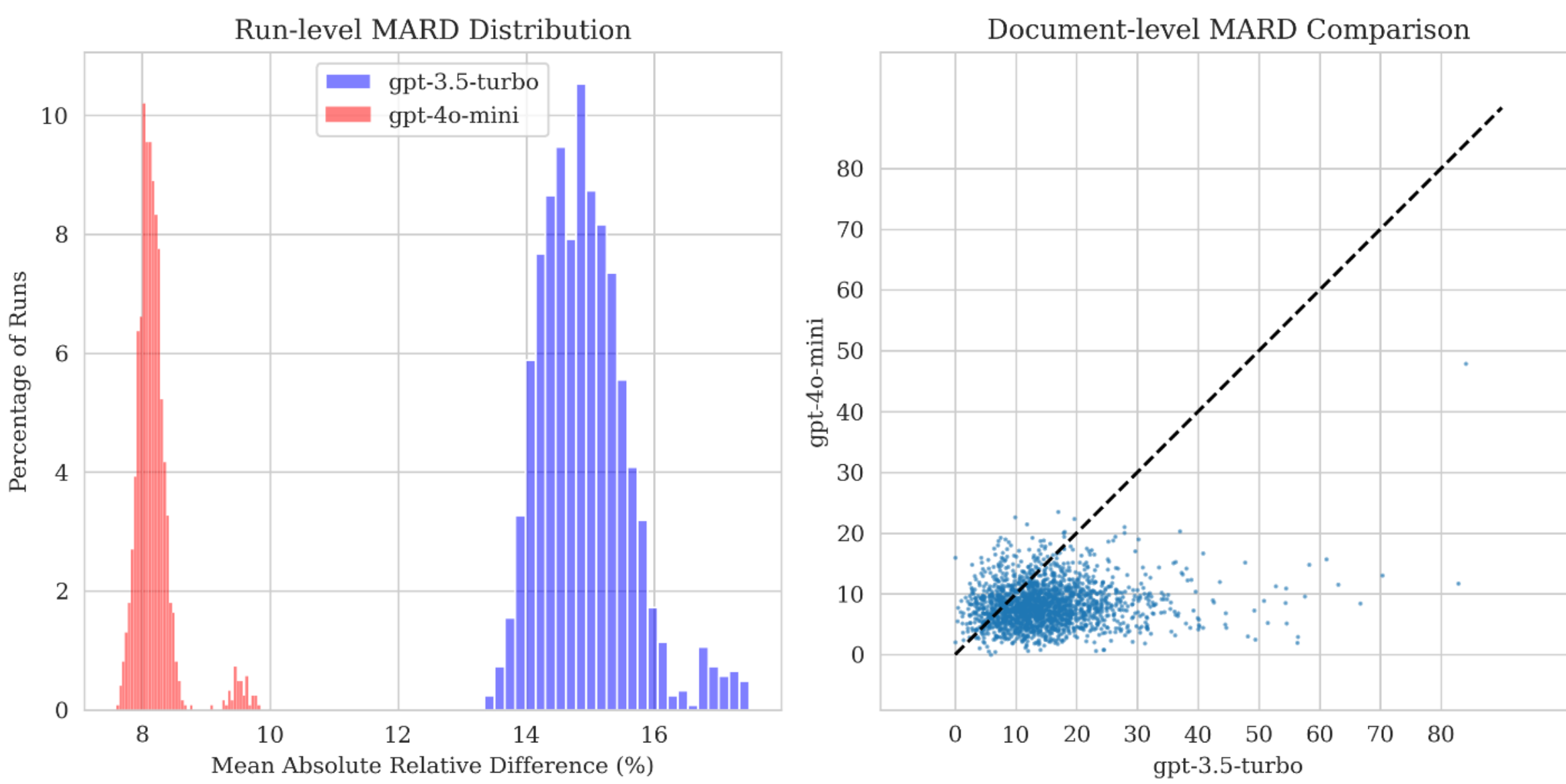

Note: The left panel shows the distribution of Mean Absolute Relative Difference (MARD) values for response lengths across multiple runs for GPT-3.5-turbo (blue) and GPT-4o-mini (red). The right panel displays document-level MARD comparisons, with each point representing a single document's MARD values for both models. Points below the diagonal indicate lower variation in GPT-4o-mini's response lengths compared to GPT-3.5-turbo for the same documents. See Appendix B for metric definitions.

**Figure B3 Consistency Analysis of Model Performance Across Multiple Runs for Prediction Task**

**Panel A: Concordance Correlation Coefficient (CCC) Distributions by Model**

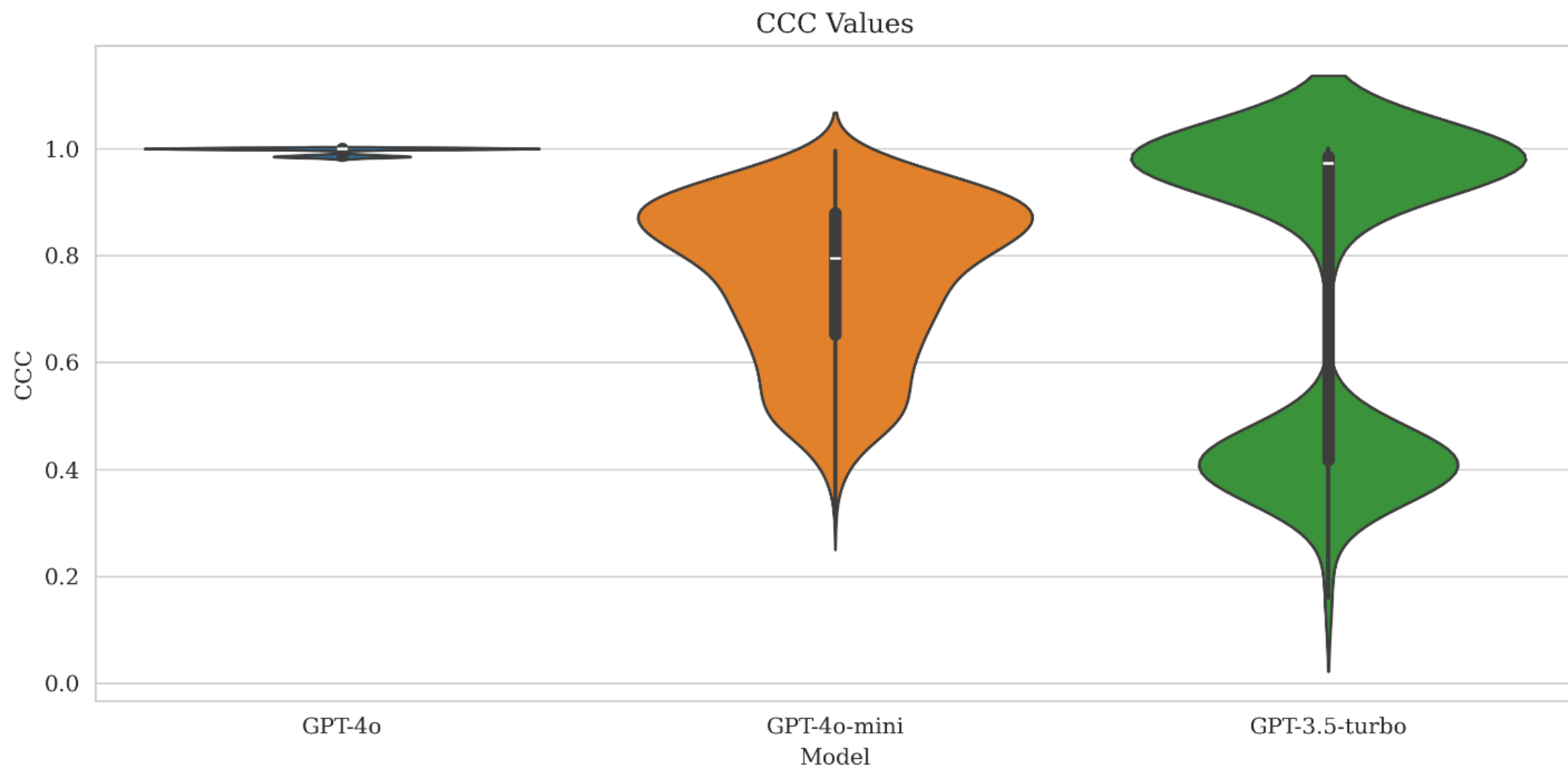

**Note:** Violin plots showing the distribution of pairwise concordance correlation coefficients between runs. GPT-4o exhibits near-perfect consistency (values clustered at 1.0), while GPT-4o-mini shows moderate variability. GPT-3.5-turbo displays a bimodal distribution, indicating two distinct performance patterns. See Appendix B for metric definitions.

**Panel B: Run-level Mean Absolute Relative Difference (MARD) Distributions**

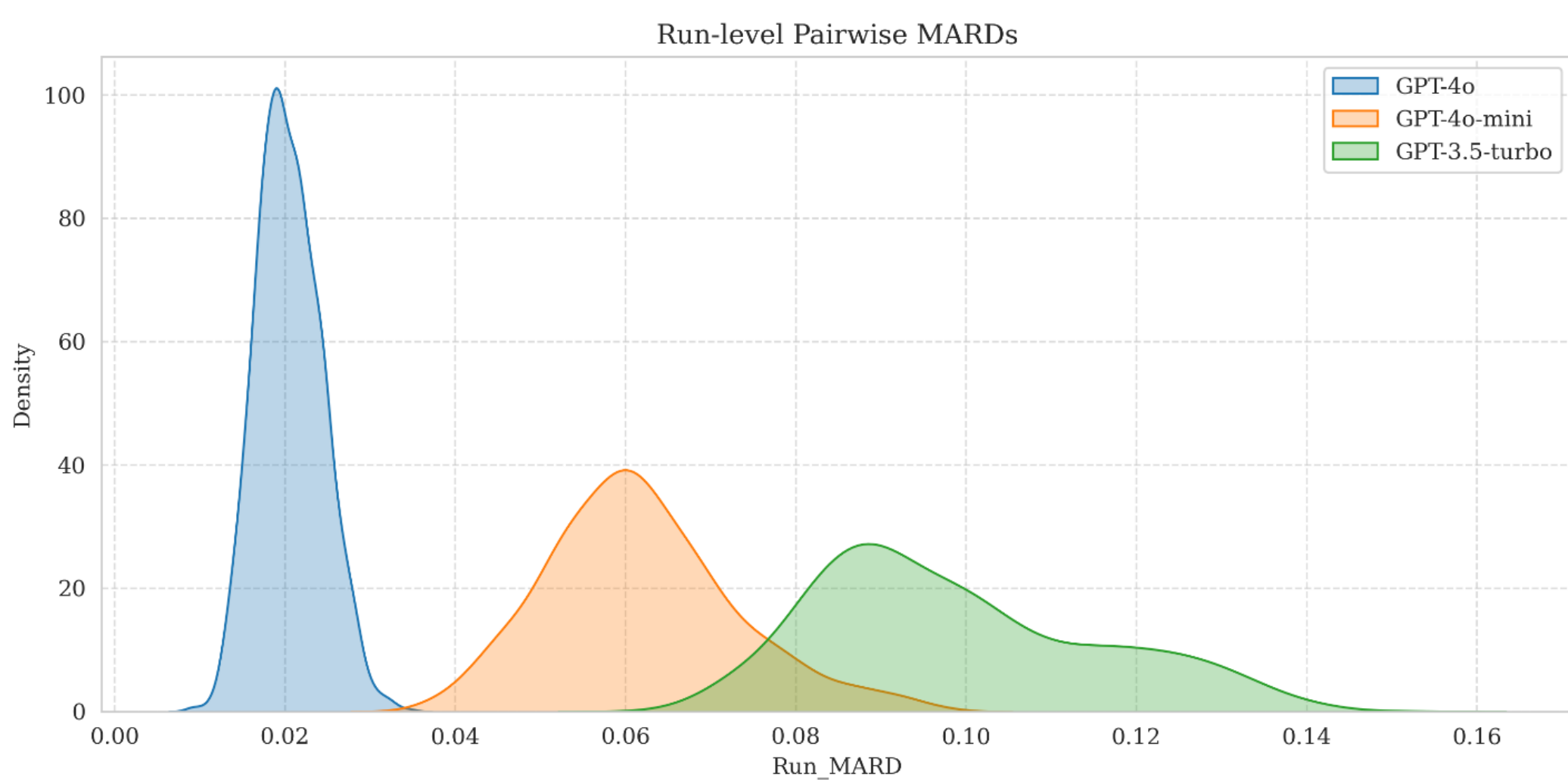

**Note:** Density plots of pairwise MARD between runs for each model. Lower MARD values indicate higher consistency. GPT-4o shows a narrow distribution centered around 2%, GPT-4o-mini around 6%, and GPT-3.5-turbo around 10%, demonstrating a clear progression in consistency with model sophistication. See Appendix B for metric definitions.

**Figure B3 Consistency Analysis of Model Performance Across Multiple Runs for Prediction (Continued)**

**Panel C: Empirical Cumulative Distribution Function (ECDF) of Document-level MARD**

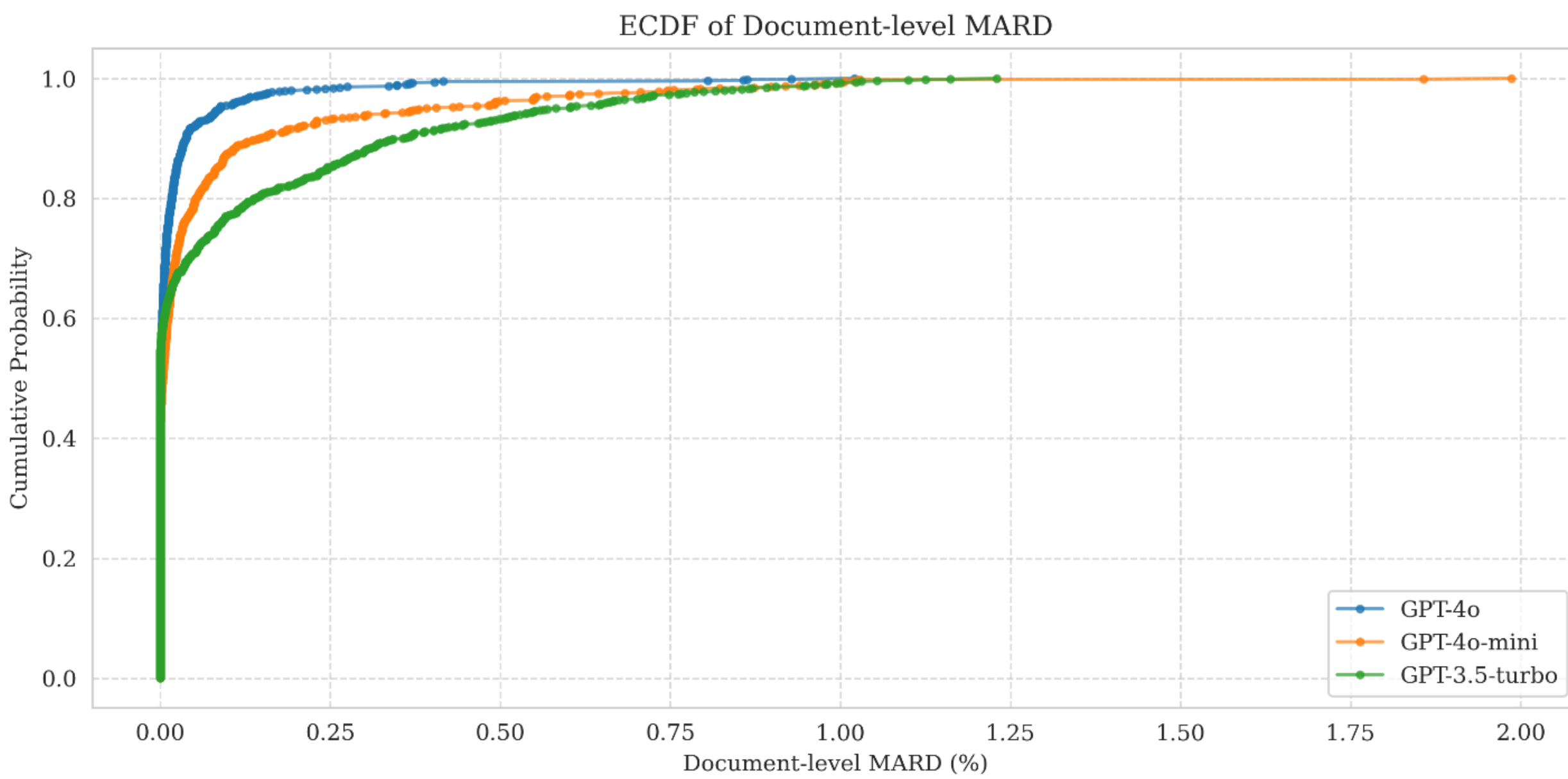

**Note:** ECDF curves showing the cumulative probability of document-level MARD values. The leftward shift of GPT-4o's curve demonstrates stochastic dominance, with approximately 80% of documents having MARD values below 10%, compared to 70% for GPT-4o-mini and 60% for GPT-3.5-turbo. See Appendix B for metric definitions.

**Table B1 Consistency Results for Binary Classification of Forward-looking Statements**

**Panel A: Overall Inter-rater and Document-level Agreements**

| Metric | GPT-4o-mini | GPT-3.5-turbo | Difference |
|---|---|---|---|
| **Sample Size** | | | |
| Number of documents | 10,000 | 10,000 | - |
| Number of runs | 50 | 50 | - |
| **Overall inter-rater agreements** | | | |
| Fleiss' Kappa | 0.97 | 0.93 | 0.04 |
| Krippendorff's Alpha | 0.97 | 0.93 | 0.04 |
| **Run-level agreements** | | | |
| Mean run-pair Cohen's Kappa | 0.97 | 0.93 | 0.04*** |
| Mean run-pair agreement (%) | 99.00 | 98.07 | 0.93*** |
| **Document-level agreements** | | | |
| Percentage of perfect agreement (%) | 96.18 | 92.17 | 4.01 |
| Mean document-wise agreement (%) | 99.00 | 98.07 | 0.93*** |
| Mean majority class strength (%) | 99.30 | 98.67 | 0.64*** |
| Mean classification uncertainty | 0.40 | 0.38 | 0.02 |
| **Class-specific Agreements** | | | |
| Forward-looking (%) | 97.36 | 94.00 | 3.36 |
| Non-forward looking (%) | 99.10 | 98.44 | 0.66 |
| **Class Distribution** | | | |
| Forward-looking (%) | 24.30 | 17.76 | 6.54 |
| Non-forward looking (%) | 75.70 | 82.24 | -6.54 |

Note: Five key metrics are tested for statistical differences between models (GPT-4o-mini minus GPT-3.5-turbo): Mean run-pair Cohen's Kappa, Mean run-pair agreement, Mean document-wise agreement, Mean majority class strength, and Mean classification uncertainty. Paired tests are used when measurements can be matched across models, while independent tests are used for aggregated metrics. Stars indicate statistical significance: *** $p<0.01$, ** $p<0.05$, * $p<0.10$. See Appendix B for metric definitions.

**Table B1 Consistency Results for Binary Classification of Forward-looking Statements (Continued)**

**Panel B: Detailed Run-level agreements**

| Model | Metric | N | Mean | Mdn | Std | Min | P25 | P75 | Max |
|---|---|---|---|---|---|---|---|---|---|
| GPT-3.5-turbo | Run-pair Cohen's Kappa score | 1225 | 0.93 | 0.93 | 0.01 | 0.90 | 0.93 | 0.94 | 0.95 |
| | Run-pair agreement (%) | 1225 | 98.07 | 98.09 | 0.20 | 97.14 | 97.98 | 98.19 | 98.52 |
| GPT-4o-mini | Run-pair Cohen's Kappa | 1225 | 0.97 | 0.97 | 0.00 | 0.96 | 0.97 | 0.97 | 0.98 |
| | Run-pair agreement (%) | 1225 | 99.00 | 99.00 | 0.09 | 98.66 | 98.93 | 99.06 | 99.25 |

Note: See Appendix B for metric definitions.

**Panel C: Detailed Document-level Agreements**

| Model | Metric | N | Mean | Mdn | Std | Min | P25 | P75 | Max |
|---|---|---|---|---|---|---|---|---|---|
| GPT-3.5-turbo | Document-wise agreement (%) | 10,000 | 98.07 | 100 | 8.11 | 48.98 | 100 | 100 | 100 |
| | Majority class strength (%) | 10,000 | 98.67 | 100 | 6.08 | 50.00 | 100 | 100 | 100 |
| | Classification uncertainty | 783 | 0.38 | 0.37 | 0.21 | 0.10 | 0.17 | 0.58 | 0.69 |
| GPT-4o-mini | Document-wise agreement (%) | 10,000 | 99.00 | 100 | 6.00 | 48.98 | 100 | 100 | 100 |
| | Majority class strength (%) | 10,000 | 99.30 | 100 | 4.47 | 50.00 | 100 | 100 | 100 |
| | Classification uncertainty | 382 | 0.40 | 0.40 | 0.21 | 0.10 | 0.17 | 0.61 | 0.69 |

Note: See Appendix B for metric definitions.

**Table B2 Consistency Results for Multi-class Classification of FOMC Statements**

**Panel A: Overall Inter-rater and Document-level Agreements**

| Metric | GPT-4o-mini | GPT-3.5-turbo | Difference |
|---|---|---|---|
| **Sample Size** | | | |
| Number of documents | 1,096 | 1,096 | - |
| Number of runs | 50 | 50 | - |
| **Overall inter-rater agreements** | | | |
| Fleiss' Kappa | 0.86 | 0.91 | -0.05 |
| Krippendorff's Alpha | 0.86 | 0.91 | -0.05 |
| **Run-level agreements** | | | |
| Mean run-pair Cohen's Kappa | 0.86 | 0.91 | -0.05*** |
| Mean run-pair agreement (%) | 88.93 | 92.96 | -4.03*** |
| **Document-level agreements** | | | |
| Percentage of perfect agreement (%) | 55.66 | 65.33 | -9.67 |
| Mean document-wise agreement (%) | 88.93 | 92.96 | -4.03*** |
| Mean majority class strength (%) | 92.26 | 95.32 | -3.06*** |
| Mean classification uncertainty | 0.40 | 0.34 | 0.06*** |
| **Class-specific Agreements** | | | |
| Hawkish (%) | 90.29 | 93.13 | -2.84 |
| Mostly Hawkish (%) | 87.19 | 92.74 | -5.55 |
| Neutral (%) | 91.83 | 93.66 | -1.83 |
| Mostly Dovish (%) | 84.90 | 91.34 | -6.44 |
| Dovish (%) | 88.61 | 90.22 | -1.61 |
| **Class Distribution** | | | |
| Hawkish (%) | 8.30 | 6.02 | 2.28 |
| Mostly Hawkish (%) | 18.70 | 23.36 | -4.65 |
| Neutral (%) | 32.57 | 34.31 | -1.73 |
| Mostly Dovish (%) | 19.07 | 25.36 | -6.30 |
| Dovish (%) | 21.35 | 10.95 | 10.40 |

Note: See the note from Table B1.

**Table B2 Consistency Results for Multi-class Classification of FOMC Statements (Continued)**

**Panel B: Detailed Run-level Agreements**

| Model | Metric | N | Mean | Mdn | Std | Min | P25 | P75 | Max |
|---|---|---|---|---|---|---|---|---|---|
| GPT-3.5-turbo | Run-pair Cohen's Kappa | 1,225 | 0.91 | 0.92 | 0.06 | 0.69 | 0.91 | 0.93 | 0.96 |
| | Run-pair agreement (%) | 1,225 | 92.96 | 94.07 | 4.28 | 77.28 | 93.52 | 94.62 | 96.81 |
| GPT-4o-mini | Run-pair Cohen's Kappa | 1,225 | 0.86 | 0.87 | 0.06 | 0.63 | 0.86 | 0.88 | 0.92 |
| | Run-pair agreement (%) | 1,225 | 88.93 | 90.15 | 4.51 | 71.81 | 89.32 | 90.78 | 93.80 |

Note: See Appendix B for metric definitions.

**Panel C: Detailed Document-level Agreements**

| Model | Metric | N | Mean | Mdn | Std | Min | P25 | P75 | Max |
|---|---|---|---|---|---|---|---|---|---|
| GPT-3.5-turbo | Document-wise agreement (%) | 1,096 | 92.96 | 100 | 13.80 | 36.24 | 92.16 | 100 | 100 |
| | Majority class strength (%) | 1,096 | 95.32 | 100 | 10.34 | 42.00 | 96.00 | 100 | 100 |
| | Classification uncertainty | 380 | 0.34 | 0.23 | 0.23 | 0.10 | 0.17 | 0.53 | 1.04 |
| GPT-4o-mini | Document-wise agreement (%) | 1,096 | 88.93 | 100 | 17.20 | 33.31 | 84.98 | 100 | 100 |
| | Majority class strength (%) | 1,096 | 92.26 | 100 | 13.49 | 42.00 | 92.00 | 100 | 100 |
| | Classification uncertainty | 486 | 0.40 | 0.37 | 0.24 | 0.10 | 0.17 | 0.63 | 1.14 |

Note: See Appendix B for metric definitions.

## Table B3 Consistency Results for Sentiment Analysis

### Panel A: Overall Inter-rater and Document-level Agreements

| | News | | MD&A | | Call Presentation | | Call Q&A | |
|---|---|---|---|---|---|---|---|---|
| **Metric** | **4oM** | **3.5T** | **4oM** | **3.5T** | **4oM** | **3.5T** | **4oM** | **3.5T** |
| **Sample Size** | | | | | | | | |
| Number of documents | 4,838 | 4,838 | 10,000 | 10,000 | 2,000 | 2,000 | 2,000 | 2,000 |
| Number of runs | 50 | 50 | 50 | 50 | 50 | 50 | 50 | 50 |
| **Overall inter-rater agreements** | | | | | | | | |
| Fleiss' Kappa | 0.96 | 0.94 | 0.97 | 0.95 | 0.98 | 0.96 | 0.97 | 0.95 |
| Krippendorff's Alpha | 0.96 | 0.94 | 0.97 | 0.95 | 0.98 | 0.96 | 0.97 | 0.95 |
| **Run-level agreements** | | | | | | | | |
| Mean run-pair Cohen's Kappa | 0.96 | 0.94 | 0.97 | 0.95 | 0.98 | 0.96 | 0.97 | 0.95 |
| Mean run-pair agreement (%) | 97.75 | 96.38 | 97.97 | 96.92 | 98.71 | 97.47 | 98.55 | 97.49 |
| **Document-level agreements** | | | | | | | | |
| Percentage of perfect agreement (%) | 88.28 | 81.09 | 92.46 | 88.53 | 95.05 | 90.65 | 93.80 | 90.10 |
| Mean document-wise agreement (%) | 97.75 | 96.38 | 97.97 | 96.92 | 98.71 | 97.47 | 98.55 | 97.49 |
| Mean majority class strength (%) | 98.48 | 97.60 | 98.57 | 97.82 | 99.09 | 98.23 | 99.01 | 98.24 |
| Mean classification uncertainty | 0.31 | 0.31 | 0.41 | 0.41 | 0.40 | 0.42 | 0.37 | 0.39 |
| **Class-specific Agreements** | | | | | | | | |
| Positive (%) | 97.42 | 95.94 | 97.30 | 95.27 | 99.15 | 97.74 | 98.80 | 97.24 |
| Neutral (%) | 96.74 | 95.44 | 97.59 | 96.62 | 96.52 | 95.21 | 97.39 | 96.97 |
| Negative (%) | 98.32 | 97.39 | 96.61 | 95.28 | 96.97 | 95.99 | 96.64 | 92.69 |
| **Class Distribution** | | | | | | | | |
| Positive (%) | 44.19 | 41.22 | 27.64 | 21.44 | 64.15 | 56.65 | 55.20 | 40.50 |
| Neutral (%) | 39.85 | 42.08 | 51.38 | 55.93 | 23.10 | 32.15 | 34.65 | 53.25 |
| Negative (%) | 15.96 | 16.70 | 20.98 | 22.63 | 12.75 | 11.20 | 10.15 | 6.25 |

Note: "News" refers to sentences from news articles, "MD&A" to sentences from MD&A sections, "Pre" to paragraphs from earnings call presentations, and "Q&A" to paragraphs from Q&A sections of earnings calls. "4oM" represents GPT-4o-mini, and "3.5T" represents GPT-3.5-turbo. For mean variables, all differences between models for the same text type are statistically significant ($p < 0.01$) except for mean classification uncertainty. See Appendix B for metric definitions.

**Table B3 Consistency Results for Sentiment Analysis (Continued)**

**Panel B: Detailed Run-level Agreements**

| Model | Metric | N | Mean | Mdn | Std | Min | P25 | P75 | Max |
|---|---|---|---|---|---|---|---|---|---|
| News - 3.5T | Run-pair Cohen's Kappa | 1,225 | 0.94 | 0.95 | 0.04 | 0.81 | 0.95 | 0.96 | 0.96 |
| | Run-pair agreement (%) | 1,225 | 96.38 | 97.04 | 2.32 | 88.07 | 96.86 | 97.19 | 97.81 |
| News - 4oM | Run-pair Cohen's Kappa | 1,225 | 0.96 | 0.97 | 0.02 | 0.88 | 0.97 | 0.97 | 0.98 |
| | Run-pair agreement (%) | 1,225 | 97.75 | 98.18 | 1.51 | 92.27 | 98.04 | 98.28 | 98.68 |
| MD&A- 3.5T | Run-pair Cohen's Kappa | 1,225 | 0.95 | 0.95 | 0.00 | 0.94 | 0.95 | 0.95 | 0.96 |
| | Run-pair agreement (%) | 1,225 | 96.92 | 96.92 | 0.18 | 96.20 | 96.81 | 97.03 | 97.43 |
| MD&A - 4oM | Run-pair Cohen's Kappa | 1,225 | 0.97 | 0.97 | 0.00 | 0.96 | 0.97 | 0.97 | 0.98 |
| | Run-pair agreement (%) | 1,225 | 97.97 | 97.97 | 0.11 | 97.57 | 97.89 | 98.05 | 98.47 |
| Pre- 3.5T | Run-pair Cohen's Kappa | 1,225 | 0.96 | 0.96 | 0.01 | 0.93 | 0.95 | 0.96 | 0.98 |
| | Run-pair agreement (%) | 1,225 | 97.47 | 97.50 | 0.46 | 96.10 | 97.15 | 97.80 | 98.65 |
| Pre- 4oM | Run-pair Cohen's Kappa | 1,225 | 0.98 | 0.97 | 0.01 | 0.96 | 0.97 | 0.98 | 0.99 |
| | Run-pair agreement (%) | 1,225 | 98.71 | 98.70 | 0.26 | 97.80 | 98.55 | 98.90 | 99.40 |
| Q&A- 3.5T | Run-pair Cohen's Kappa | 1,225 | 0.95 | 0.96 | 0.01 | 0.92 | 0.95 | 0.96 | 0.98 |
| | Run-pair agreement (%) | 1,225 | 97.49 | 97.60 | 0.55 | 95.40 | 97.25 | 97.85 | 98.65 |
| Q&A - 4oM | Run-pair Cohen's Kappa | 1,225 | 0.97 | 0.98 | 0.01 | 0.94 | 0.97 | 0.98 | 0.99 |
| | Run-pair agreement (%) | 1,225 | 98.55 | 98.60 | 0.38 | 96.75 | 98.40 | 98.80 | 99.60 |

Note: "News" refers to sentences from news articles, "MD&A" to sentences from MD&A sections, "Pre" to paragraphs from earnings call presentations, and "Q&A" to paragraphs from Q&A sections of earnings calls. "4oM" represents GPT-4o-mini, and "3.5T" represents GPT-3.5-turbo. N=1,225 represents all possible run pairs ((50 × 49)/2). P25 and P75 represent the 25th and 75th percentiles, respectively. See Appendix B for metric definitions.

**Table B3 Consistency Results for Sentiment Analysis (Continued)**

**Panel C: Detailed Document-level Agreements**

| Model | Metric | N | Mean | Mdn | Std | Min | P25 | P75 | Max |
|---|---|---|---|---|---|---|---|---|---|
| News-3.5T | Document-wise agreement (%) | 4,838 | 96.38 | 100 | 10.12 | 45.88 | 100 | 100 | 100 |
| | Majority class strength (%) | 4,838 | 97.60 | 100 | 7.53 | 50.00 | 100 | 100 | 100 |
| | Classification uncertainty | 915 | 0.31 | 0.23 | 0.20 | 0.10 | 0.17 | 0.47 | 0.83 |
| News-4oM | Document-wise agreement (%) | 4,838 | 97.75 | 100 | 8.31 | 48.98 | 100 | 100 | 100 |
| | Majority class strength (%) | 4,838 | 98.48 | 100 | 6.23 | 50.00 | 100 | 100 | 100 |
| | Classification uncertainty | 567 | 0.31 | 0.17 | 0.21 | 0.10 | 0.17 | 0.53 | 0.69 |
| MD&A-3.5T | Document-wise agreement (%) | 10,000 | 96.92 | 100 | 10.32 | 32.33 | 100 | 100 | 100 |
| | Majority class strength (%) | 10,000 | 97.82 | 100 | 7.92 | 38.00 | 100 | 100 | 100 |
| | Classification uncertainty | 1,147 | 0.41 | 0.44 | 0.22 | 0.10 | 0.23 | 0.61 | 1.09 |
| MD&A-4oM | Document-wise agreement (%) | 10,000 | 97.97 | 100 | 8.49 | 32.00 | 100 | 100 | 100 |
| | Majority class strength (%) | 10,000 | 98.57 | 100 | 6.47 | 34.00 | 100 | 100 | 100 |
| | Classification uncertainty | 754 | 0.41 | 0.44 | 0.21 | 0.10 | 0.23 | 0.61 | 1.10 |
| Pre-3.5T | Document-wise agreement (%) | 2,000 | 97.47 | 100 | 9.45 | 43.10 | 100 | 100 | 100 |
| | Majority class strength (%) | 2,000 | 98.23 | 100 | 7.10 | 50.00 | 100 | 100 | 100 |
| | Classification uncertainty | 187 | 0.42 | 0.44 | 0.22 | 0.10 | 0.17 | 0.62 | 0.90 |
| Pre-4oM | Document-wise agreement (%) | 2,000 | 98.71 | 100 | 6.86 | 33.14 | 100 | 100 | 100 |
| | Majority class strength (%) | 2,000 | 99.09 | 100 | 5.27 | 42.00 | 100 | 100 | 100 |
| | Classification uncertainty | 99 | 0.40 | 0.40 | 0.23 | 0.10 | 0.17 | 0.62 | 1.08 |
| Q&A-3.5T | Document-wise agreement (%) | 2,000 | 97.49 | 100 | 9.29 | 48.98 | 100 | 100 | 100 |
| | Majority class strength (%) | 2,000 | 98.24 | 100 | 7.07 | 50.00 | 100 | 100 | 100 |
| | Classification uncertainty | 198 | 0.39 | 0.40 | 0.22 | 0.10 | 0.17 | 0.61 | 0.69 |
| Q&A-4oM | Document-wise agreement (%) | 2,000 | 98.55 | 100 | 7.09 | 32.49 | 100 | 100 | 100 |
| | Majority class strength (%) | 2,000 | 99.01 | 100 | 5.31 | 38.00 | 100 | 100 | 100 |
| | Classification uncertainty | 124 | 0.37 | 0.33 | 0.23 | 0.10 | 0.17 | 0.59 | 1.09 |

Note: “News” refers to sentences from news articles, “MD&A” to sentences from MD&A sections, “Pre” to paragraphs from earnings call presentations, and “Q&A” to paragraphs from Q&A sections of earnings calls. “4oM” represents GPT-4o-mini, and “3.5T” represents GPT-3.5-turbo. Classification uncertainty is calculated only for documents with disagreements. P25 and P75 represent the 25th and 75th percentiles, respectively. See Appendix B for metric definitions.

**Table B4 Semantic Similarity of Generated Summaries**

| Text | Metric | N | Mean | Mdn | Std | Min | P25 | P75 | Max |
|---|---|---|---|---|---|---|---|---|---|
| MD&A | Run-pair Similarity | 1,225 | 0.98 | 0.98 | 0.00 | 0.97 | 0.98 | 0.98 | 0.98 |
| MD&A | Document-level Similarity | 1,000 | 0.98 | 0.98 | 0.01 | 0.90 | 0.97 | 0.99 | 1.00 |
| Call presentations | Run-pair Similarity | 1,225 | 0.98 | 0.98 | 0.00 | 0.97 | 0.98 | 0.98 | 0.98 |
| Call presentations | Document-level Similarity | 1,000 | 0.98 | 0.98 | 0.02 | 0.88 | 0.97 | 0.99 | 1.00 |

Note: "Run-pair similarity" represents the similarity score between output from different model runs. Since each run consists of 1,000 documents, it is calculated as the average similarity across these 1,000 documents. "Document-level similarity" measures the average similarity between summaries of the same document across 1,225 run pairs. All similarity scores are calculated using cosine similarity of jina-embeddings-v3 embeddings. Higher values indicate greater semantic consistency between summaries. See Appendix B for metric definitions.

**Table B5 Consistency Results for Summary Lengths of MD&As and Earnings Call Presentations**

| Metric | MD&As | Earnings Call Presentations |
|---|---|---|
| **Run-level metrics** | | |
| ICC2 | 0.66 | 0.75 |
| Mean concordance correlation | 0.43 | 0.56 |
| Median concordance correlation | 0.43 | 0.56 |
| Std dev of concordance correlation | 0.03 | 0.03 |
| Mean Pearson correlation | 0.66 | 0.75 |
| Median Pearson correlation | 0.66 | 0.75 |
| Std dev of Pearson correlation | 0.02 | 0.02 |
| Mean Spearman correlation | 0.64 | 0.76 |
| Median Spearman correlation | 0.64 | 0.76 |
| Std dev of Spearman correlation | 0.02 | 0.02 |
| Run-pair MARD (%) | 8.51 | 8.37 |
| Median run-pair MARD (%) | 8.48 | 8.37 |
| Std dev of run-pair MARD (%) | 0.28 | 0.35 |
| **Document-level metrics** | | |
| Documents with length (%) | 0.00 | 0.10 |
| Document-wise MARD (%) | 8.51 | 8.37 |
| Median document-wise MARD (%) | 8.20 | 7.85 |
| Std dev of document-wise MARD (%) | 3.38 | 4.34 |

Note: This table reports consistency results for summary lengths at two levels: (1) run-level metrics assess overall consistency across different model executions, and (2) document-level metrics evaluate consistency of summary lengths for individual documents across runs. See Appendix B for metric definitions.

**Table B6 Consistency Results for Tone of Summaries**

**Panel A: Summaries of MD&As**

| Metric | FinBERT | L&M | Difference |
|---|---|---|---|
| **Sample Size** | | | |
| Number of documents | 1,000 | 1,000 | |
| Number of runs | 50 | 50 | |
| **Overall inter-rater agreements** | | | |
| Fleiss' Kappa | 0.68 | 0.55 | -0.13 |
| Krippendorff's Alpha | 0.68 | 0.55 | -0.13 |
| **Run-level agreements** | | | |
| Mean run-pair Cohen's Kappa | 0.68 | 0.55 | -0.13*** |
| Mean run-pair agreement (%) | 84.40 | 73.15 | -11.25*** |
| **Document-level agreements** | | | |
| Percentage of perfect agreement (%) | 48.10 | 28.40 | -19.70 |
| Mean document-wise agreement (%) | 84.40 | 73.15 | -11.25*** |
| Mean majority class strength (%) | 89.03 | 80.67 | -8.35*** |
| Mean classification uncertainty | 0.48 | 0.62 | 0.14*** |
| **Class-specific Agreements** | | | |
| Positive (%) | 88.39 | 77.93 | -10.45 |
| Neutral (%) | 76.15 | 72.30 | -3.85 |
| Negative (%) | 81.19 | 74.13 | -7.05 |
| **Class Distribution** | | | |
| Positive (%) | 67.50 | 54.80 | -12.70 |
| Neutral (%) | 6.50 | 9.10 | 2.60 |
| Negative (%) | 26.00 | 36.10 | 10.10 |

Note: FinBERT indicates the tone is measured by using FinBERT, whereas L&M indicates that the tone is measured by using the Loughran & McDonald sentiment dictionary. See Table B1 for additional notes, and see Appendix B for metric definitions.

**Table B6 Consistency Results for Tone of Summaries (Continued)**

## Panel B: Summaries of Earnings Call Presentations

| Metric | FinBERT | L&M | Difference |
|---|---|---|---|
| **Sample Size** | | | |
| Number of documents | 1,000 | 1,000 | |
| Number of runs | 50 | 50 | |
| **Overall inter-rater agreements** | | | |
| Fleiss' Kappa | 0.65 | 0.59 | -0.07 |
| Krippendorff's Alpha | 0.65 | 0.59 | -0.07 |
| **Run-level agreements** | | | |
| Mean run-pair Cohen's Kappa | 0.65 | 0.59 | -0.07*** |
| Mean run-pair agreement (%) | 95.55 | 91.21 | -4.34*** |
| **Document-level agreements** | | | |
| Percentage of perfect agreement (%) | 82.20 | 70.30 | -11.90 |
| Mean document-wise agreement (%) | 95.55 | 91.21 | -4.34*** |
| Mean majority class strength (%) | 96.92 | 93.88 | -3.04*** |
| Mean classification uncertainty | 0.39 | 0.50 | 0.10*** |
| **Class-specific Agreements** | | | |
| Positive (%) | 96.88 | 93.86 | -3.02 |
| Neutral (%) | 84.67 | 79.03 | -5.64 |
| Negative (%) | 82.66 | 73.11 | -9.55 |
| **Class Distribution** | | | |
| Positive (%) | 93.90 | 90.70 | -3.20 |
| Neutral (%) | 1.70 | 4.10 | 2.40 |
| Negative (%) | 4.40 | 5.20 | 0.80 |

Note: FinBERT indicates the tone is measured by using FinBERT, whereas L&M indicates that the tone is measured by using the Loughran & McDonald sentiment dictionary. See Table B1 for additional notes, and see Appendix B for metric definitions.

**Table B7 Semantic Similarity of Q&A Answers**

| Model | Metric | N | Mean | Mdn | Std | Min | P25 | P75 | Max |
|---|---|---|---|---|---|---|---|---|---|
| GPT-3.5-turbo | Run-pair Similarity | 1,225 | 0.94 | 0.95 | 0.00 | 0.93 | 0.94 | 0.95 | 0.95 |
| GPT-3.5-turbo | Document-level Similarity | 2,000 | 0.94 | 0.95 | 0.04 | 0.46 | 0.93 | 0.97 | 1.00 |
| GPT-4o-mini | Run-pair Similarity | 1,225 | 0.97 | 0.97 | 0.00 | 0.96 | 0.97 | 0.97 | 0.97 |
| GPT-4o-mini | Document-level Similarity | 1,000 | 0.97 | 0.97 | 0.02 | 0.76 | 0.96 | 0.98 | 1.00 |

Note: The similarity measures are based on cosine similarity and are calculated in the same way as Table B4. The difference between the two models is statistically significant for both run-pair similarity and document-level similarity. The former is tested using an independent t-test and the latter using a paired t-test. See Appendix B for metric definitions.

**Table B8 Consistency Results for Lengths of Q&A Answers**

| Metric | GPT-4o-mini | GPT-3.5-turbo | Difference |
|---|---|---|---|
| **Run-level metrics** | | | |
| ICC2 | 0.89 | 0.76 | 0.13 |
| Mean concordance correlation | 0.79 | 0.58 | 0.21 |
| Median concordance correlation | 0.79 | 0.57 | 0.22 |
| Std dev of concordance correlation | 0.01 | 0.03 | -0.02 |
| Mean Pearson correlation | 0.89 | 0.76 | 0.13 |
| Median Pearson correlation | 0.89 | 0.76 | 0.13 |
| Std dev of Pearson correlation | 0.01 | 0.02 | -0.01 |
| Mean Spearman correlation | 0.89 | 0.78 | 0.10 |
| Median Spearman correlation | 0.89 | 0.78 | 0.10 |
| Std dev of Spearman correlation | 0.01 | 0.01 | -0.007 |
| Mean Run-pair MARD (%) | 8.18 | 14.91 | -6.73 |
| Median run-pair MARD (%) | 8.13 | 14.84 | -6.71 |
| Std dev of run-pair MARD (%) | 0.33 | 0.70 | -0.37 |
| **Document-level metrics** | | | |
| Documents with identical length (%) | 0.05 | 0.10 | -0.05 |
| Mean document-wise MARD (%) | 8.18 | 14.91 | -6.73 |
| Median document-wise MARD (%) | 7.83 | 13.56 | -5.72 |
| Std dev of document-wise MARD (%) | 3.52 | 8.50 | -4.99 |

Note: The “Difference” column shows the difference between “GPT-4o-mini” and “GPT-3.5-turbo”. See Appendix B for metric definitions.

**Table B9 Consistency Results for Tone of Q&A Answers**

**Panel A: Tone measured by FinBERT**

| Metric | GPT-4o-mini | GPT-3.5-turbo | Difference |
|---|---|---|---|
| **Sample Size** | | | |
| Number of documents | 2,000 | 2,000 | |
| Number of runs | 50 | 50 | |
| **Overall inter-rater agreements** | | | |
| Fleiss' Kappa | 0.66 | 0.65 | 0.01 |
| Krippendorff's Alpha | 0.66 | 0.65 | 0.01 |
| **Run-level agreements** | | | |
| Mean run-pair Cohen's Kappa | 0.66 | 0.65 | 0.01*** |
| Mean run-pair agreement (%) | 90.64 | 85.34 | 5.30*** |
| **Document-level agreements** | | | |
| Percentage of perfect agreement (%) | 67.00 | 49.55 | 17.45 |
| Mean document-wise agreement (%) | 90.64 | 85.34 | 5.30*** |
| Mean majority class strength (%) | 93.40 | 89.67 | 3.73*** |
| Mean classification uncertainty | 0.45 | 0.46 | -0.01 |
| **Class-specific Agreements** | | | |
| Positive (%) | 93.26 | 89.02 | 4.24 |
| Neutral (%) | 81.18 | 81.22 | -0.04 |
| Negative (%) | 79.00 | 76.97 | 2.03 |
| **Class Distribution** | | | |
| Positive (%) | 85.70 | 73.65 | 12.05 |
| Neutral (%) | 9.45 | 22.60 | -13.15 |
| Negative (%) | 4.85 | 3.75 | 1.10 |

Note: See Table B1 for additional notes, and see Appendix B for metric definitions.

**Table B9 Consistency Results for Tone of Q&A Answers (Continued)**

**Panel B: Tone Measured by L&M Dictionary**

| Metric | GPT-4o-mini | GPT-3.5-turbo | Difference |
|---|---|---|---|
| **Sample Size** | | | |
| Number of documents | 2,000 | 2,000 | |
| Number of runs | 50 | 50 | |
| **Overall inter-rater agreements** | | | |
| Fleiss' Kappa | 0.57 | 0.56 | 0.00 |
| Krippendorff's Alpha | 0.57 | 0.56 | 0.00 |
| **Run-level agreements** | | | |
| Mean run-pair Cohen's Kappa | 0.57 | 0.56 | 0.00** |
| Mean run-pair agreement (%) | 81.98 | 77.48 | 4.50*** |
| **Document-level agreements** | | | |
| Percentage of perfect agreement (%) | 44.10 | 33.70 | 10.40 |
| Mean document-wise agreement (%) | 81.98 | 77.48 | 4.50*** |
| Mean majority class strength (%) | 87.30 | 83.85 | 3.45*** |
| Mean classification uncertainty | 0.54 | 0.56 | -0.02 |
| **Class-specific Agreements** | | | |
| Positive (%) | 86.96 | 83.21 | 3.75 |
| Neutral (%) | 76.83 | 75.66 | 1.17 |
| Negative (%) | 73.49 | 71.52 | 1.96 |
| **Class Distribution** | | | |
| Positive (%) | 78.40 | 68.35 | 10.05 |
| Neutral (%) | 7.55 | 14.95 | -7.40 |
| Negative (%) | 14.05 | 16.70 | -2.65 |

Note: See Table B1 for additional notes, and see Appendix B for metric definitions.

**Table B10 Consistency Results for Directional Predictions of Future Earnings**

| Metric | GPT-4o | GPT-4o-mini | GPT-3.5-turbo |
|---|---|---|---|
| **Sample Size** | | | |
| Number of documents | 1,000 | 1,000 | 1,000 |
| Number of runs | 50 | 50 | 50 |
| **Overall inter-rater agreements** | | | |
| Fleiss' Kappa | 0.98 | 0.99 | 0.97 |
| Krippendorff's Alpha | 0.98 | 0.99 | 0.97 |
| **Run-level agreements** | | | |
| Mean run-pair Cohen's Kappa | 0.98 | 0.99 | 0.97 |
| Mean run-pair agreement (%) | 99.19 | 99.33 | 98.54 |
| **Document-level agreements** | | | |
| Percentage of perfect agreement (%) | 94.90 | 97.30 | 93.70 |
| Mean document-wise agreement (%) | 99.19 | 99.33 | 98.54 |
| Mean majority class strength (%) | 99.49 | 99.51 | 99.05 |
| Mean classification uncertainty | 0.27 | 0.39 | 0.37 |
| **Class-specific Agreements** | | | |
| Increase (%) | 99.09 | 98.95 | 98.32 |
| Decrease (%) | 98.95 | 99.13 | 98.10 |
| **Class Distribution** | | | |
| Increase (%) | 54.20 | 45.10 | 52.80 |
| Decrease (%) | 45.80 | 54.90 | 47.20 |

Note: See Table B1 for additional notes, and see Appendix B for metric definitions.

**Table B11 Consistency Results for Point Estimates of Future Earnings**

| Metric | GPT-4o | GPT-4o-mini) | GPT-3.5-turbo |
|---|---|---|---|
| **Run-level metrics** | | | |
| ICC2 | 0.997 | 0.864 | 0.85 |
| Mean concordance correlation | 0.995 | 0.759 | 0.75 |
| Median concordance correlation | 1.000 | 0.795 | 0.97 |
| Std dev of concordance correlation | 0.007 | 0.149 | 0.28 |
| Mean Pearson correlation | 0.997 | 0.868 | 0.88 |
| Median Pearson correlation | 1.000 | 0.892 | 0.99 |
| Std dev of Pearson correlation | 0.004 | 0.087 | 0.14 |
| Mean Spearman correlation | 0.999 | 0.971 | 0.97 |
| Median Spearman correlation | 0.999 | 0.971 | 0.97 |
| Std dev of Spearman correlation | 0.001 | 0.007 | 0.01 |
| Mean Run-pair MARD (%) | 2.053 | 6.176 | 9.89 |
| Median run-pair MARD (%) | 2.020 | 6.058 | 9.57 |
| Std dev of run-pair MARD (%) | 0.381 | 1.107 | 1.66 |
| **Document-level metrics** | | | |
| Documents with identical point estimate (%) | 29.200 | 32.800 | 10.90 |
| Mean document-wise MARD (%) | 2.053 | 6.421 | 9.89 |
| Median document-wise MARD (%) | 0.123 | 0.375 | 0.00 |
| Std dev of document-wise MARD (%) | 7.746 | 18.104 | 20.69 |

Note: See Table B1 for additional notes, and see Appendix B for metric definitions.